\RequirePackage{ifpdf}
\ifpdf
	\documentclass[twocolumn, prb, letterpaper, showkeys, showpacs, pdftex]{revtex4}
\else
	\documentclass[twocolumn, prb, letterpaper, showkeys, showpacs, dvips]{revtex4}
\fi

\usepackage{amsmath}
\usepackage[nodvipsnames]{color}

\ifpdf
  \usepackage[pdftex]{graphicx}
  \usepackage{subfigure}
  \graphicspath{{pdf-figures/}{pdf-figures/old/}}
  \usepackage[%
    pdftex,%
    pdftitle = {{Effect of lattice mismatch-induced strains on coupled diffusive and displacive phase transformations}},%
	pdfsubject = {{Materials which can undergo extremely fast displacive transformations as well as very slow diffusive transformations are studied using a phase-field framework to understand the physics behind microstructure formation and TTT diagrams.}},%
	pdfauthor = {{Mathieu Bouville and Rajeev Ahluwalia}},%
	pdfkeywords = {{austenite, martensite, pearlite, bainite, steel, spinodal decomposition, shape memory, phase transformation, TTT diagram, microstructures, phase-field, elastic compatibility}},%
    hyperindex,%
    bookmarksopen,%
    bookmarksopenlevel=2,%
	breaklinks,%
  ]{hyperref}
\else
  \usepackage[dvips]{graphicx}
  \usepackage{subfigure}
  \usepackage[
	dvips,
    pdftitle = {{Effect of lattice mismatch-induced strains on coupled diffusive and displacive phase transformations}},%
	pdfsubject = {{Materials which can undergo extremely fast displacive transformations as well as very slow diffusive transformations are studied using a phase-field framework to understand the physics behind microstructure formation and TTT diagrams.}},%
	pdfauthor = {{Mathieu Bouville and Rajeev Ahluwalia}},%
	pdfkeywords = {{austenite, martensite, pearlite, bainite, steel, spinodal decomposition, shape memory, phase transformation, TTT diagram, microstructures, phase-field, elastic compatibility}},%
	letterpaper,%
    hyperindex,%
    bookmarksopen,%
    bookmarksopenlevel=2,%
	breaklinks,%
]{hyperref}
\fi

\newcommand{\p}{\ensuremath{_\text{P}}}
\newcommand{\m}{\ensuremath{_\text{M}}}

\begin{document}
\title{Effect of lattice mismatch-induced strains on\\coupled diffusive and displacive phase transformations}

\author{Mathieu Bouville}
	\email{m-bouville@imre.a-star.edu.sg}
	\affiliation{Institute of Materials Research and Engineering, Singapore 117602}
	
	\author{Rajeev Ahluwalia}%
	\email{a-rajeev@imre.a-star.edu.sg}
	\affiliation{Institute of Materials Research and Engineering, Singapore 117602}
	
\date{\today}
\begin{abstract}
\addcontentsline{toc}{section}{Abstract}
Materials which can undergo slow diffusive transformations as well as fast displacive transformations are studied using the phase-field method. The model captures the essential features of the time--temperature--transformation (TTT) diagrams, continuous cooling transformation (CCT) diagrams, and microstructure formation of these alloys. In some materials systems there can exist an intrinsic volume change associated with these transformations. We show that these coherency strains can stabilize mixed microstructures (such as retained austenite--martensite and pearlite--martensite mixtures) by an interplay between diffusive and displacive mechanisms, which can alter TTT and CCT diagrams. Depending on the conditions there can be competitive or cooperative nucleation of the two kinds of phases. The model also shows that small differences in volume changes can have noticeable effects on the early stages of martensite formation and on the resulting microstructures.
\end{abstract}
\keywords{Ginzburg--Landau, martensite, pearlite, spinodal decomposition, shape memory, microstructures, TTT diagram, CCT diagram, elastic compatibility}
\pacs{
81.30.-t, 
81.30.Kf, 
81.40.-z, 
64.70.Kb, 
82.20.Wt} 
\maketitle

\section{Introduction}

Solid-to-solid phase transformations have been traditionally classified as diffusive and displacive, depending on the underlying kinetics.~\cite{Christian-book-02}
Diffusive phase transformations, such as spinodal decomposition and precipitation in alloys, are slow because they require long-range motion of the atoms.\cite{Gunton-83} In displacive phase transformations, on the other hand, the crystal structure changes through a unit cell distortion. Since atoms move over very short distances, these transformations can be very rapid. First order displacive transformations \mbox{---also} referred to as martensitic transformations--- are responsible for the shape-memory effect and for the pseudoelastic behavior.\cite{Bhattacharya-book,Otsuka-book} They are usually accompanied by a spontaneous strain. The high temperature cubic phase (austenite) transforms to a low temperature tetragonal, orthorhombic or monoclinic phase (martensite).\cite{Bhattacharya-book}

Diffusive and displacive phase transformations may interact or compete with each other. The best known material system where this interplay is observed is eutectoid steel. The high temperature austenite phase typically decomposes into pearlite, i.e.\ ferrite plus cementite (iron carbide, Fe$_3$C), by a diffusive process. 
For fast cooling rates, a transformation from austenite (face-centered cubic) to metastable martensite (base-centered tetragonal) may instead take place.

Interplay between diffusive and displacive transformations has also been observed in other materials systems, for instance Ti--Al--Nb,\cite{Ren-acta_mater-01} Cu--Al--Ag,\cite{Adorno-J_alloys_Comp-01} Cu--Zn--Al,\cite{Pons-Mater_Trans-93} and Pu--Ga alloys.\cite{Hecker} In NiTi shape-memory alloys, the formation of precipitates by a diffusive phase transformation influences the mechanical response and the stress-induced martensitic transformations.\cite{Gall-acta_mater-02}

Information on diffusive and displacive phase transformations and their kinetics can be conveniently represented on time--transformation--temperature (TTT) and continuous cooling transformation (CCT) diagrams: the volume fractions of the different phases or microstructures are plotted as a function of time at different temperatures. These diagrams account for meta\-stable phases such as martensite which do not appear in phase diagrams.

In order to describe the full microstructural complexity generated by the transformations, it is necessary to use numerical, rather than analytical, treatments.
The phase-field method (also known as the time-dependent Ginzburg--Landau method) is a powerful technique to describe the microstructures and kinetics of phase transformations. In particular, one needs not make any assumption regarding the arrangement of the different phases: microstructures are an output of the simulations.
It has been extensively used to study microstructural evolution in diffusive~\cite{Cahn-acta_met-61, Halperin-Rev_Mod_Phys-77, Bray-94, Onuki-PRL-01} and martensitic~\cite{Falk80, Artemev97, Wang97, Chen_Shen98, Onuki-99, Ahluwalia-prl-03, Ahluwalia-acta_mater-04, Ahluwalia-acta_mater-06, Vedantam-05} transformations. However, apart from one work on a TTT diagram for ferrite and martensite in iron,~\cite{Rao-prl-05} diffusive and displacive transformations have never been studied together using the phase-field method.

In a previous work\cite{Bouville-PRL-06} we introduced a phase-field model to study the interplay between diffusive and displacive transformations. TTT diagrams and microstructures were obtained. If the atomic volume of martensite is different from that of austenite or if there is a lattice mismatch due to phase separation, phase transformations create hydrostatic strain. An important conclusion from our previous work was that these intrinsic volume changes can stabilize mixed microstructures. 
 In the present article we systematically study the effect of such strains on microstructures, TTT diagrams, and CCT diagrams.

Section~\ref{sec-model} presents a phase-field model suitable for a system which can undergo phase separation as well as a square-\linebreak[3]to-\linebreak[3]rectangle martensitic transformation. In Sec.~\ref{sec-no_x12} the model is used to obtain TTT diagrams and microstructures in the absence of volume change. Sections~\ref{sec-x12} and~\ref{sec-x1c} focus on the effect on isothermal transformations of a volume change associated with martensite and pearlite formations, respectively. Section~\ref{CCT} presents continuous cooling results.

\section{\label{sec-model}Phase-field model}
Since the system can undergo both a phase separation and a  martensitic transformation, we consider three `phases': austenite, martensite (of which there exists two variants, elongated along $[0\,1]$ and along $[1\,0]$, respectively), and pearlite (of which there are two kinds, with compositions higher and lower than that of austenite). We therefore label `pearlite' {\itshape any} phase-separated region, even if the microstructure does not correspond to what one would typically call pearlite. Further, although the vocabulary used to designate phases may be reminiscent of steel, this is simply to avoid cumbersome periphrases. The model presented here is not specific to steel, neither is it claimed that it is a realistic representation of steel.

\subsection{Free energy}
We use two kinds of variables: composition and strains. $c$ is the composition, it is conserved. Since only its variations are relevant to diffusion, $c$ can be defined to a constant: we set $c=0$ as the composition of austenite and pearlite corresponds to $c = \pm c_0$.
$e_1$, $e_2$, and $e_3$ are non-conserved variables. $e_1$ is the hydrostatic strain, $e_2$ is the deviatoric strain, and $e_3$ is the shear strain:
\begin{subequations}
\label{def-ei}
\begin{align}
	e_1 &= (\varepsilon_{xx}+\varepsilon_{yy})/\sqrt{2},
	\label{def-e1}\\
	e_2 &= (\varepsilon_{xx}-\varepsilon_{yy})/\sqrt{2},
	\label{def-e2}\\
	e_3 &= \varepsilon_{xy}.
	\label{def-e3}
\end{align}
\end{subequations}
\noindent The $\{\varepsilon_{ij}\}$ are the linearized strain tensor components.

The free energy of the system is expressed as\cite{Bouville-PRL-06}
\begin{equation*}
	G=\int \left(g_\text{el} + g_\text{ch} + g_\text{cpl} + \frac{k_c}{2} \|\nabla c\|^2 + \frac{k_{e2}}{2} \|\nabla e_2\|^2 \right) \mathrm{d}\mathbf{r},
\end{equation*}
\noindent where $g_\text{el}$ is the non-linear elastic free energy density for a square-\linebreak[3]to-\linebreak[3]rectangle martensitic transition:\cite{Falk80, Onuki-99}
\begin{eqnarray}
	&g_\text{el} = &\dfrac{A_{22}}{2}\dfrac{T\!-T\m}{T\m}(e_2)^2 - \dfrac{A_{24}}{4}(e_2)^4 + \dfrac{A_{26}}{6}(e_2)^6 +\nonumber\\
	& & \dfrac{A_1}{2} \left\{ e_1 - \left[ x_{1c}\,c + x_{12}\,(e_2)^2  \right]\right\}^2 + \dfrac{A_3}{2} (e_3)^2.\quad
\label{eq-g_el}
\end{eqnarray}
\noindent Here $T$ is the dimensionless temperature and $T\m$ and $T\p$ are constants pertaining to the austenite--martensite and austenite--pearlite phase transformations respectively. The parameters $x_{12}$ and $x_{1c}$ are related to the volumetric strains coming from martensite and pearlite formations, respectively.~\footnote{In stress-free conditions, at equilibrium, $e_1 = x_{12}\,(e_2)^2 + x_{1c}\,c$. This expresses the fact that during the transformations, volume changes may be introduced, either due to a lattice mismatch between the phase separating components ($x_{1c}$) or due to the transformation strains ($x_{12}$). Substituting this value of $e_1$ into Eq.~(\ref{eq-g_el}) makes the $A_1$ term disappear. Therefore the transition temperatures are not affected by the parameters $x_{12}$ and $x_{1c}$. There is no `renormalization' of the transformation temperatures.}

The chemical free energy is given by\cite{Cahn-acta_met-61}
\begin{equation}
	g_\text{ch} = \dfrac{B_2}{2}\dfrac{T\!-T\p}{T\p}c^2 + \dfrac{B_4}{4}\, c^4
\label{eq-g_chem}
\end{equation}
\noindent and a coupling between elastic distortions and composition is introduced as
\begin{equation}
	g_\text{cpl} = x_{2c}\, c^2 (e_2)^2.
\label{eq-g_cpl}
\end{equation}

\begin{figure}
\centering
\setlength{\unitlength}{1cm}
\begin{picture}(8.5,3.15)(.2,0)
\subfigure
{
    \label{energy(a)}
    \includegraphics[height=3.15cm]{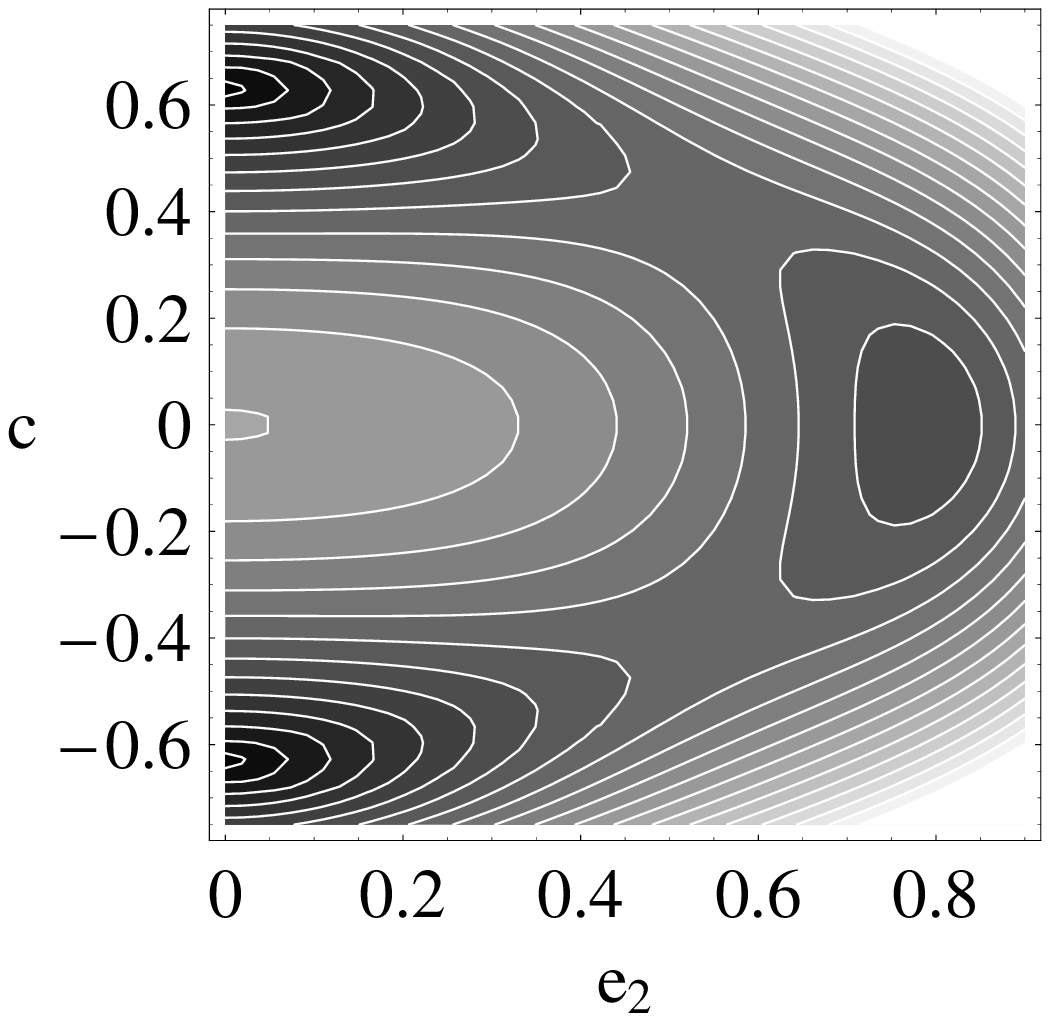}
	\put(-0.65, 2.75){\bf(a)}
}\subfigure{
    \label{energy(b)}
    \includegraphics[height=3.15cm]{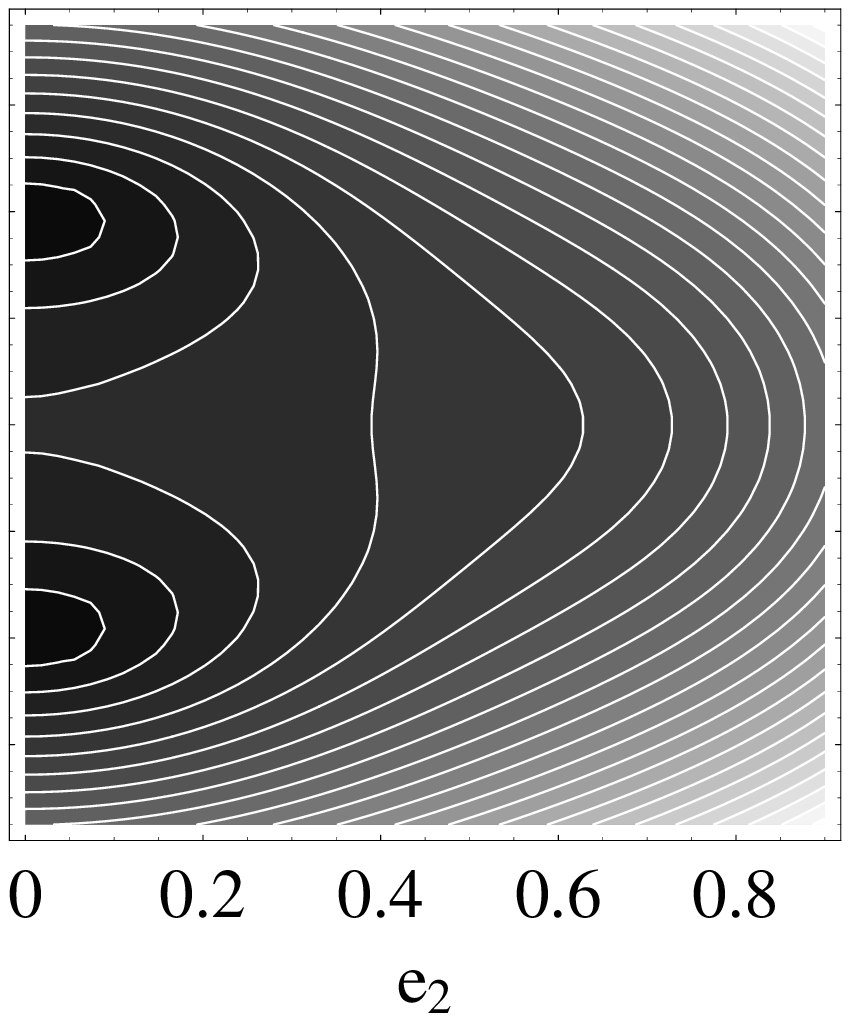}
	\put(-0.65, 2.75){\bf(b)}
}\subfigure{
    \label{energy(c)}
    \includegraphics[height=3.15cm]{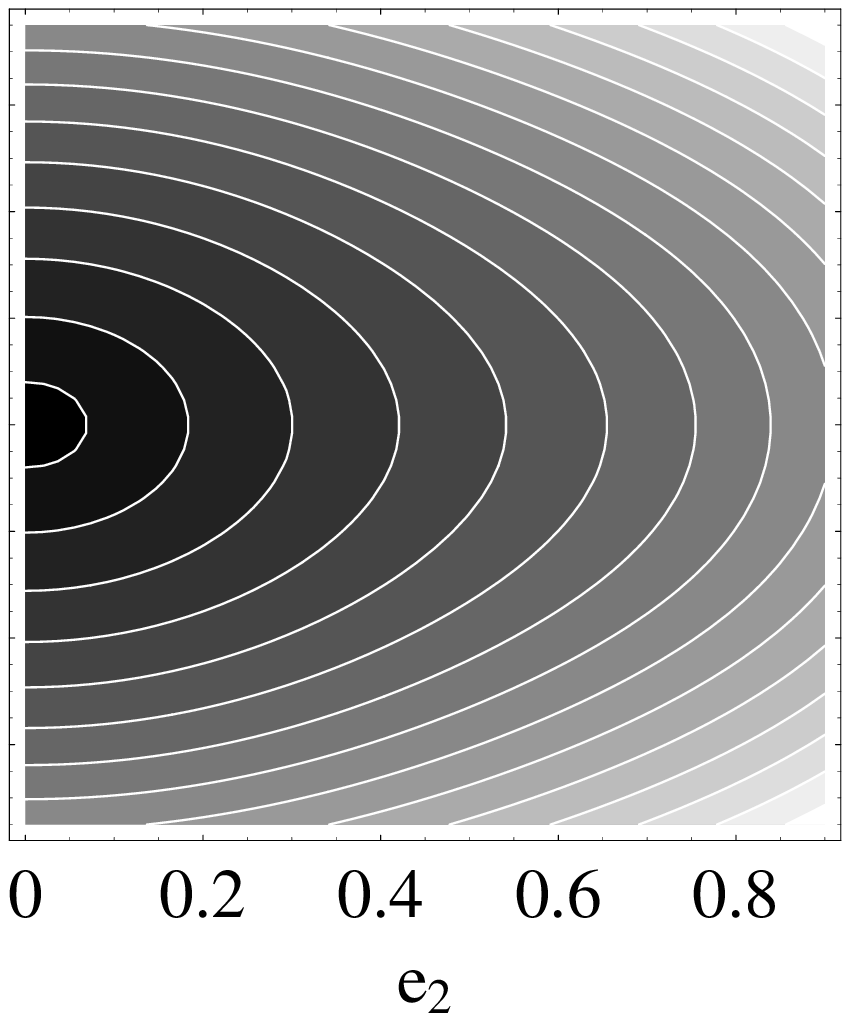}
	\put(-0.65, 2.75){\bf(c)}
}
\end{picture}
\caption{\label{energy}The bulk energy as a function of $e_2$ and $c$ at (a) $T = 0.2$, (b) $T = 0.7$, and (c) $T = 1.5$. Darker areas correspond to lower energies.}
\end{figure}

Since we are interested in a qualitative understanding of the physical mechanisms we choose simple values for the parameters. Yet we ensure that the energy of pearlite is always lower than that of martensite and that pearlite formation is slower than martensitic transformation by orders of magnitude. $A_1=1$, $A_{22}=2$, $A_{24}=4$, $A_{26}=9.6$, $A_3=1$, $B_2=6$, $B_4=12$, $x_{2c}=5$, $T\m=0.5$, $T\p = 1$, $k_c=2$, and $k_{e2}=0.1$.
The coupling constants $x_{12}$ and $x_{1c}$ are varied in different cases to understand the effect of volume changes: in Sec.~\ref{sec-no_x12} both are set to zero, Sec.~\ref{sec-x12} focuses on $x_{12} \ne 0$ and Sec.~\ref{sec-x1c} on $x_{1c} \ne 0$.

The homogeneous part of the free energy is depicted in Fig.~\ref{energy} as a function of $e_2$ and $c$ at different temperatures. 
The different phases can be identified as follows: austenite corresponds to $c=0$ and $e_2=0$, martensite to $c=0$ and $e_2\ne0$, and pearlite corresponds to $c\ne0$ and $e_2=0$. Above $T\p$ only austenite is stable, Fig.~\ref{energy(c)}. Between $T\m$ and $T\p$ austenite and martensite are unstable and pearlite is the ground state, Fig.~\ref{energy(b)}. Below $T\m$, pearlite is the ground state and martensite is metastable, Fig.~\ref{energy(a)}.

\begin{figure*}
\centering
\setlength{\unitlength}{1cm}
\begin{picture}(17.8, 11.7)(-.35,0)
\shortstack[c]{
\subfigure{
	\put(-0.4, 2.6){\rotatebox{90}{$x_{1c}=0$}}
	\put(4, 5.5){{$x_{12}=0$}}
    \label{128-0_0-TTT}
    \includegraphics[width=8.5cm]{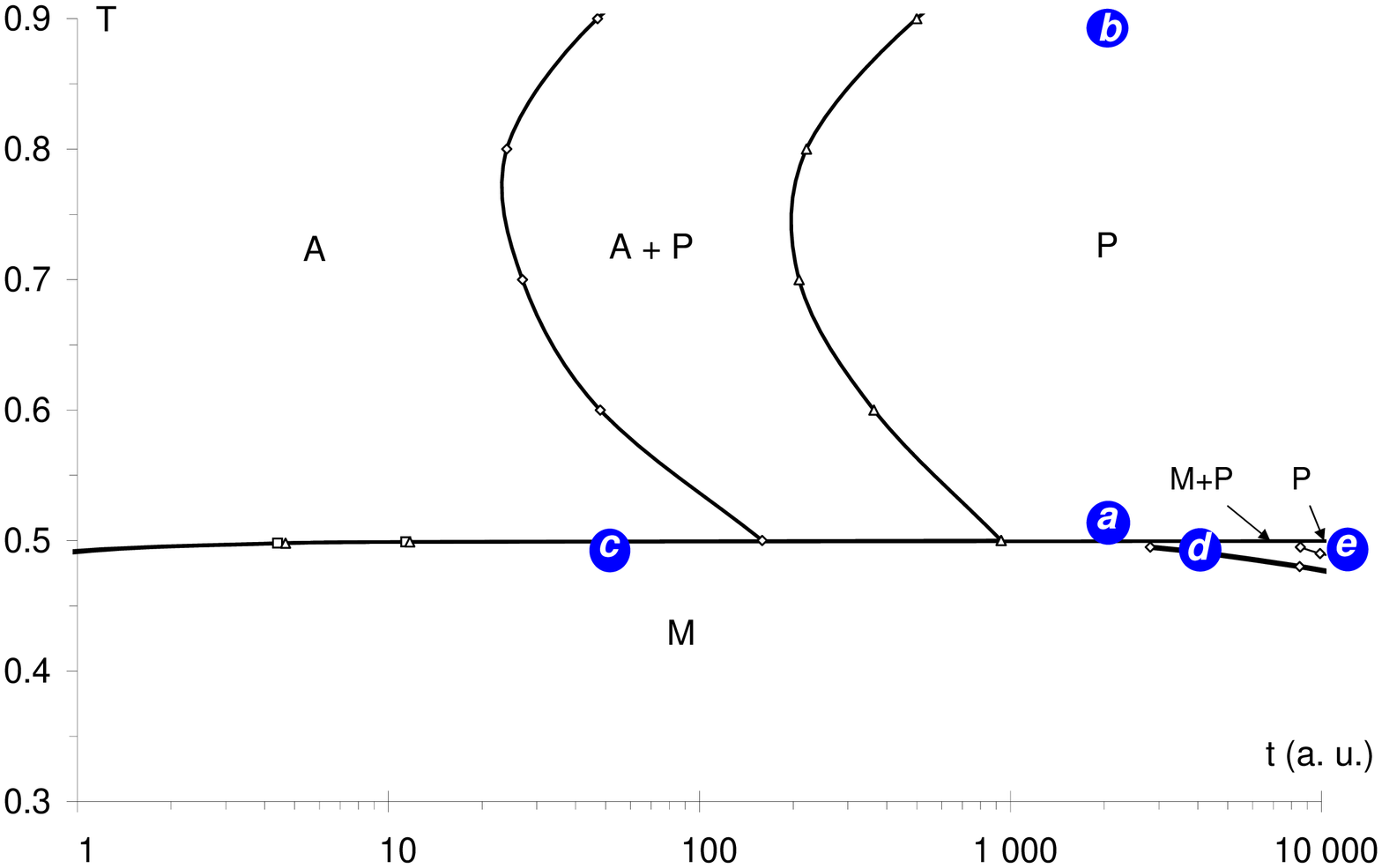}
	\put(-0.8, 4.9){\bf(a)}
}\quad
\subfigure{
	\put(4, 5.5){{$x_{12}=1$}}
    \label{128-1_0-TTT}
    \includegraphics[width=8.5cm]{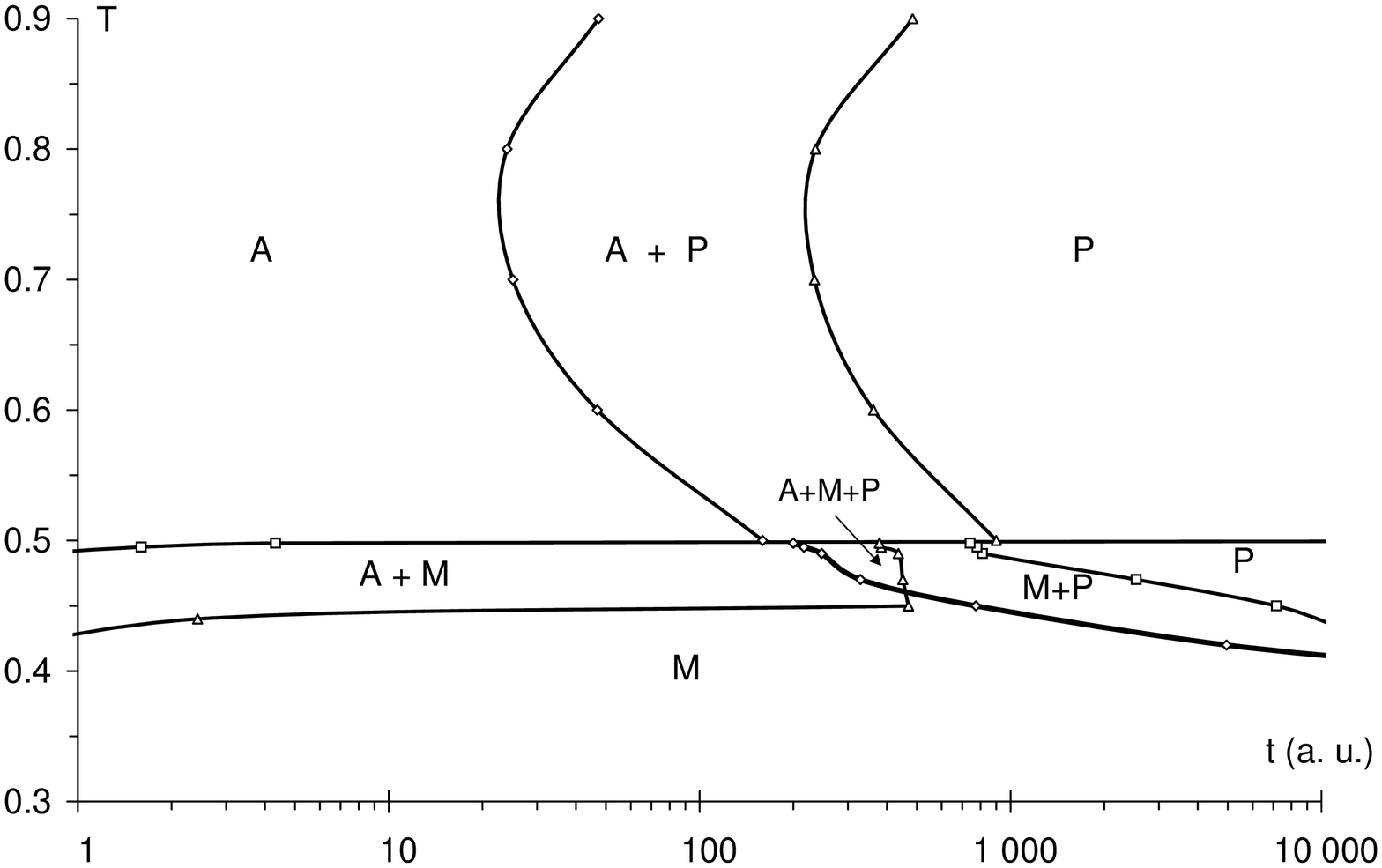}
	\put(-0.8, 4.9){\bf(b)}
}\\
\subfigure{
	\put(-0.4, 2.6){\rotatebox{90}{$x_{1c}=1$}}
    \label{128-0_1-TTT}
    \includegraphics[width=8.5cm]{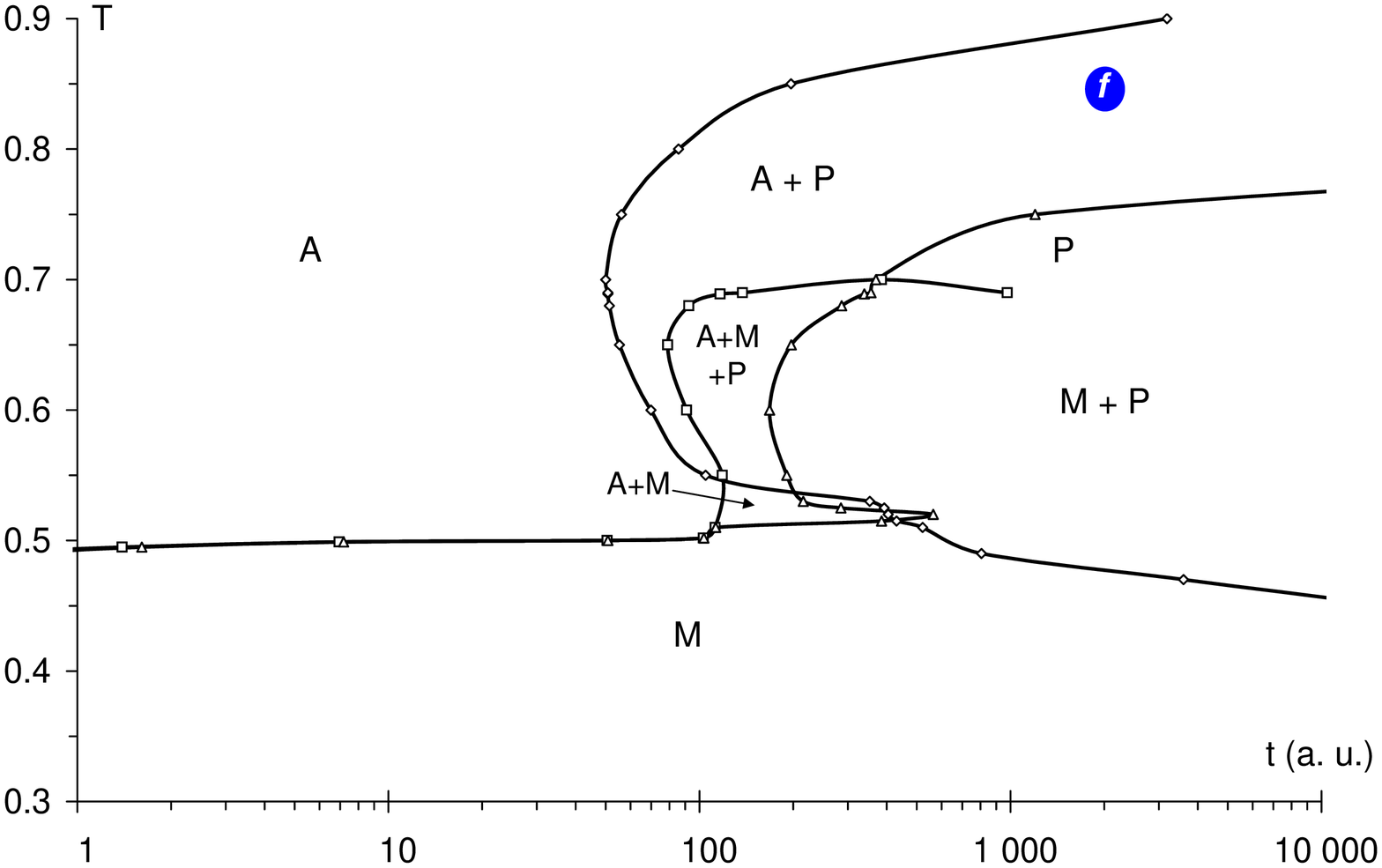}
	\put(-0.8, 4.9){\bf(c)}
}\quad
\subfigure{
    \label{128-1_1-TTT}
    \includegraphics[width=8.5cm]{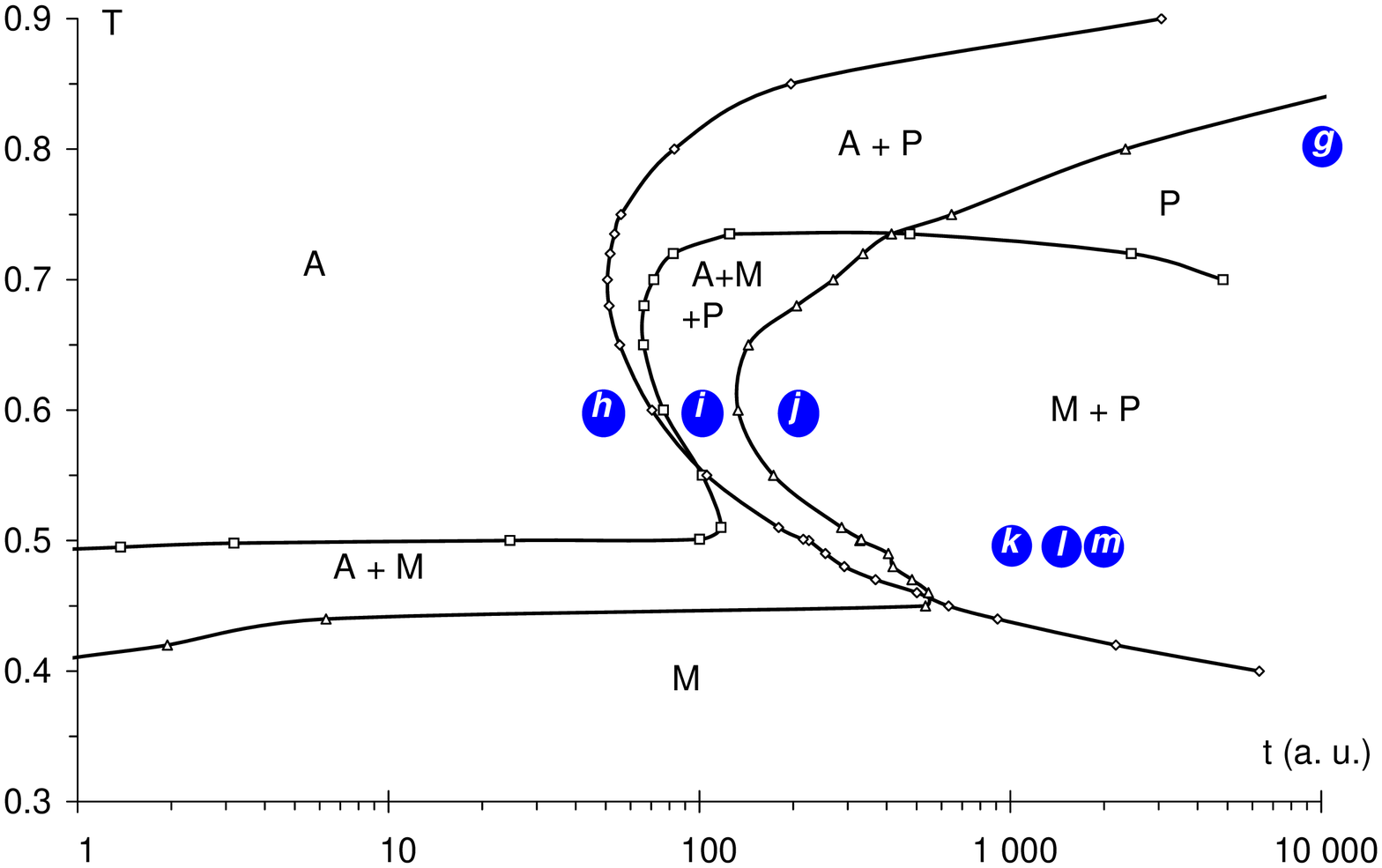}
	\put(-0.8, 4.9){\bf(d)}
}}
\end{picture}
	\caption{\label{128-TTT}Time--temperature--transformation diagrams for several values of $x_{12}$ 
and $x_{1c}$. 
 A: at least 10\% austenite, M: at least 10\% martensite, and P: at least 10\% pearlite. Points a--m correspond to the conditions at which the microstructures in Figs.~\ref{pearlites}, \ref{128-0_0-T0490}, \ref{128-1_1}, and~\ref{128-1_1-T0600} are obtained.}
\end{figure*}

\subsection{Evolution equations}
The evolution of the composition is described by the Cahn--Hillard equation~\cite{Cahn-acta_met-61}
\begin{equation}
	\frac{\partial\, c(\mathbf{r}, t)}{\partial\, t} = M\, \mathbf{\nabla}^2 \frac{\delta\, G}{\delta\, c(\mathbf{r}, t)},
\end{equation}
\noindent where $\delta$ is the functional derivative and $M$ is the temperature-dependent mobility,
\begin{equation}
	M = M_0 \,\exp(-Q/T).
\end{equation}
\noindent This temperature dependence of the mobility critically influences the TTT diagrams.

The evolution of the displacements is described by~\cite{landau-lifschitz}
\begin{subequations}
\label{du_dt2}
\begin{align}
	\rho \frac{\partial^2\, u_i(\mathbf{r}, t)}{\partial\, t^2} &= 
\sum_j\frac{\partial\, \sigma_{ij}(\mathbf{r}, t)}{\partial\, r_j} +
\eta \,\mathbf{\nabla}^2 v_i(\mathbf{r}, t),
\label{du_dt}\\
	\sigma_{ij}(\mathbf{r}, t) &= \frac{\delta\, G}{\delta\, \varepsilon_{ij}(\mathbf{r}, t)},
\end{align}
\end{subequations}
\noindent where $\rho$ is a density, $\{\sigma_{ij}\}$ are stresses, and $\mathbf{v}$ is the time derivative of the displacements, $\mathbf{u}$. The second term on the right-hand side in Eq.~(\ref{du_dt}) is a viscous damping term; it is a simplification of the more general damping of Ref.~\onlinecite{Jacobs-PRB-03}.

After the displacements are calculated from Eq.~(\ref{du_dt2}), the strains are obtained as their derivatives. Then $e_1$, $e_2$, and $e_3$ are obtained {\it via} Eq.~(\ref{def-ei}) and are used in Eqs.~(\ref{eq-g_el}), (\ref{eq-g_chem}), and~(\ref{eq-g_cpl}) to calculate the energy. The energy will in turn be used to evolve the displacements through Eq.~(\ref{du_dt2}) at the next time step.

All simulations are two-dimensional ($128 \times 128$ lattice) with periodic boundaries.\footnote{Larger systems are out of reach due to the time scales required by diffusive processes. Since (apart from Fig.~\ref{128-x_1-T0680}) the periodicity of the microstructures is reasonably smaller than the system size, one does not expect finite size effects to be dominant. We have reproduced some of the results on larger grids and found no noticeable difference.} A finite difference scheme with $\Delta x=1$ and $\Delta t= 0.2$ is used. We use $\eta = 0.01$, $\rho =1$, $M_0 = 2$, and $Q = 5$.
The initial system, made of 100\% austenite, includes random fluctuations around $c=0$ and $\mathbf{u}=0$. The system is quenched instantaneously to temperature $T$ and held at this temperature. For each value of $T$ we record the times at which 10\% martensite and 10\% pearlite form, as well as the time at which the austenite content drops below 10\%.

\section{\label{sec-no_x12}In the absence of volume change} 
In this section we consider the case of $x_{12}=0$ and $x_{1c}=0$, i.e.\ that there is no hydrostatic strain associated with either martensite or pearlite formation.

\subsection{TTT diagram}
Figure~\ref{128-0_0-TTT} shows the resulting TTT diagram. The austenite--pearlite phase transformation requires diffusion and therefore time. At low temperature, diffusion is slow and so is pearlite formation. At temperatures close to $T\p=1$ the driving force is small and pearlite formation again is slow. Consequently there exists an intermediate temperature at which pearlite formation is the fastest, this accounts for the C-curve around $T = 0.8$ in Fig.~\ref{128-0_0-TTT}. This is a typical feature of experimental TTT diagrams.

Above $T\m=0.5$, only pearlite is stable. Therefore it will necessarily form, albeit slowly. Below $T\m$, although pearlite is still the ground state, martensite is metastable. Unlike pearlite, martensite forms through a displacive mechanism, which does not require long-range motion of atoms. The phase transformations are then controlled by kinetics and the fast martensitic transformation takes place instead of the thermodynamically favorable (but slow) pearlite formation: Fig.~\ref{128-0_0-TTT} shows that below $T \approx 0.5$ (the `martensite start temperature') austenite transforms to martensite.
Notice that eventually pearlite will form even below $T=0.5$ as it is the ground state. Pearlite formation below the martensite start temperature has been observed experimentally.\cite{Boyer-book-77}

\begin{figure}
\centering
\setlength{\unitlength}{1cm}
\begin{picture}(8.5,2.05)(.2,0)
\subfigure
{
    \label{fine-pearlite}
    \includegraphics[height=2.05cm]{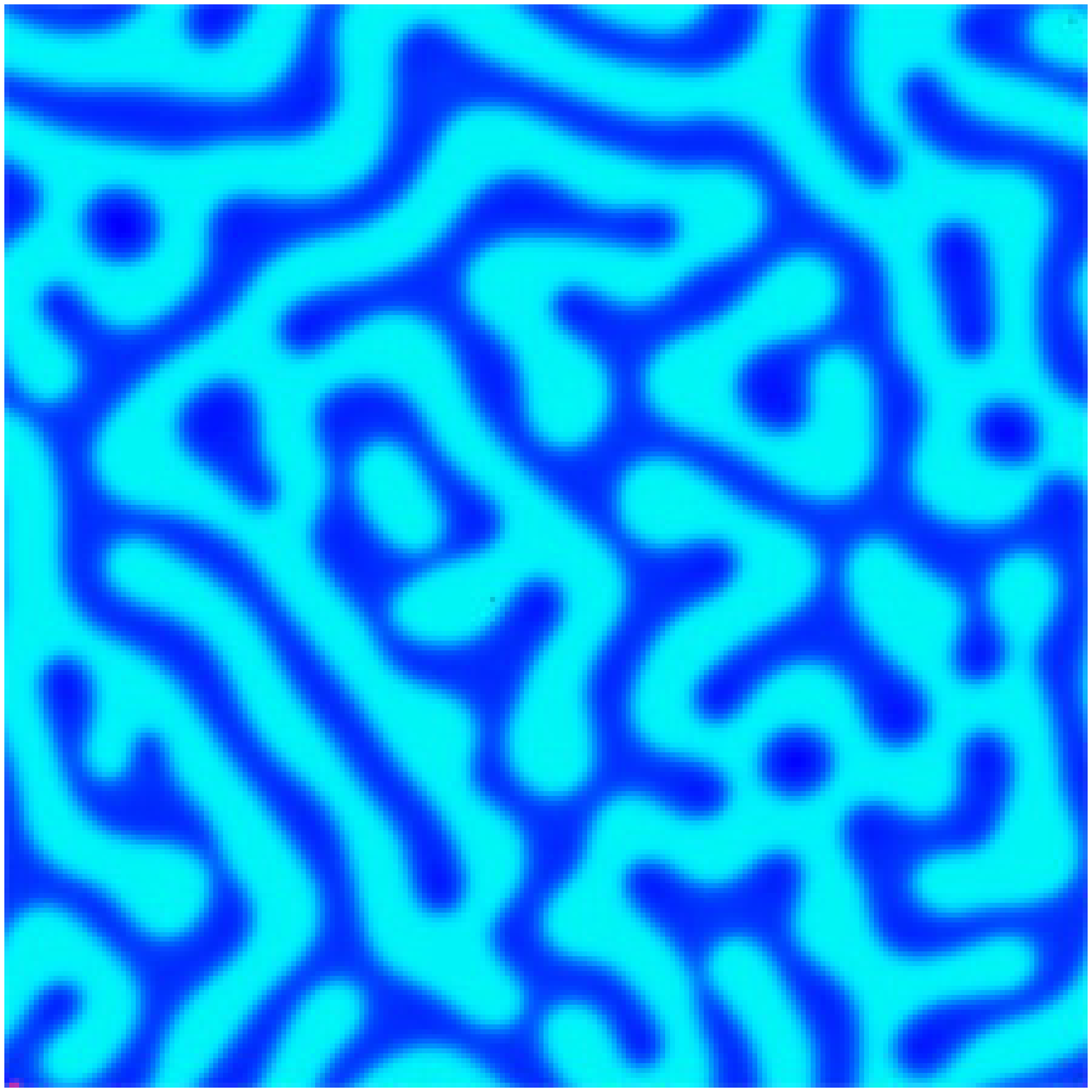}
	\put(-0.65, 1.6){\color{white}\bf(a)}
}\subfigure{
    \label{coarse-pearlite}
    \includegraphics[height=2.05cm]{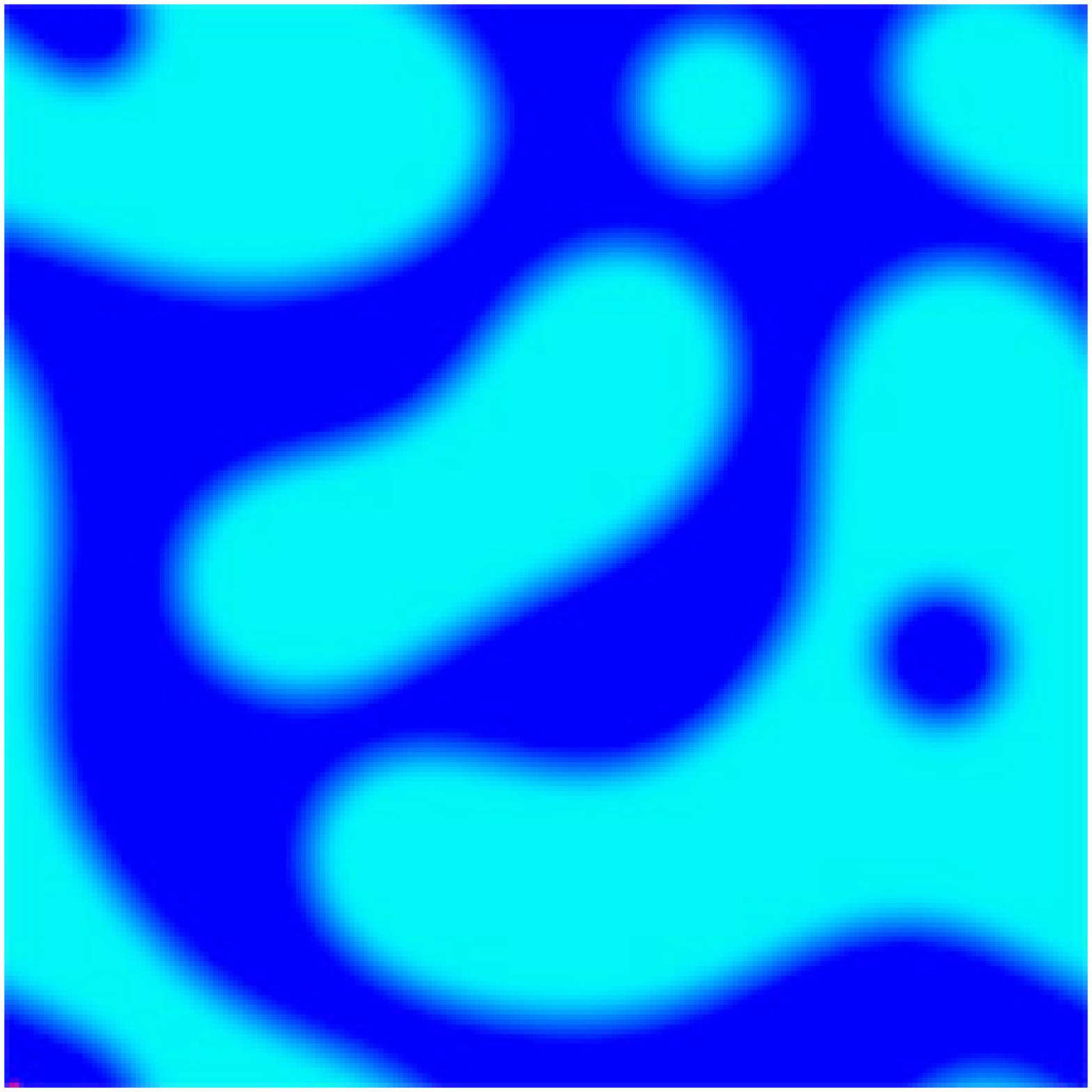}
	\put(-0.65, 1.6){\color{white}\bf(b)}
}\subfigure{
    \label{128-0_1-T0850_t002000} 
    \includegraphics[height=2.05cm]{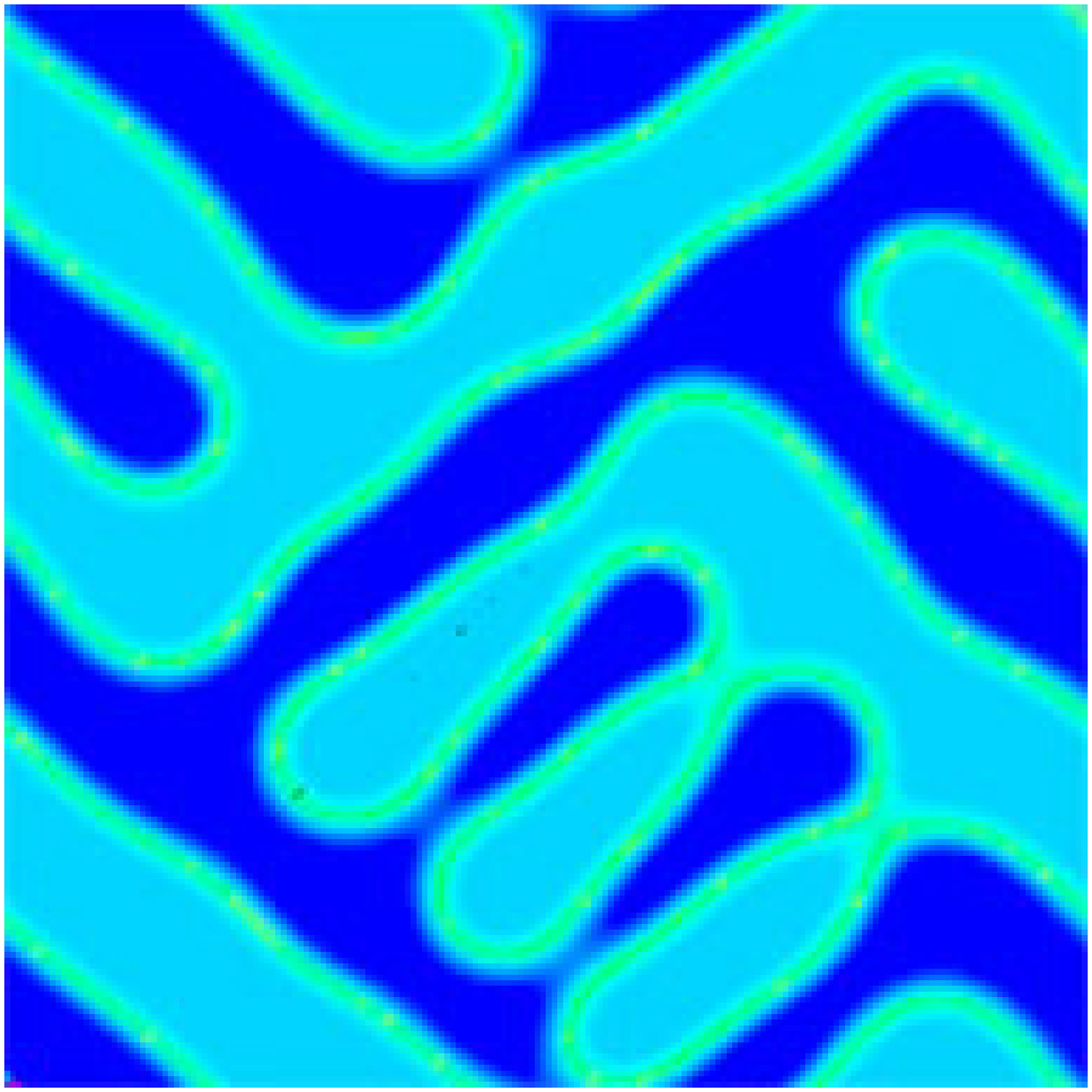}
	\put(-0.65, 1.6){\color{white}\bf(c)}
}\subfigure{
    \label{128-1_1-T0800_t010000}
    \includegraphics[height=2.05cm]{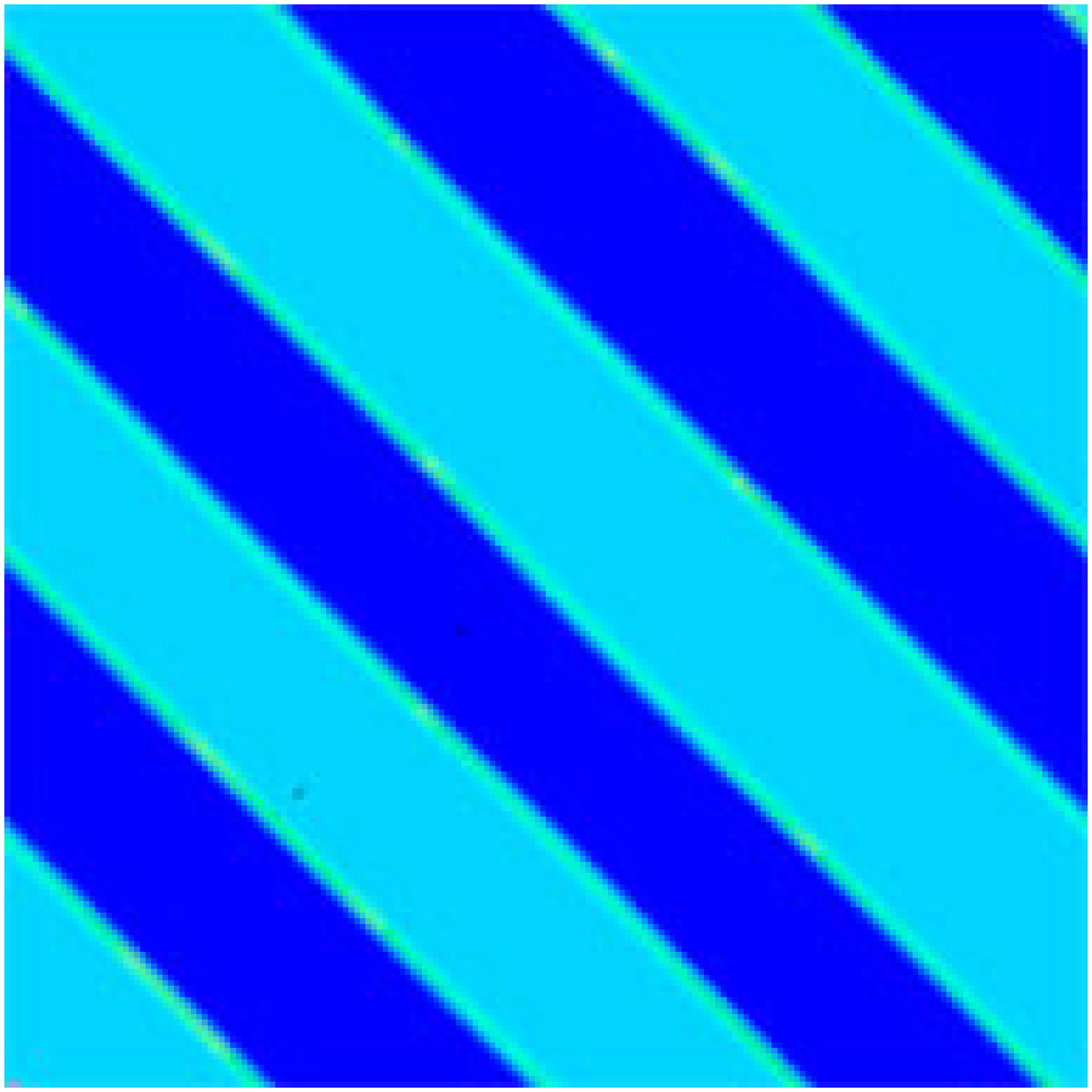}
	\put(-0.65, 1.6){\color{white}\bf(d)}
}
\end{picture}
	\caption{\label{pearlites}(color online) Pearlite microstructures (light blue: `cementite' and dark blue: `ferrite').
(a): $x_{12} = x_{1c} = 0$ at $T=0.5$ and $t=2~000$, point a in Fig.~\ref{128-0_0-TTT}. 
(b): $x_{12} = x_{1c} = 0$ at $T=0.9$ and $t=2~000$, point b in Fig.~\ref{128-0_0-TTT}.
(c): $x_{12} = 0$, $x_{1c} = 1$ at $T=0.85$ and $t=2~000$, point f in Fig.~\ref{128-0_1-TTT}.
(d): $x_{12} = x_{1c} = 1$ at $T=0.8$ and $t=10~000$, point g in Fig.~\ref{128-1_1-TTT}.}
\end{figure}

\subsection{Microstructure evolution}
The TTT diagram in Fig.~\ref{128-0_0-TTT} shows the existence of various microstructures at different temperatures and times.
The typical length scale of pearlite depends on the diffusion length and therefore on temperature: at low temperatures pearlite is fine \mbox{---Fig.~\ref{fine-pearlite}---} and it is coarser close to $T\p$ ---Fig.~\ref{coarse-pearlite}\mbox{---,} consistent with experimental observations. Note that the amounts of the two components of pearlite are equal because, as Fig.~\ref{energy} shows, their compositions are symmetric with respect to that of austenite.

\begin{figure}
\centering
\setlength{\unitlength}{1cm}
\begin{picture}(8.5,2.75)(.1,0)
\subfigure{
    \label{martensite}
    \includegraphics[height=2.75cm]{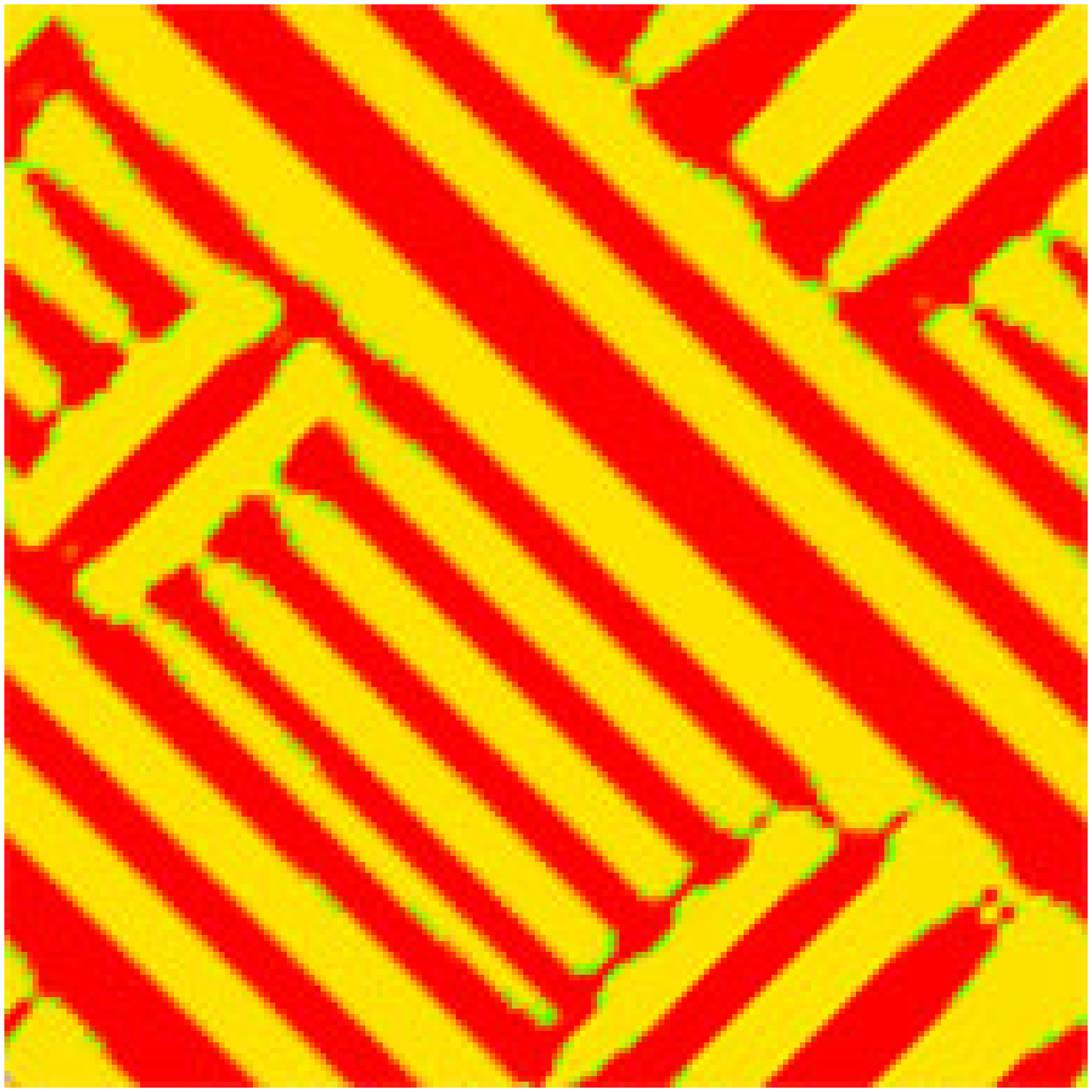}
	\put(-0.6, 2.3){\bf(a)}
}\subfigure{
    \label{128-0_0-T0490_t004000}
    \includegraphics[height=2.75cm]{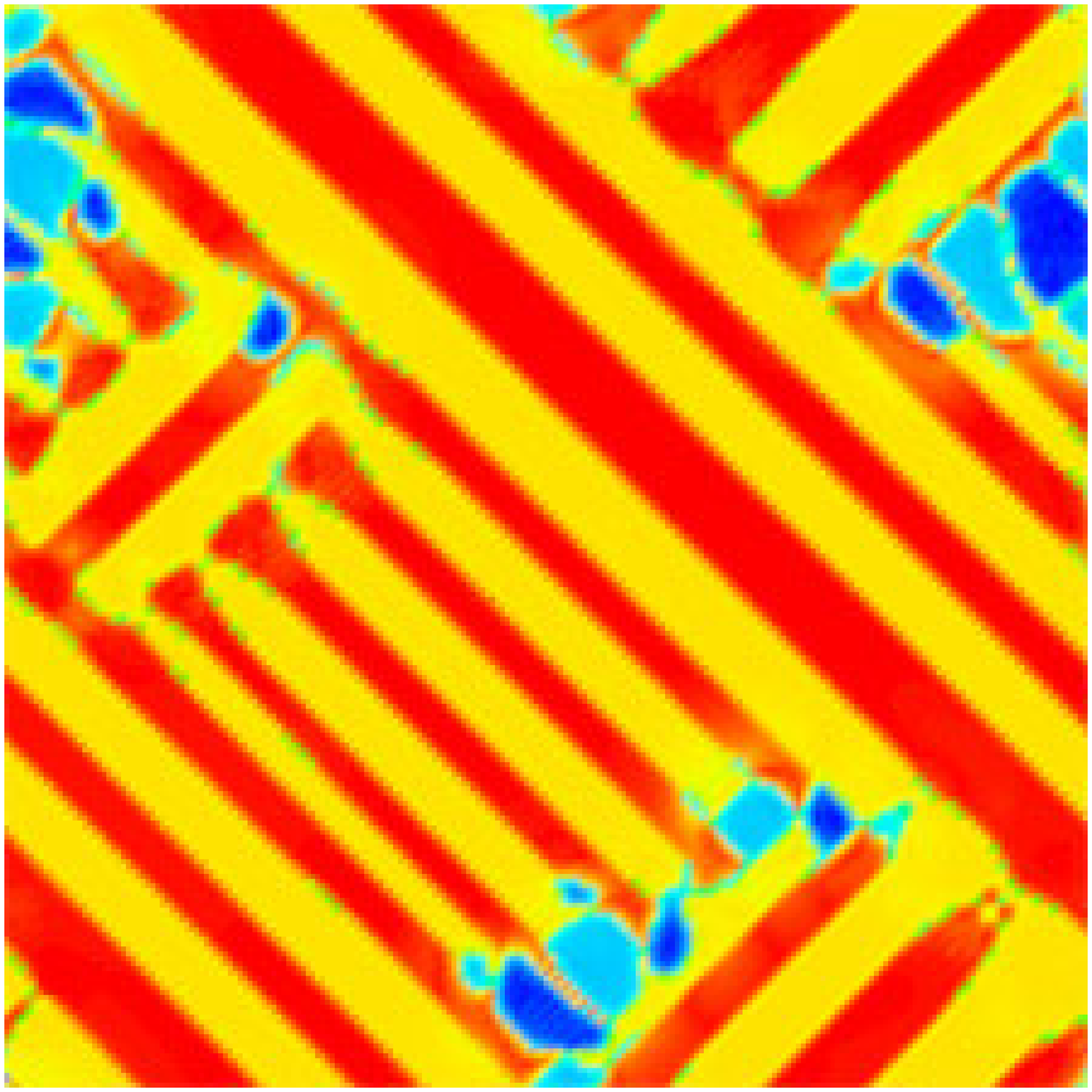}
	\put(-0.6, 2.3){\bf(b)}
}\subfigure{
    \label{128-0_0-T0490_t012000}
    \includegraphics[height=2.75cm]{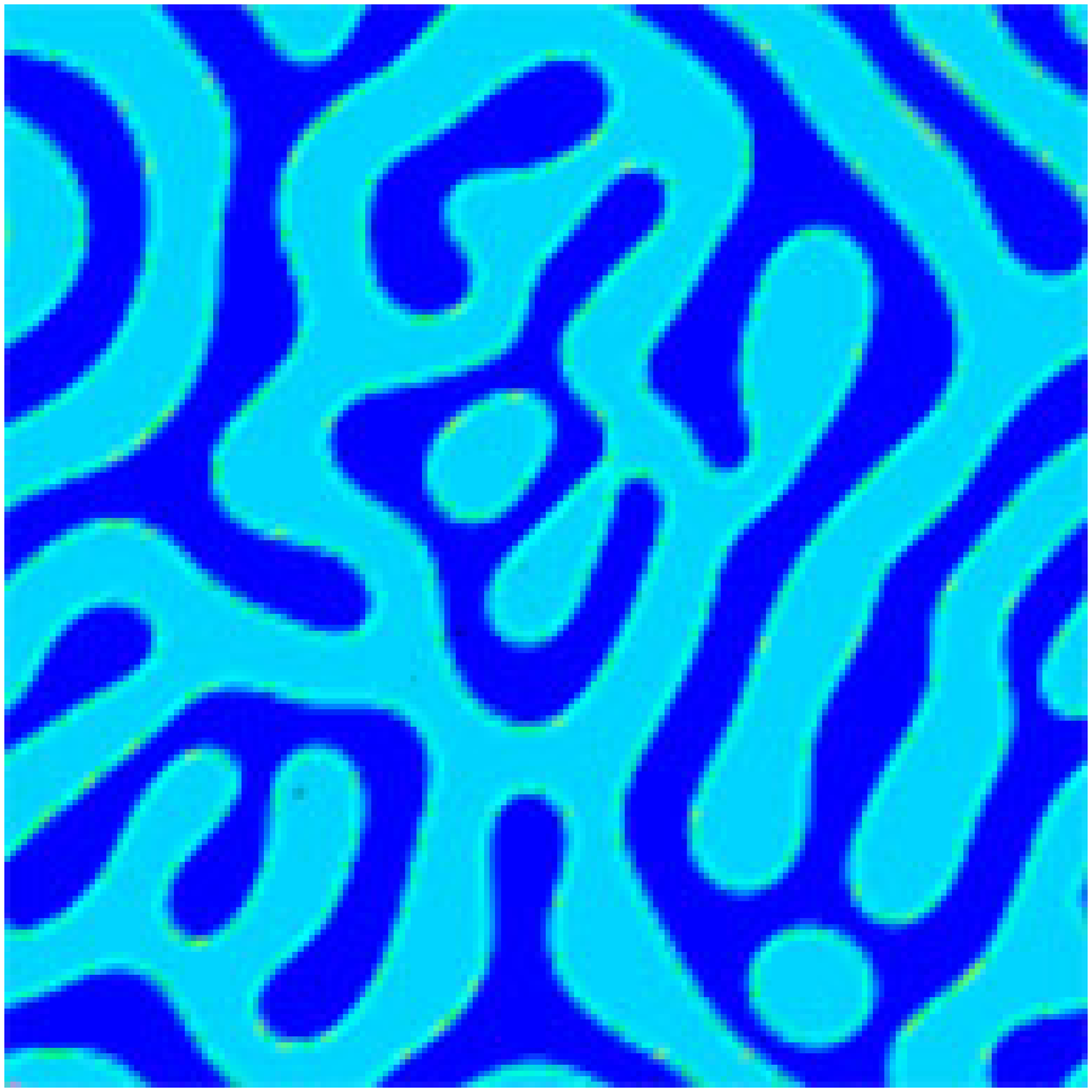}
	\put(-0.65, 2.3){\color{white}\bf(c)}
}
\end{picture}
	\caption{\label{128-0_0-T0490}(color online) Microstructures for $x_{12} = x_{1c} = 0$ at $T=0.49$ and (a) $t=50$, (b) $t=4~000$, and (c) $t=12~000$. They correspond to points~c, d, and e in Fig.~\ref{128-0_0-TTT}. Red and yellow: martensite; light and dark blue: pearlite.}
\end{figure}

Figure~\ref{128-0_0-T0490} shows the microstructure evolution at $T=0.49$ (just below the martensite start temperature). At $t=50$ only martensite is found, Fig.~\ref{martensite}. At $t=4~000$, pearlite has already started forming at the interfaces between the martensite variants aligned along $[1\,1]$ and those along $[1\,\bar{1}]$, Fig.~\ref{128-0_0-T0490_t004000}, that is at interfaces which are relatively high in energy. For longer times pearlite grows at the expense of martensite, which finally disappears [Fig.~\ref{128-0_0-T0490_t012000}]. This appearance of pearlite in the martensite region occurs at progressively later times at lower temperatures. Figure~\ref{128-0_0-T0490_t012000} (obtained at $T=0.49$) is similar to Fig.~\ref{fine-pearlite}, which is obtained at $T=0.5$. However, the two microstructures were the outcomes of very different routes.

As discussed earlier, below $T=0.5$, pearlite is the ground state and martensite is only metastable. Thus, the martensite forms only because pearlite formation is slow at this temperature. It is therefore expected that pearlite can eventually form in a martensitic system as this will decrease the energy. However there exists an energy barrier because ---unlike austenite which is unstable below $T=1$--- martensite is metastable. This makes pearlite nucleation off martensite slower than pearlite nucleation off austenite and accounts for the discontinuity of the 10\% pearlite line across $T=0.5$ in Fig.~\ref{128-0_0-TTT}.\footnote{The model presented here cannot be expected to provide an accurate description of this nucleation process, as this would require atomistic data. The time necessary for pearlite to form may therefore be misestimated.}

\section{\label{sec-x12}Effect of martensite volume change \texorpdfstring{($x_{12} \ne 0$)}{(x{12} > 0)}}
In the previous section, we assumed that there was no strain associated with pearlite or martensite formation, i.e.\ $x_{12}=x_{1c}=0$ in Eq.~(\ref{eq-g_el}). We now relax this constraint and look at the effect of the term $x_{12}\,(e_2)^2$ in the energy expression, keeping $x_{1c}$ set to zero. This term couples deviatoric strain (i.e.\ martensite) and hydrostatic strain: $x_{12} \ne 0$ associates a net volume change with the martensitic transformation. 

\subsection{TTT diagram}
Figure~\ref{128-1_0-TTT} shows the TTT diagram for $x_{12}=1$ and   $x_{1c}=0$. There are two noticeable differences compared to the TTT diagram obtained with $x_{12}=x_{1c}=0$, Fig.~\ref{128-0_0-TTT}: there exists a martensite start temperature as well as a martensite finish temperature and there is hardly any discontinuity of the 10\% pearlite line across $T=0.5$.

The martensite start and finish temperatures  are due to the presence of retained austenite between $T\approx 0.45$ and $T=0.5$. The hydrostatic strain associated with the martensitic transformation results in coherency stresses. Therefore martensite growth in the austenite matrix is arrested before it is complete. The martensite/austenite ratio increases for decreasing temperature and the system is purely martensitic at very low temperature. This trend is consistent with experimental results: in Cu-17.0\% Zn-13.7 at.\% Al, the volume change is small and the difference between martensite start and finish temperatures is 30~$^\circ$C,\cite{Vives-PRB-95} whereas it is around 100~$^\circ$C in steel, where the volume change is larger.

When $x_{12}=x_{1c}=0$, below $T=0.5$ pearlite can nucleate heterogeneously at the interface between martensite variants (Fig.~\ref{128-0_0-T0490}); this process is rather slow. When $x_{12}=1$, even below $T=0.5$ some austenite remains and pearlite formation proceeds in this retained austenite. Since austenite is unstable there is no nucleation barrier for pearlite formation and there is hardly any discontinuity of the 10\% pearlite line. This is unlike in Fig.~\ref{128-0_0-T0490_t004000} where pearlite must nucleate at the boundary between martensite variants as there is no retained austenite. The pearlite nucleation is slower at lower $T$ only because the diffusivity is lower.

\begin{figure}
\centering
\setlength{\unitlength}{1cm}
\begin{picture}(8.5,4.5)(-.1,0)
\shortstack[c]{
\subfigure{
	\put(.6,2.1){$x_{12}=0$}
	\put(-0.25, 0.45){\rotatebox{90}{$x_{1c}=0$}}
    \label{128-0_0-T0490_t000050}
    \includegraphics[height=2.cm]{128-0_0-T0490_t000050-sm}
	\put(-0.6, 1.55){\bf(a)}
}\subfigure{
	\put(.4,2.1){$x_{12}=0.85$}
    \label{128-085_0-T0490_t000050}
    \includegraphics[height=2.cm]{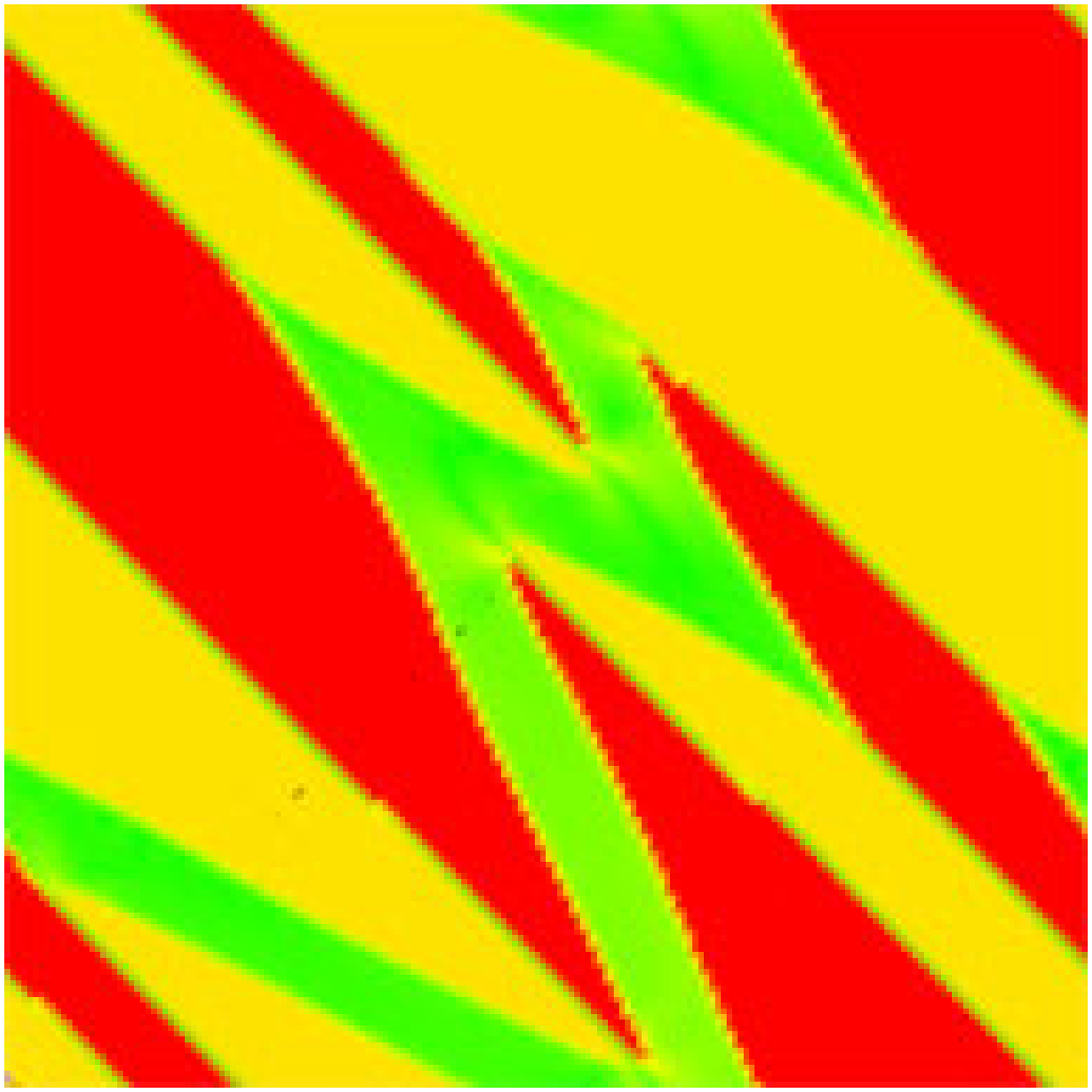}
	\put(-0.65, 1.55){\bf(b)}
}\subfigure{
	\put(.6,2.1){$x_{12}=1$}
    \label{128-1_0-T0490_t000050}
    \includegraphics[height=2.cm]{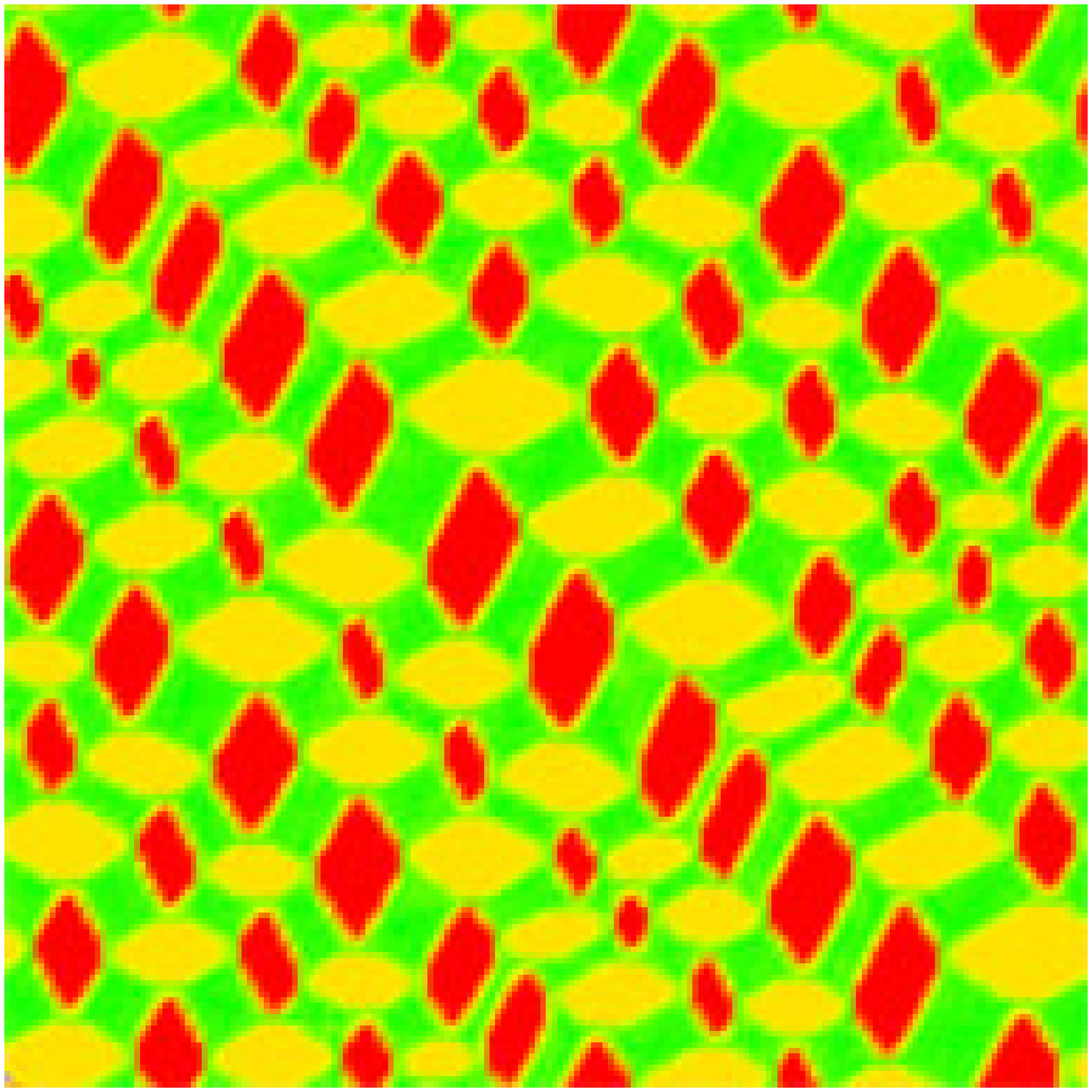}
	\put(-0.65, 1.55){\bf(c)}
}\subfigure{
	\put(.45,2.1){$x_{12}=1.5$}
    \label{128-15_0-T0490_t000050}
    \includegraphics[height=2.cm]{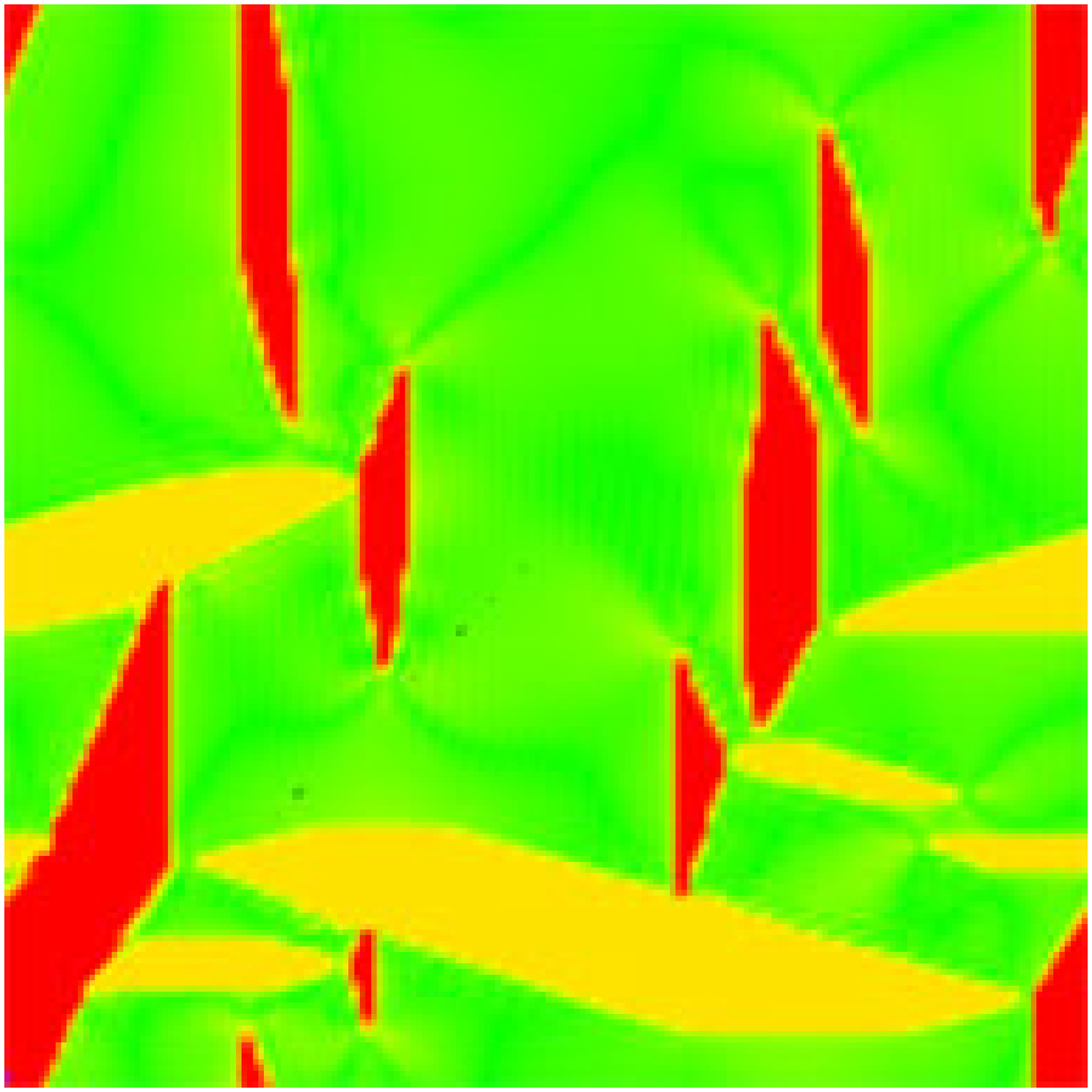}
	\put(-0.65, 1.55){\bf(d)}
}\vspace{-.25cm}\\
\subfigure{
	\put(-0.25, 0.4){\rotatebox{90}{$x_{1c}=1$}}
    \label{128-0_1-T0490_t000050}
    \includegraphics[height=2.cm]{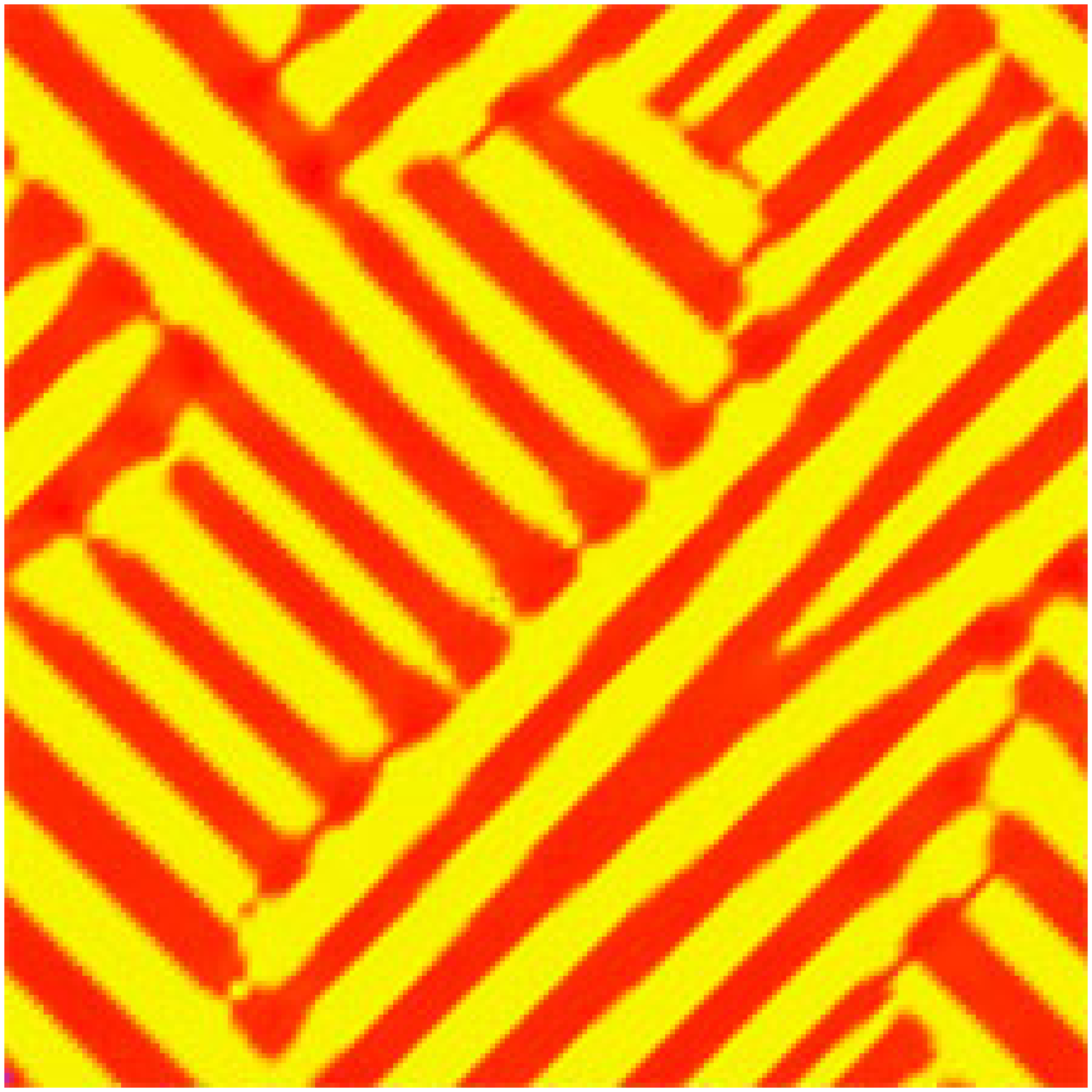}
	\put(-0.65, 1.55){\bf(e)}
}\subfigure{
    \label{128-085_1-T0490_t000050}
    \includegraphics[height=2.cm]{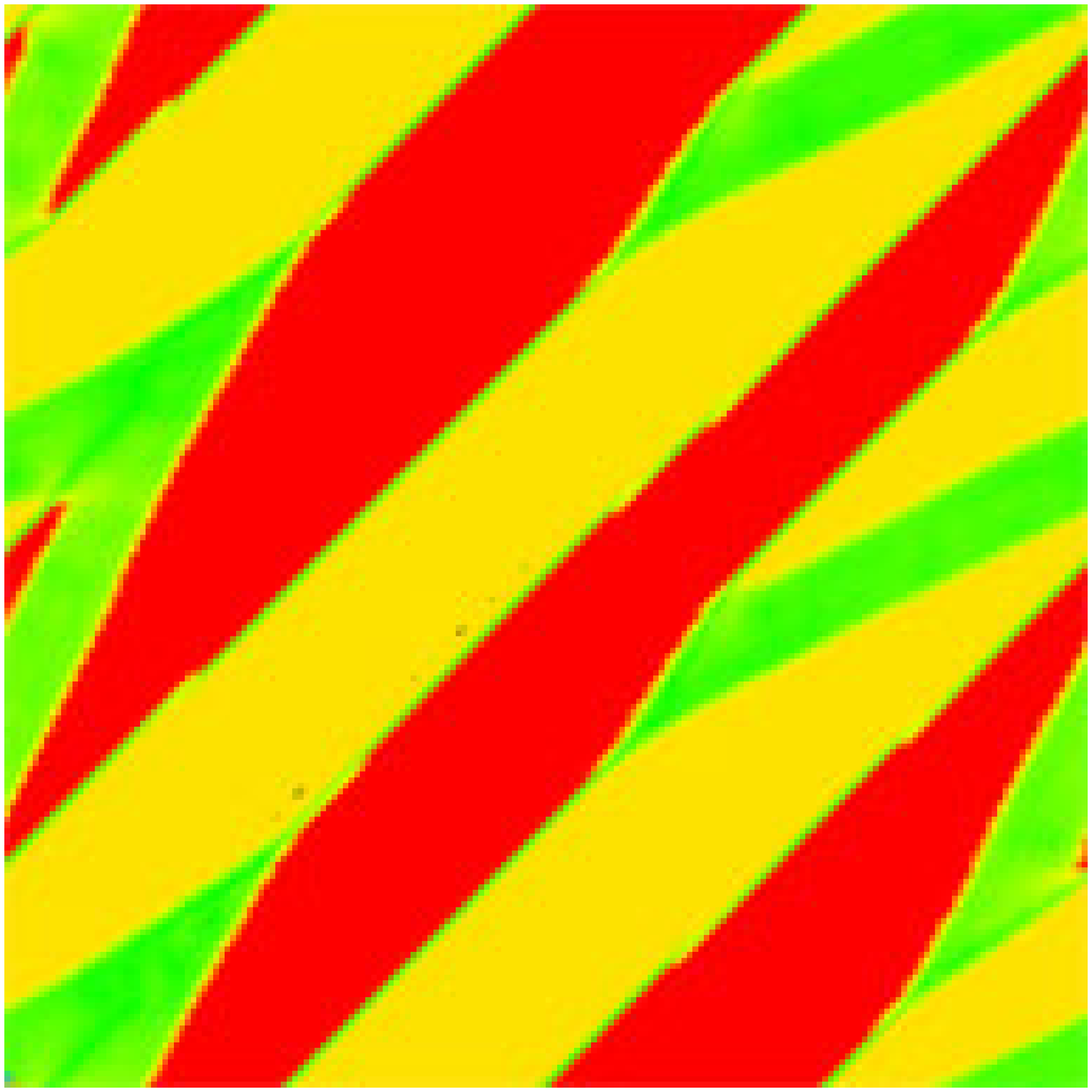}
	\put(-0.65, 1.55){\bf(f)}
}\subfigure{
    \label{128-1_1-T0490_t000050}
    \includegraphics[height=2.cm]{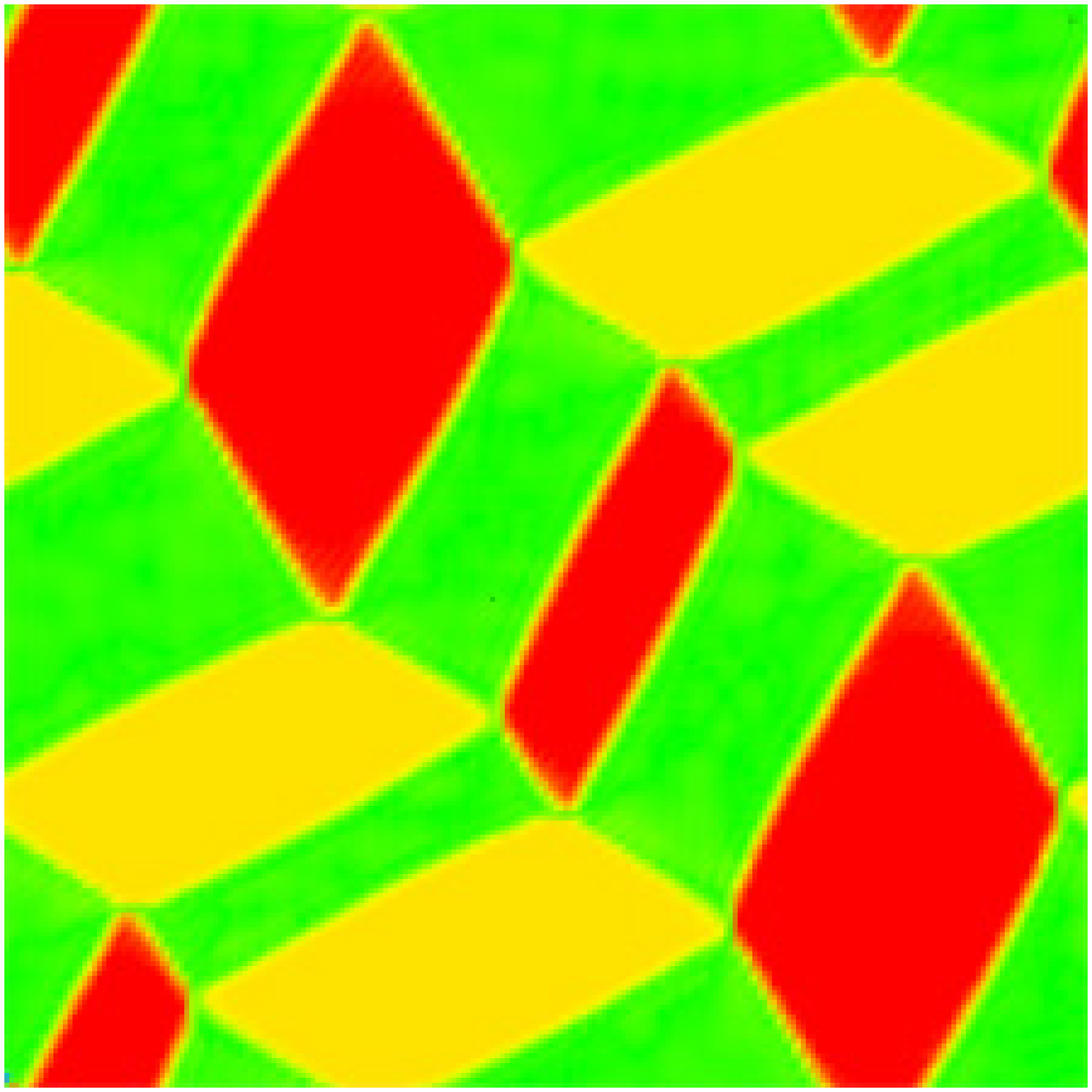}
	\put(-0.65, 1.55){\bf(g)}
}\subfigure{
    \label{128-15_1-T0490_t000050}
    \includegraphics[height=2.cm]{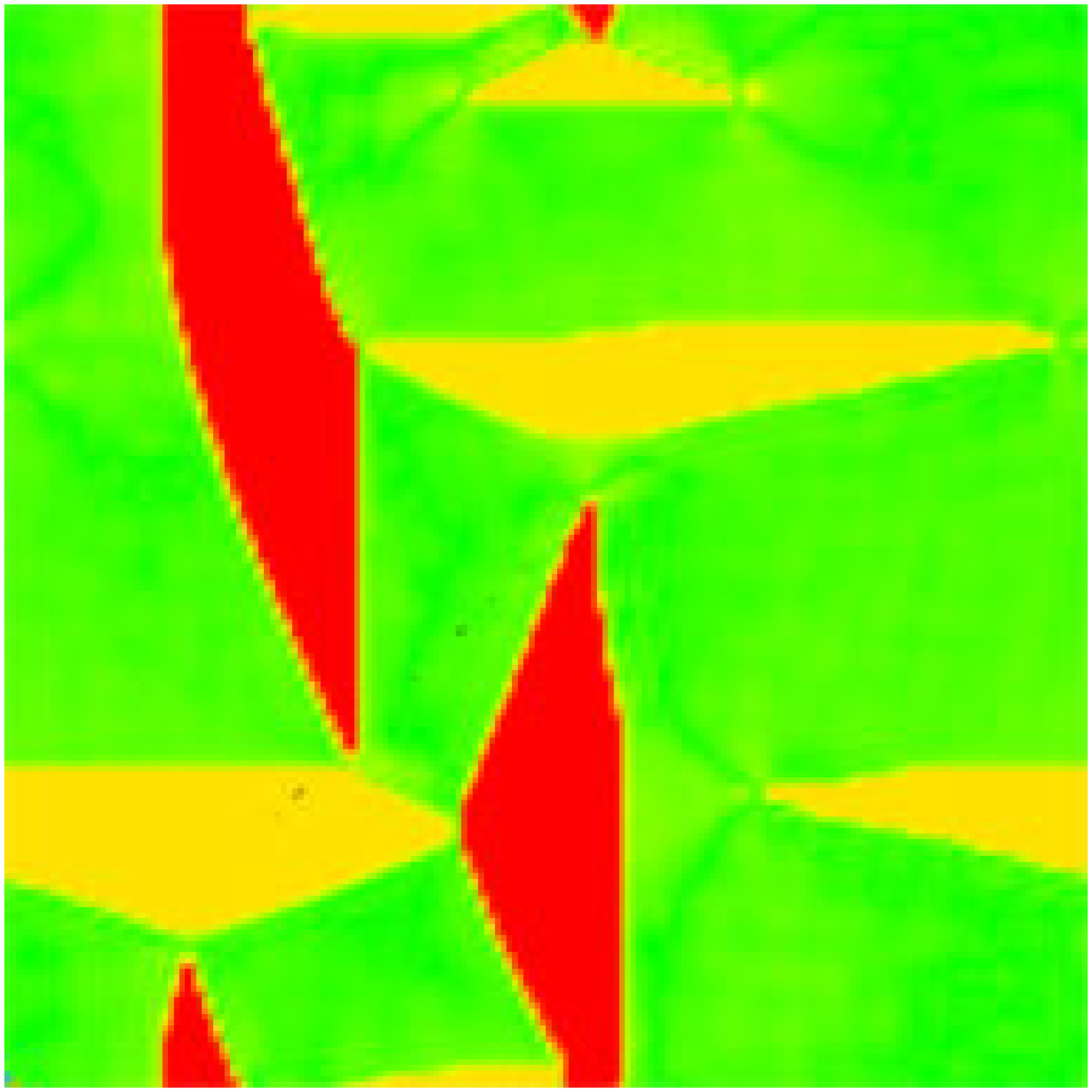}
	\put(-0.65, 1.55){\bf(h)}
}}
\end{picture}
	\caption{\label{128-T0490_t000050}(color online) Microstructures at $T=0.49$ and $t=50$. From left to right: $x_{12} = 0$, $x_{12} = 0.85$, $x_{12} = 1$, and $x_{12} = 1.5$. Top: $x_{1c} = 0$ and bottom: $x_{1c} = 1$. Red and yellow: martensite; light and dark blue: pearlite.}
\end{figure}

\subsection{Microstructures}
The effect of $x_{12}$ on the microstructure at $T=0.49$ can be seen in Figs.~\ref{128-0_0-T0490_t000050}--\ref{128-15_0-T0490_t000050}. 
If $x_{12} \ne 0$, there is a competition between the driving force favoring the transformation and the coherency stress arising out of the volume change: the larger $x_{12}$, the larger the stress and the lower the equilibrium martensite fraction. 
This gives rise to several microstructures for increasing $x_{12}$. If $x_{12}$ is large the system is mostly austenitic, Fig.~\ref{128-15_0-T0490_t000050}. At intermediate values the system looks like a checkerboard of alternating martensite and austenite, Fig.~\ref{128-1_0-T0490_t000050} (this kind of microstructure has been observed experimentally by \citet{LeBouar-98}). At lower $x_{12}$ the system is mostly martensite with small austenitic regions, Fig.~\ref{128-085_0-T0490_t000050}. For even lower values of $x_{12}$ the stress is not sufficient to make a somewhat relaxed austenite--martensite mixture more favorable than stressed martensite and the system is made of pure martensite.

Comparison of Figs.~\ref{128-0_1-T0490_t000050}--\ref{128-15_1-T0490_t000050} to Figs.~\ref{128-0_0-T0490_t000050}--\ref{128-15_0-T0490_t000050} shows that, at $T=0.49$ and $t=50$, the microstructures do not depend on the value of $x_{1c}$ nearly as much as on the value of $x_{12}$: the mixed austenite--martensite microstructures are controlled by the value of $x_{12}$. This is because $x_{1c}$ is important only in the presence of pearlite.

\begin{figure}
\centering
\setlength{\unitlength}{1cm}
\begin{picture}(8.5,2.75)(.1,0)
\subfigure{
    \label{128-085_0-T0490_t001000}
    \includegraphics[height=2.75cm]{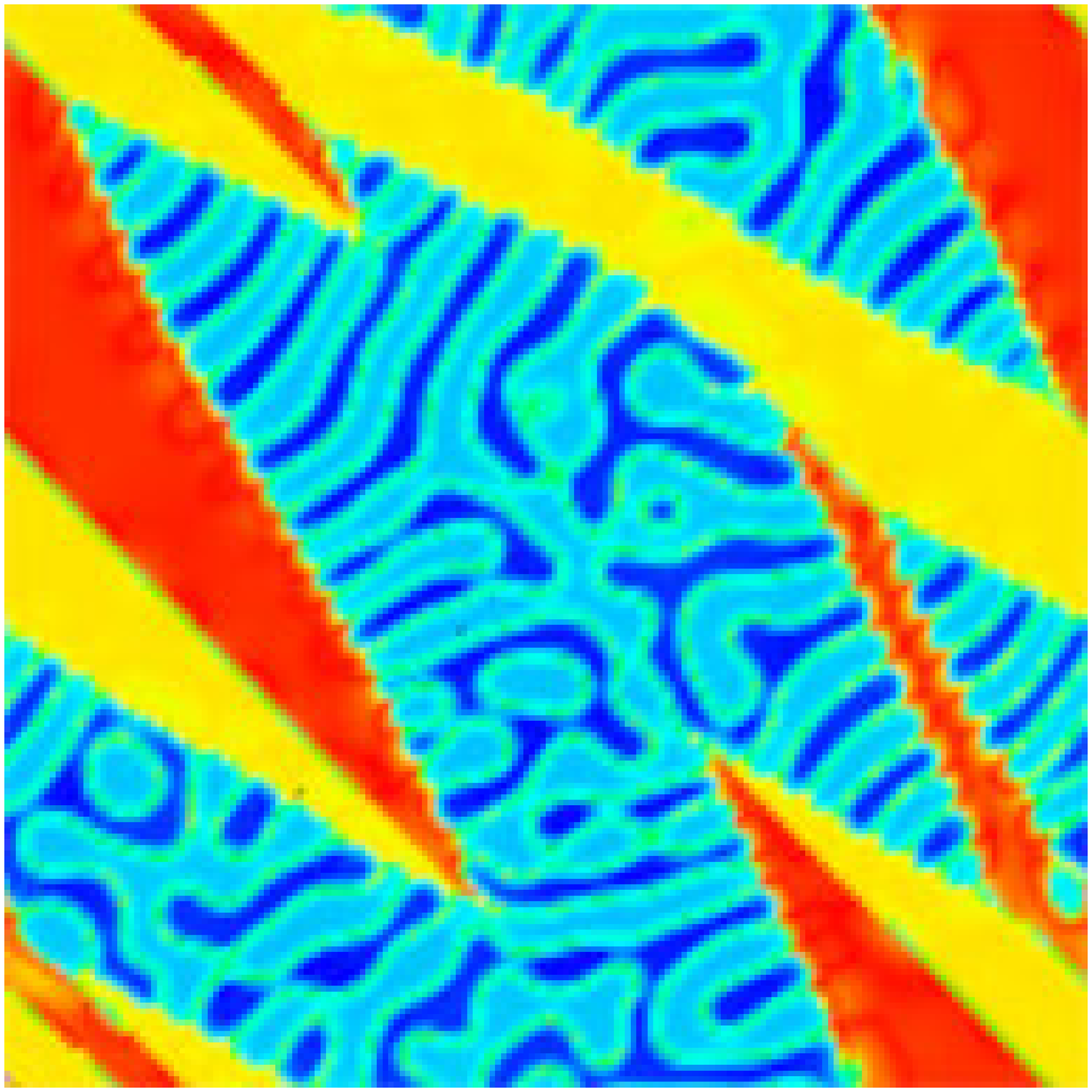}
	\put(-0.65,2.3){\color{white}\bf(a)}
}\subfigure{
    \label{128-085_0-T0490_t002000}
    \includegraphics[height=2.75cm]{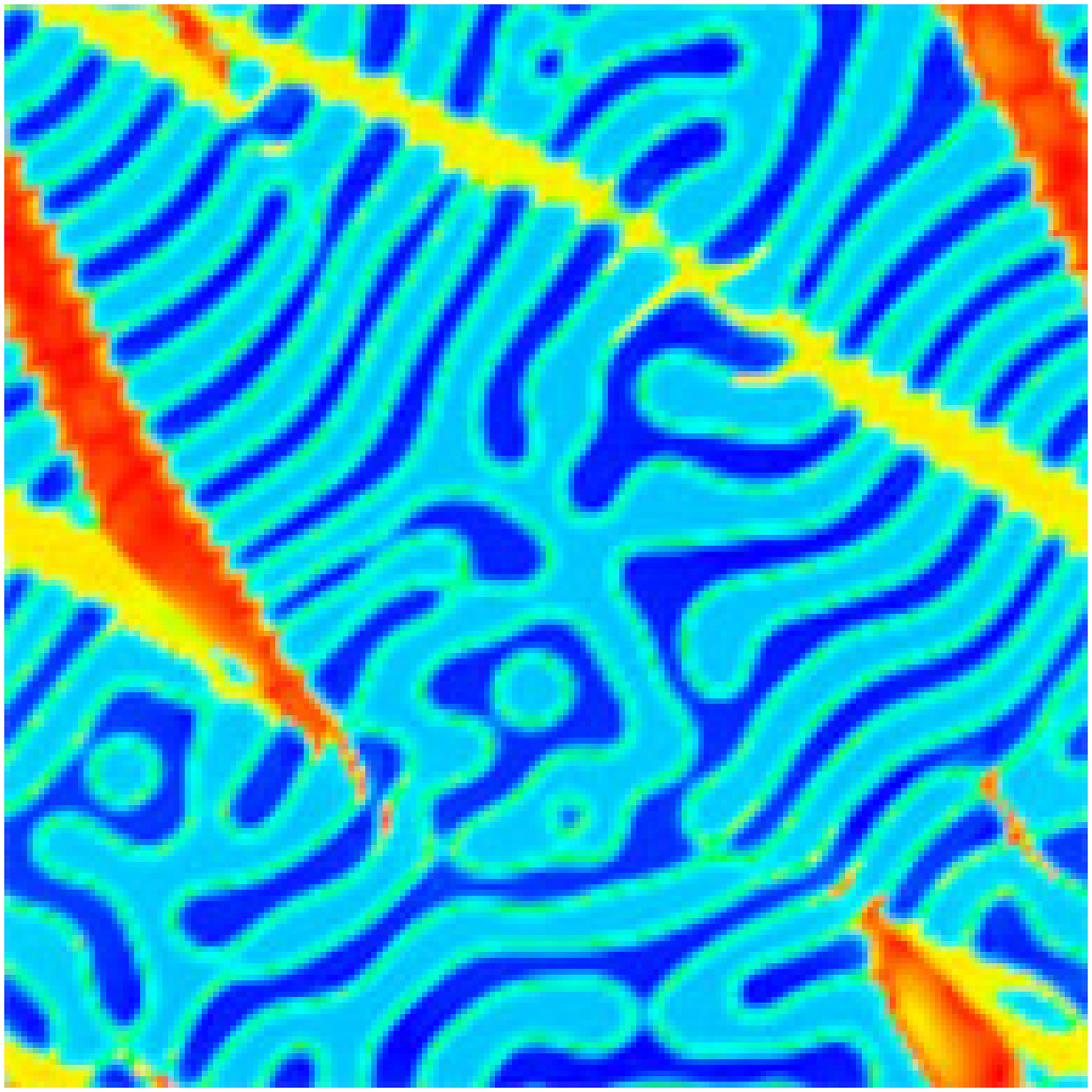}			
	\put(-0.65,2.3){\color{white}\bf(b)}
}\subfigure{
    \label{128-085_0-T0490_t003500}
    \includegraphics[height=2.75cm]{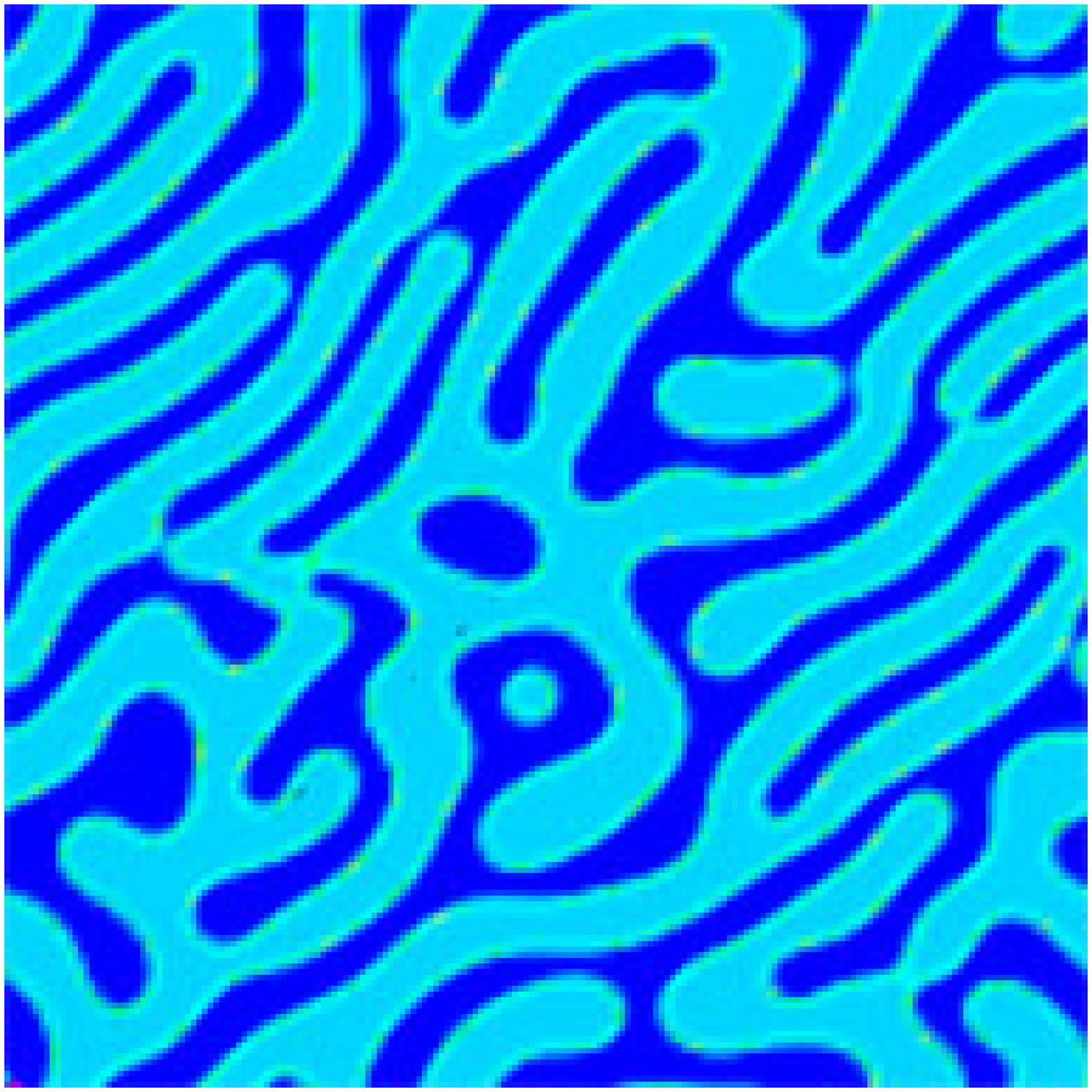}
	\put(-0.65,2.3){\color{white}\bf(c)}
}
\end{picture}
	\caption{\label{128-085_0-T0490}(color online) Microstructures for $x_{12}=0.85$ and $x_{1c}=0$ at $T=0.49$: (a) $t=1~000$, (b) $t=2~000$, and (c) $t=3~500$. Red and yellow: martensite; light and dark blue: pearlite.}
\end{figure}

Along with Fig.~\ref{128-085_0-T0490_t000050}, Fig.~\ref{128-085_0-T0490} shows the microstructure evolution at $T=0.49$ for $x_{12}=0.85$ and $x_{1c}=0$. Unlike pearlite nucleated from austenite, Fig.~\ref{fine-pearlite}, pearlite here exhibits some texture along $[1\,1]$, Fig.~\ref{128-085_0-T0490_t003500}. This texture is due to the way pearlite grew, not to thermodynamics; it is therefore different from Figs.~\ref{128-0_1-T0850_t002000} and~\ref{128-1_1-T0800_t010000} (see 
Sec.~\ref{x1c-microstructures}). This alignment of pearlite perpendicularly to the interface where it nucleated has been extensively described.

\subsection{\label{sec-orientation}Interface orientation}
As expected from elastic compatibility, martensite--martensite interfaces are oriented along $\langle 1\,1 \rangle$, as seen in Fig.~\ref{128-0_0-T0490_t000050}.\cite{Sapriel75} However, the orientation of martensite--austenite interfaces is different, as shown in Figs.~\ref{128-085_0-T0490_t000050}--\ref{128-15_0-T0490_t000050} (the same holds for martensite--pearlite interfaces). 

We consider an interface (martensite--martensite, martensite--pearlite, or martensite--austenite) at an angle $\theta$ with the y-axis and determine what values of $\theta$ are allowed by elastic compatibility. $e_1$, $e_2$, and $e_3$ depend only on $x \cos\theta + y \sin\theta$ and, for $i=1,2,3$,
\begin{equation}
	\frac{\partial^2 e_i}{\partial x^2} = e_i'' \cos^2\theta
	\text{\quad and\quad}
	\frac{\partial^2 e_i}{\partial y^2} = e_i'' \sin^2\theta.
	\label{derivatives}
\end{equation}
\noindent As the only contribution of $e_3$ to the free energy is through the $(e_3)^2$ term, energy minimization with respect to $e_3$ gives $e_3=0$. Elastic compatibility then requires\cite{Kartha95}
\begin{equation}
	\frac{\partial^2 e_1}{\partial x^2} + \frac{\partial^2 e_1}{\partial y^2} - \frac{\partial^2 e_2}{\partial x^2} + \frac{\partial^2 e_2}{\partial y^2}=0.
	\label{compatibility}
\end{equation}
\noindent If $x_{1c}\,c + x_{12}\,(e_2)^2=0$ in Eq.~(\ref{eq-g_el}), energy minimization gives $e_1=0$. This occurs if (i) $x_{12}=0$ for a martensite--martensite or a martensite--austenite interface, (ii) ${x_{1c}=0}$ for a pearlite--austenite interface, or (iii) $x_{12}=x_{1c}=0$. Equations~(\ref{derivatives}) and~(\ref{compatibility}) then give
\begin{equation}
	\cos2\theta =0.
	\label{compatibility2}
\end{equation}
\noindent In order to satisfy compatibility,  interfaces must be aligned along $\langle 1\,1 \rangle$, as observed for martensite--martensite interfaces in Figs.~\ref{128-0_0-T0490_t000050} and~\ref{128-0_1-T0490_t000050} for instance.

\begin{figure}
\centering
\setlength{\unitlength}{1cm}
\begin{picture}(8.5,11.3)(-.1,0)
\shortstack[c]{
\subfigure{
	\put(-0.25, 0.5){\rotatebox{90}{$x_{12}=0$}}
	\put(0.55, 2.1){{$t=0.85$}}
    \label{early-0_0-850}
    \includegraphics[height=2.cm]{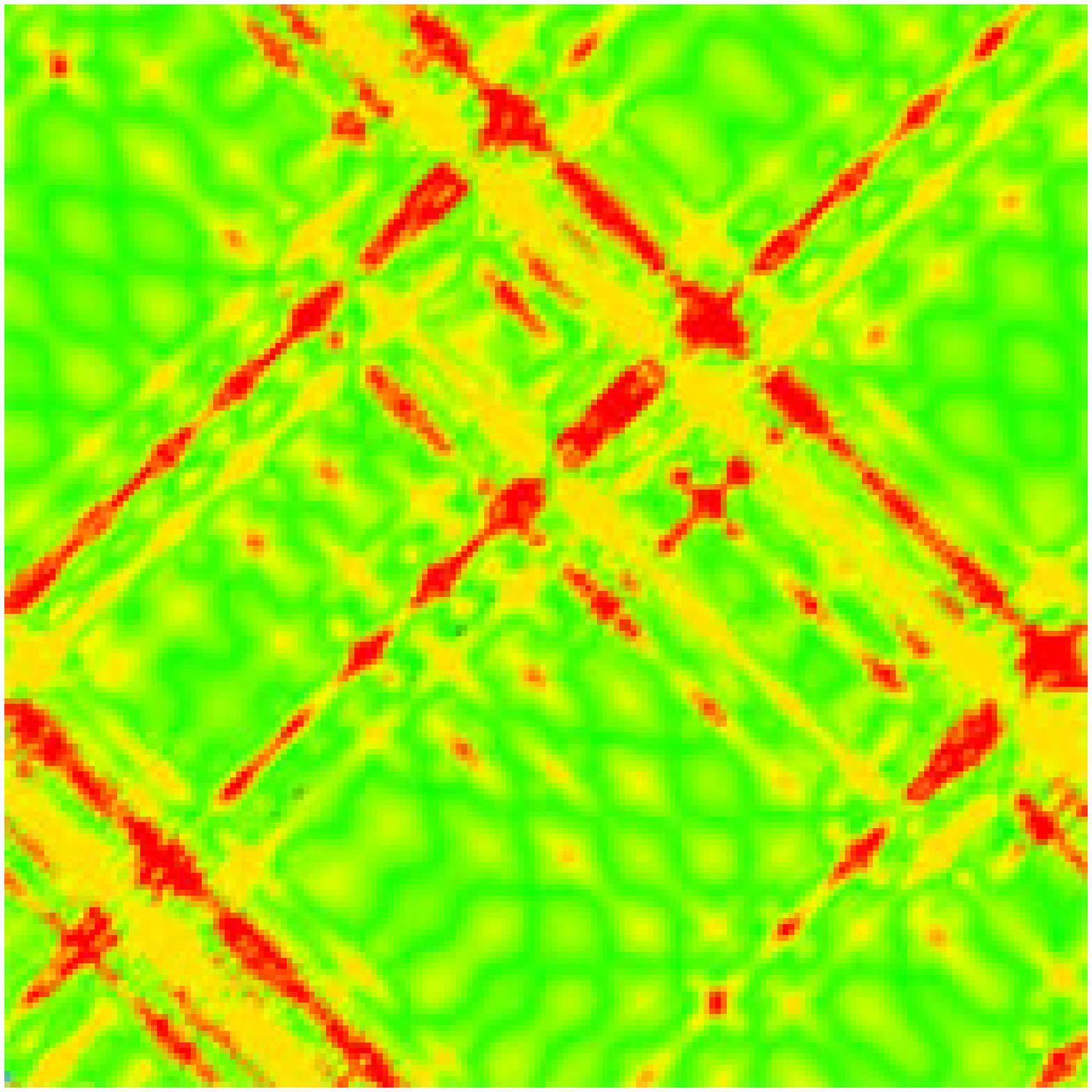}
	\put(-0.65, 1.55){\bf(a)}
}\subfigure{
	\put(0.6, 2.1){{$t=0.9$}}
    \label{early-0_0-900}
    \includegraphics[height=2.cm]{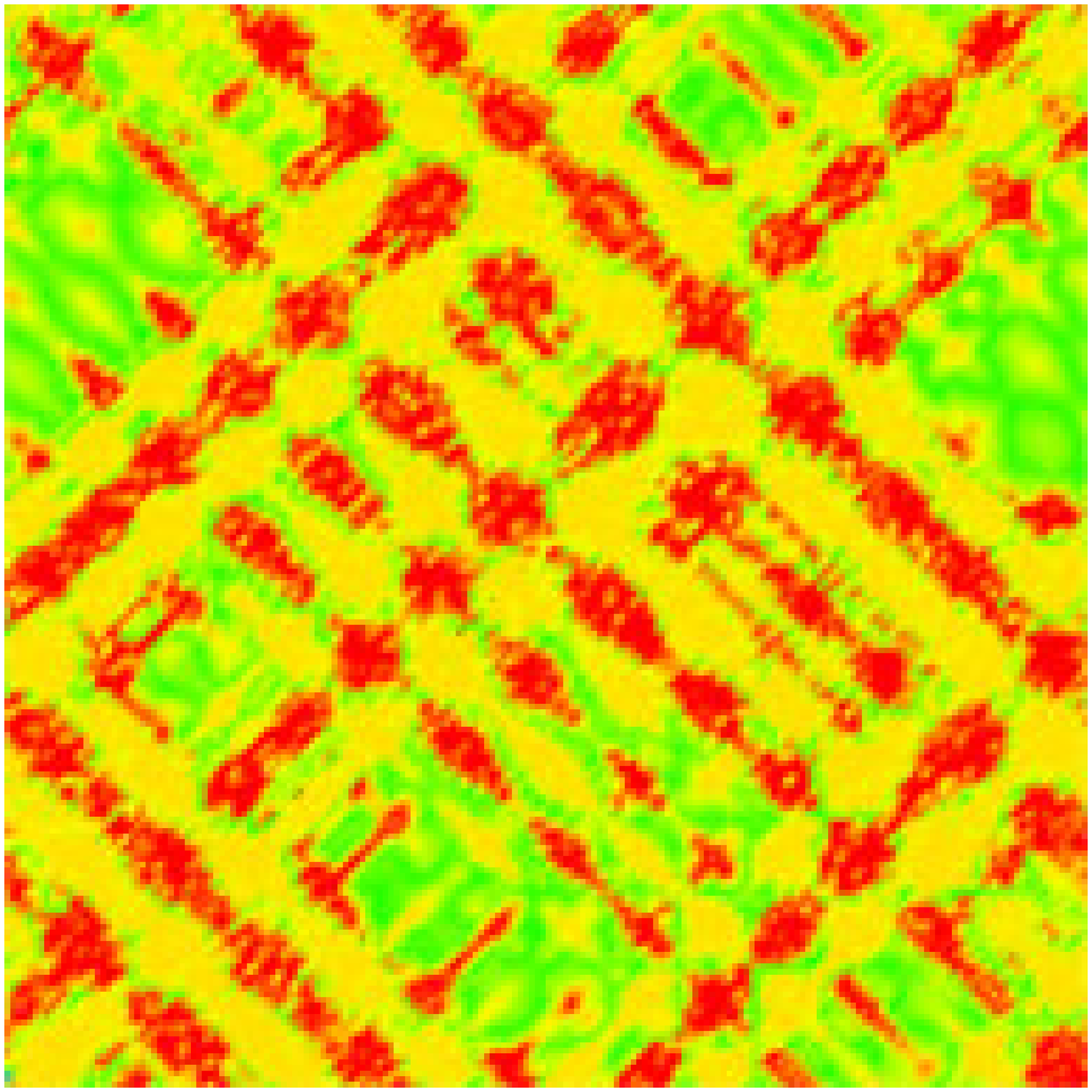}
	\put(-0.65, 1.55){\bf(b)}
}\subfigure{
	\put(0.75, 2.1){{$t=1$}}
    \label{early-0_0-1000}
    \includegraphics[height=2.cm]{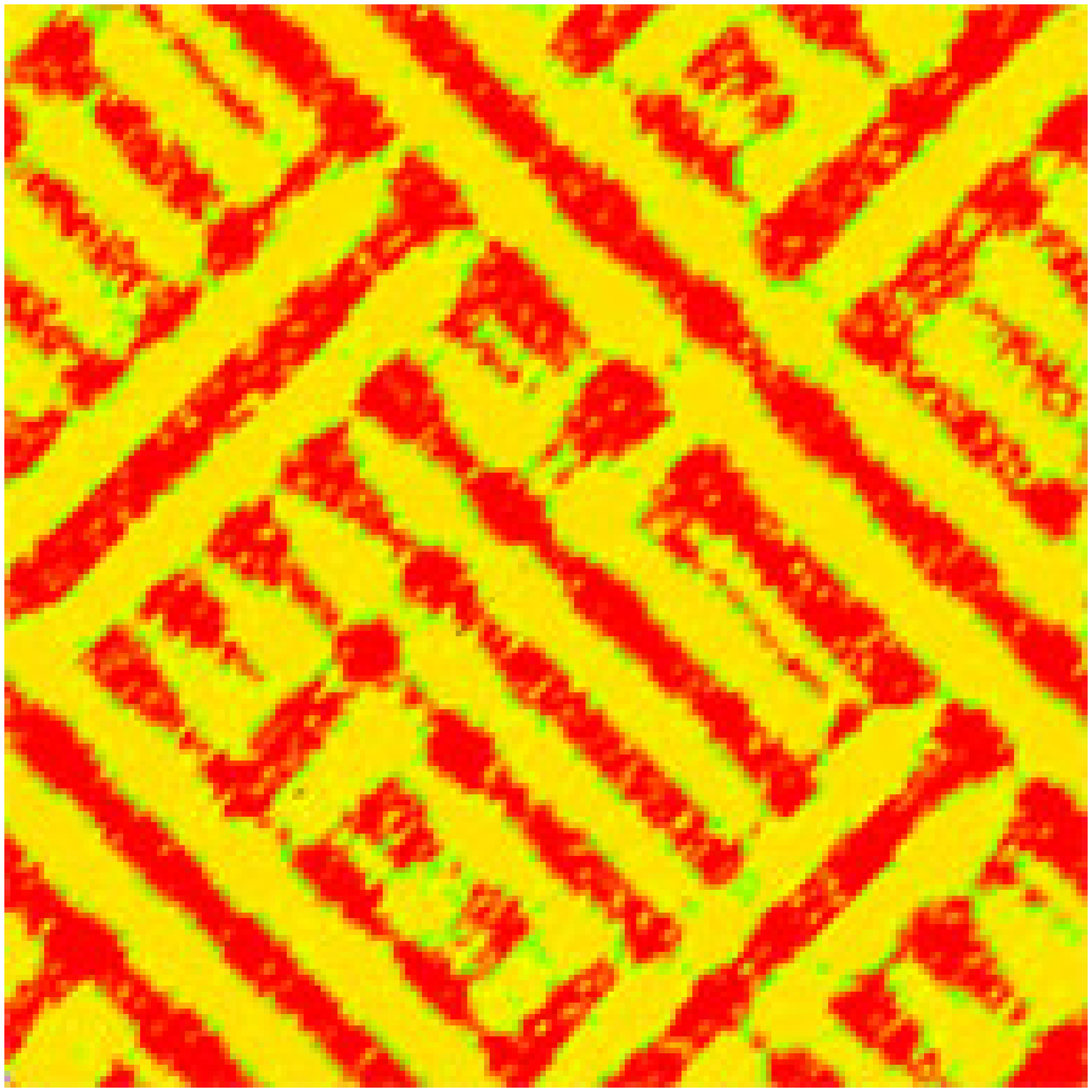}
	\put(-0.65, 1.55){\bf(c)}
}\subfigure{
	\put(0.6, 2.1){{$t=1.5$}}
    \label{early-0_0-1500}
    \includegraphics[height=2.cm]{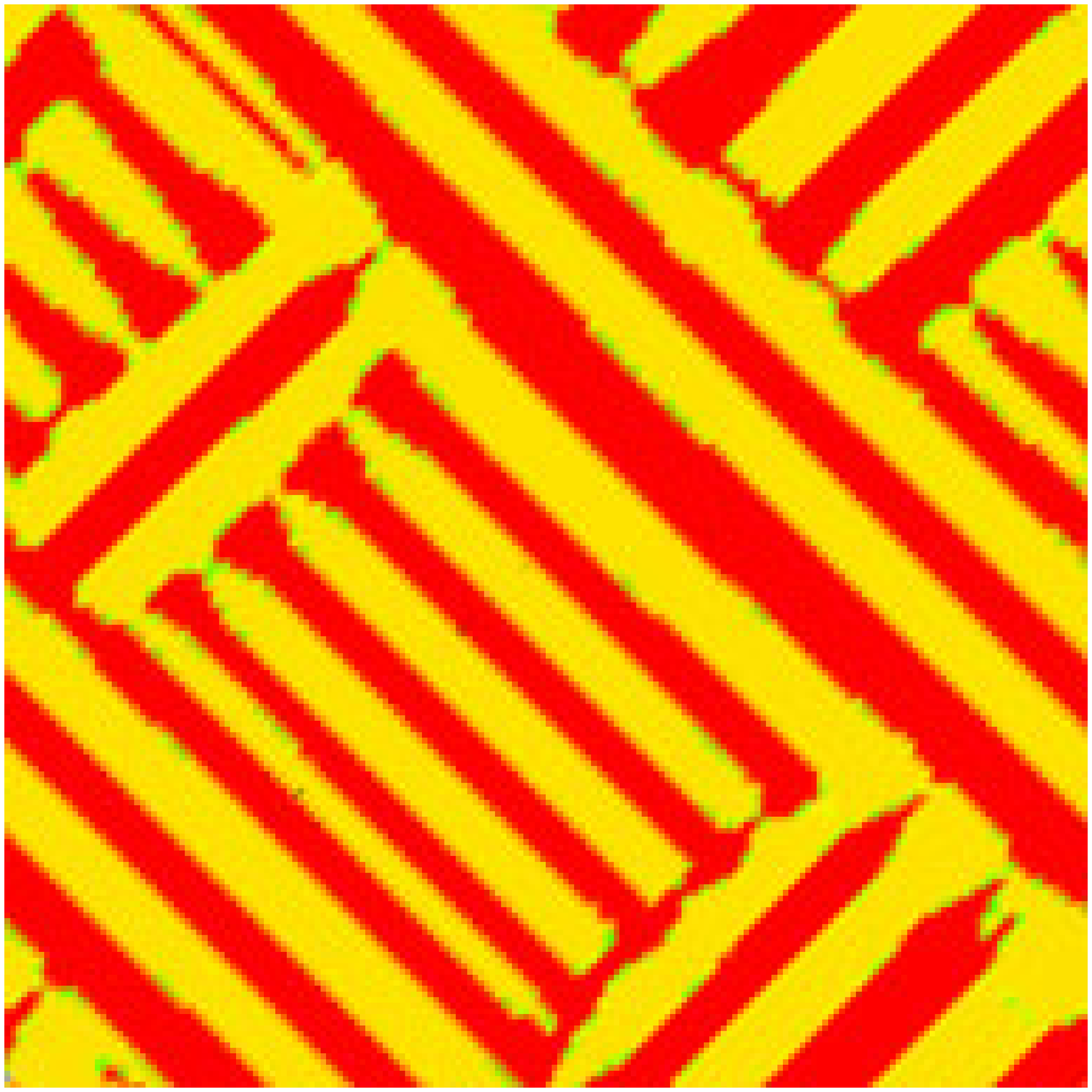}
	\put(-0.65, 1.55){\bf(d)}
}\vspace{-.2cm}\\
\subfigure{
	\put(-0.2, 0.2){\rotatebox{90}{$x_{12}=0.75$}}
    \label{early-075_0-850}
    \includegraphics[height=2.cm]{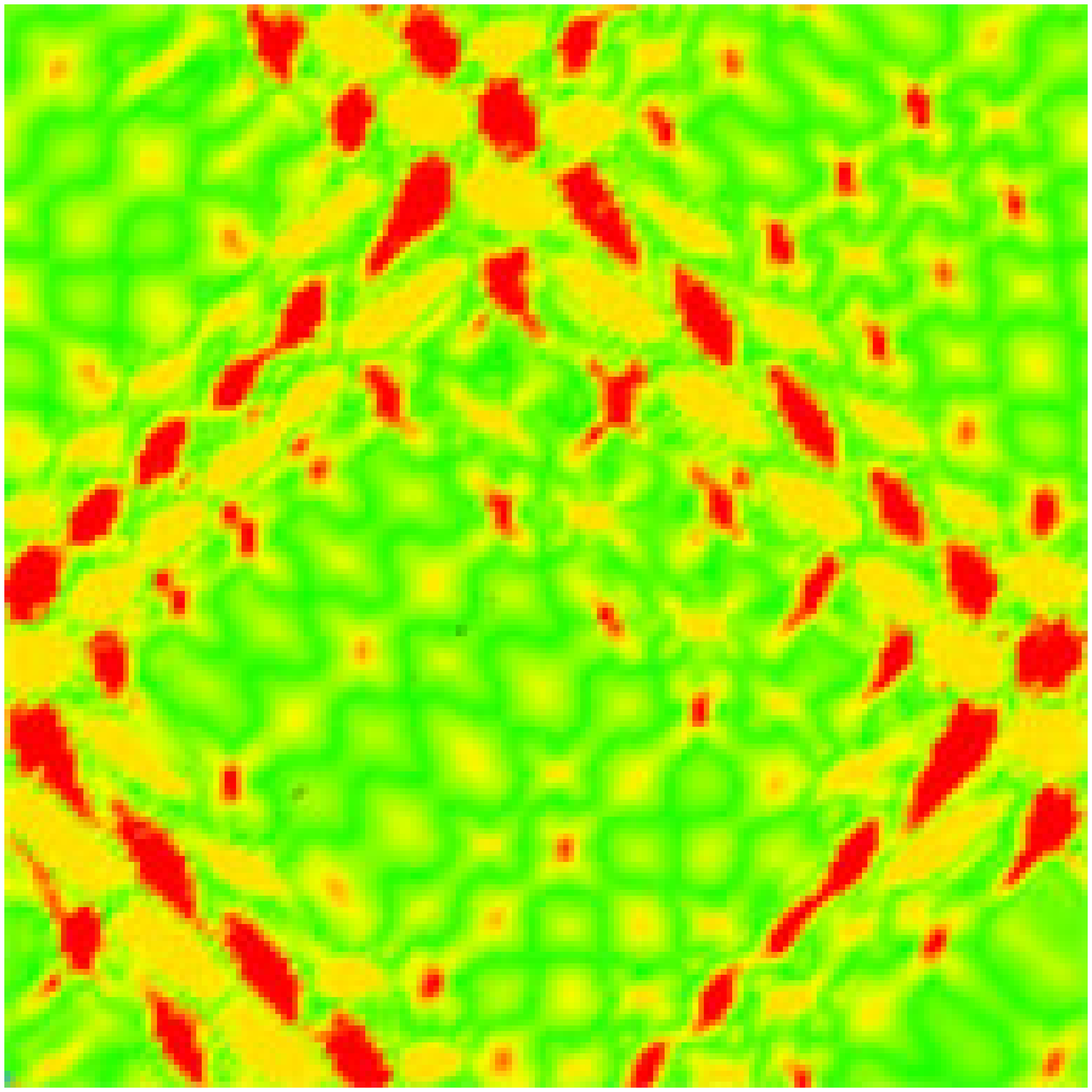}
	\put(-0.65, 1.55){\bf(e)}
}\subfigure{
    \label{early-075_0-900}
    \includegraphics[height=2.cm]{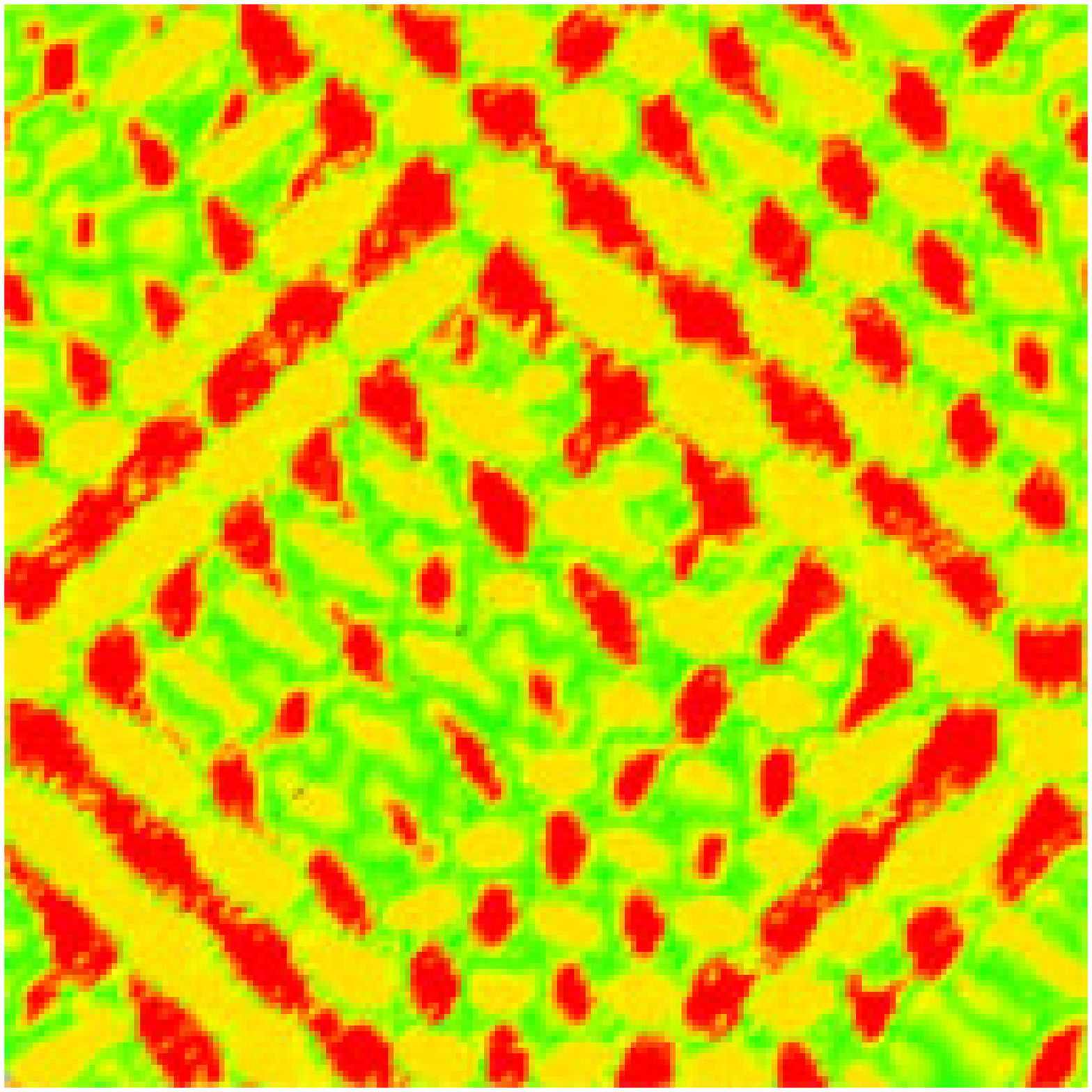}
	\put(-0.65, 1.55){\bf(f)}
}\subfigure{
    \label{early-075_0-1000}
    \includegraphics[height=2.cm]{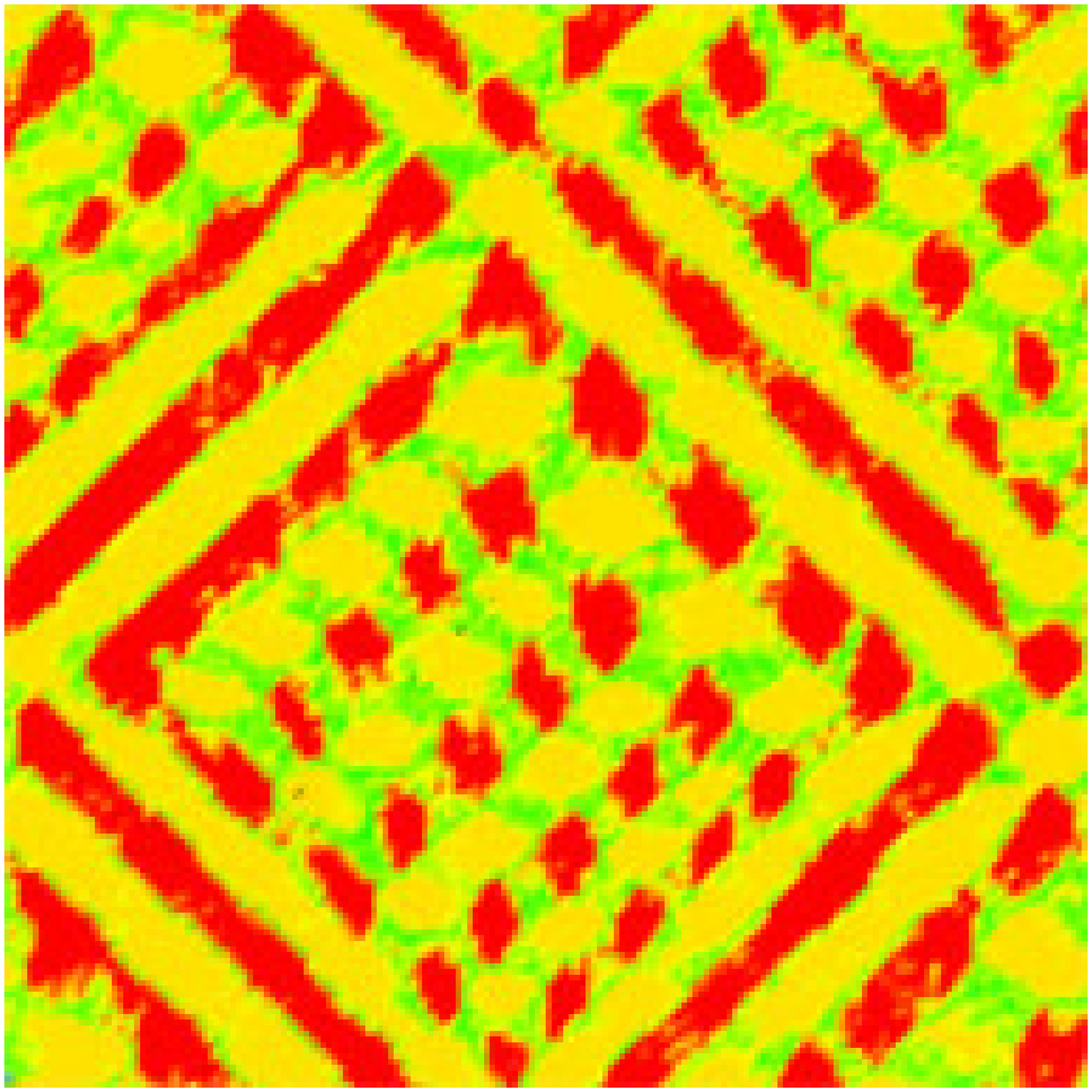}
	\put(-0.65, 1.55){\bf(g)}
}\subfigure{
    \label{early-075_0-1500}
    \includegraphics[height=2.cm]{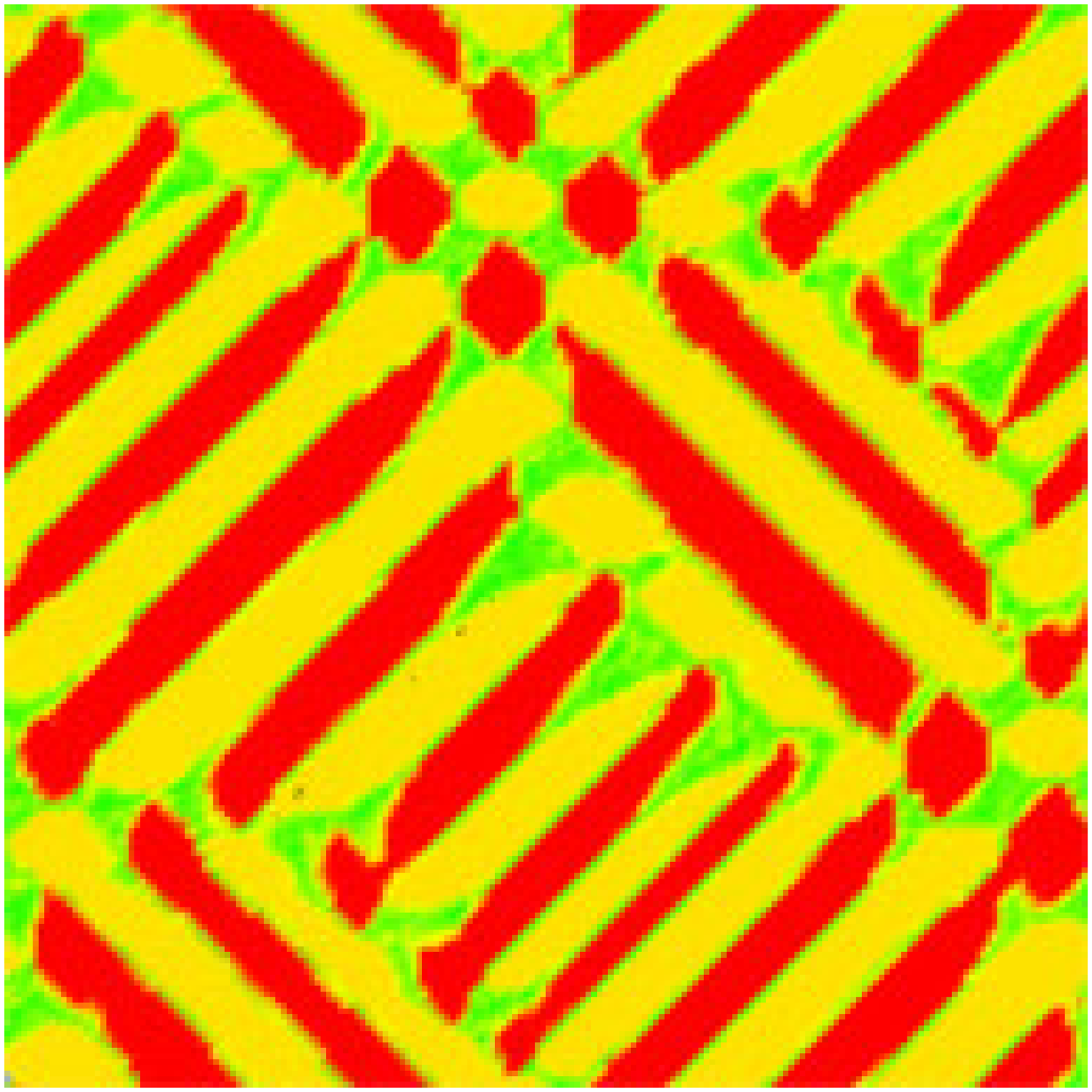}
	\put(-0.65, 1.55){\bf(h)}
}\vspace{-.2cm}\\
\subfigure{
	\put(-0.2, 0.2){\rotatebox{90}{$x_{12}=0.85$}}
    \label{early-085_0-850}
    \includegraphics[height=2.cm]{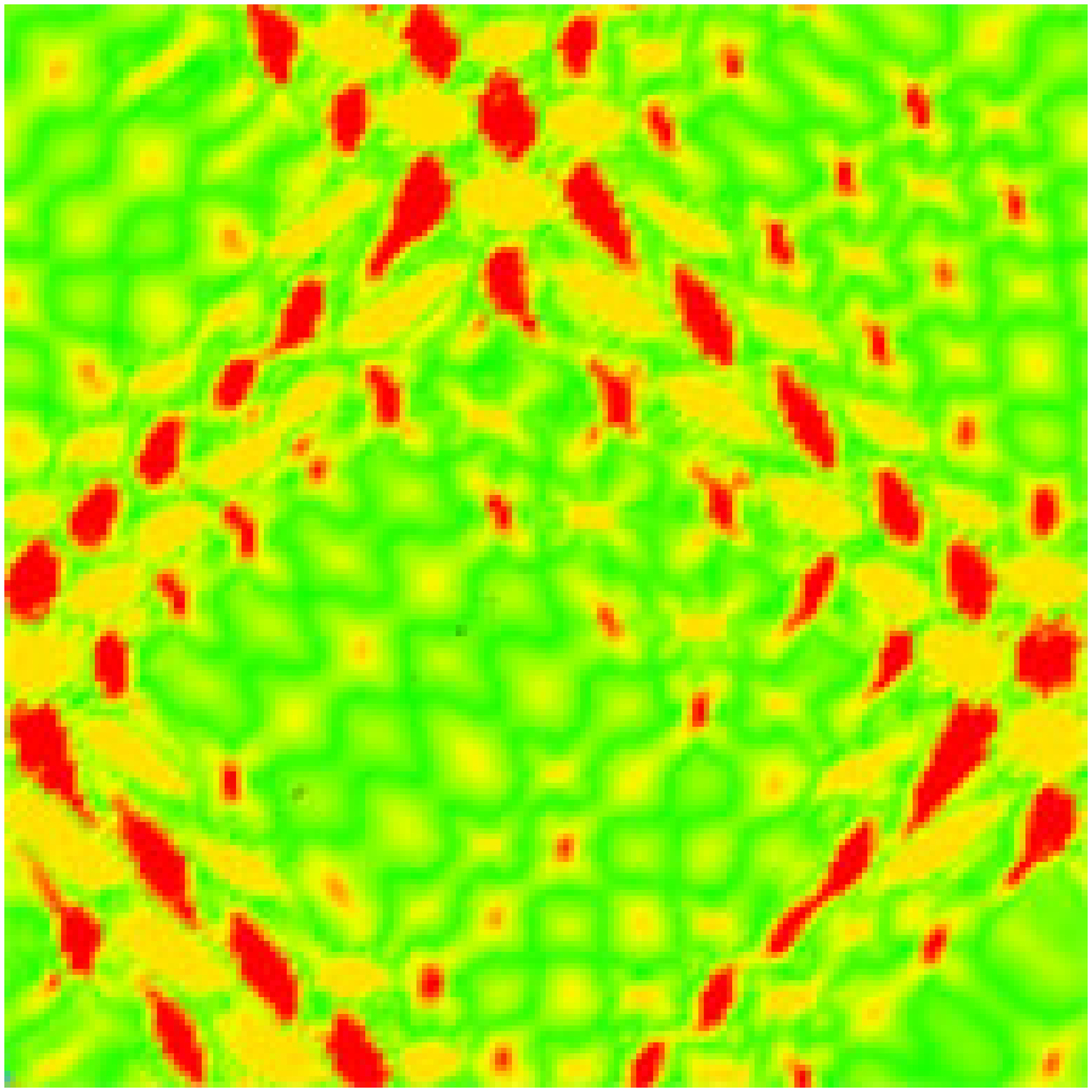}
	\put(-0.65, 1.55){\bf(i)}
}\subfigure{
    \label{early-085_0-900}
    \includegraphics[height=2.cm]{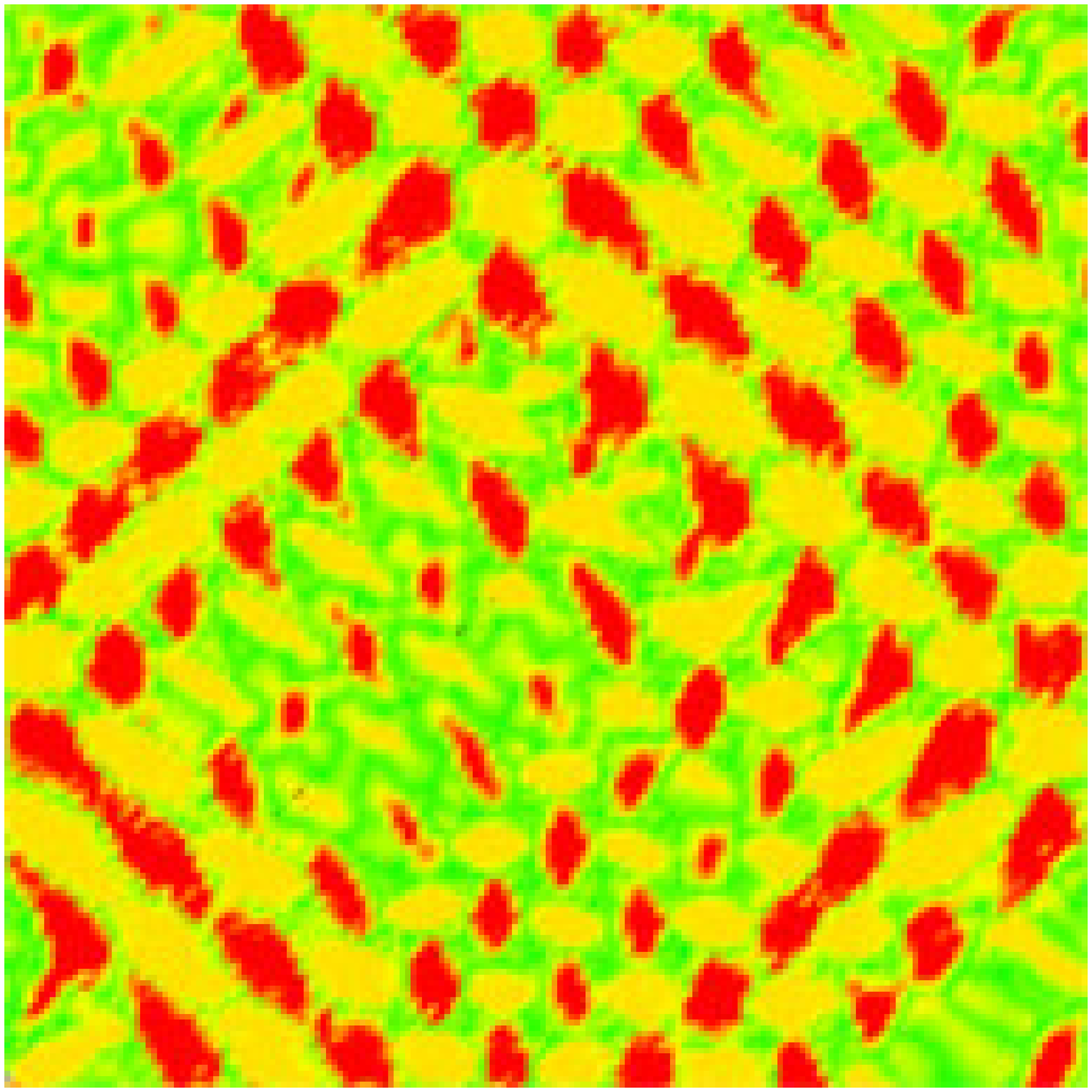}
	\put(-0.65, 1.55){\bf(j)}
}\subfigure{
    \label{early-085_0-1000}
    \includegraphics[height=2.cm]{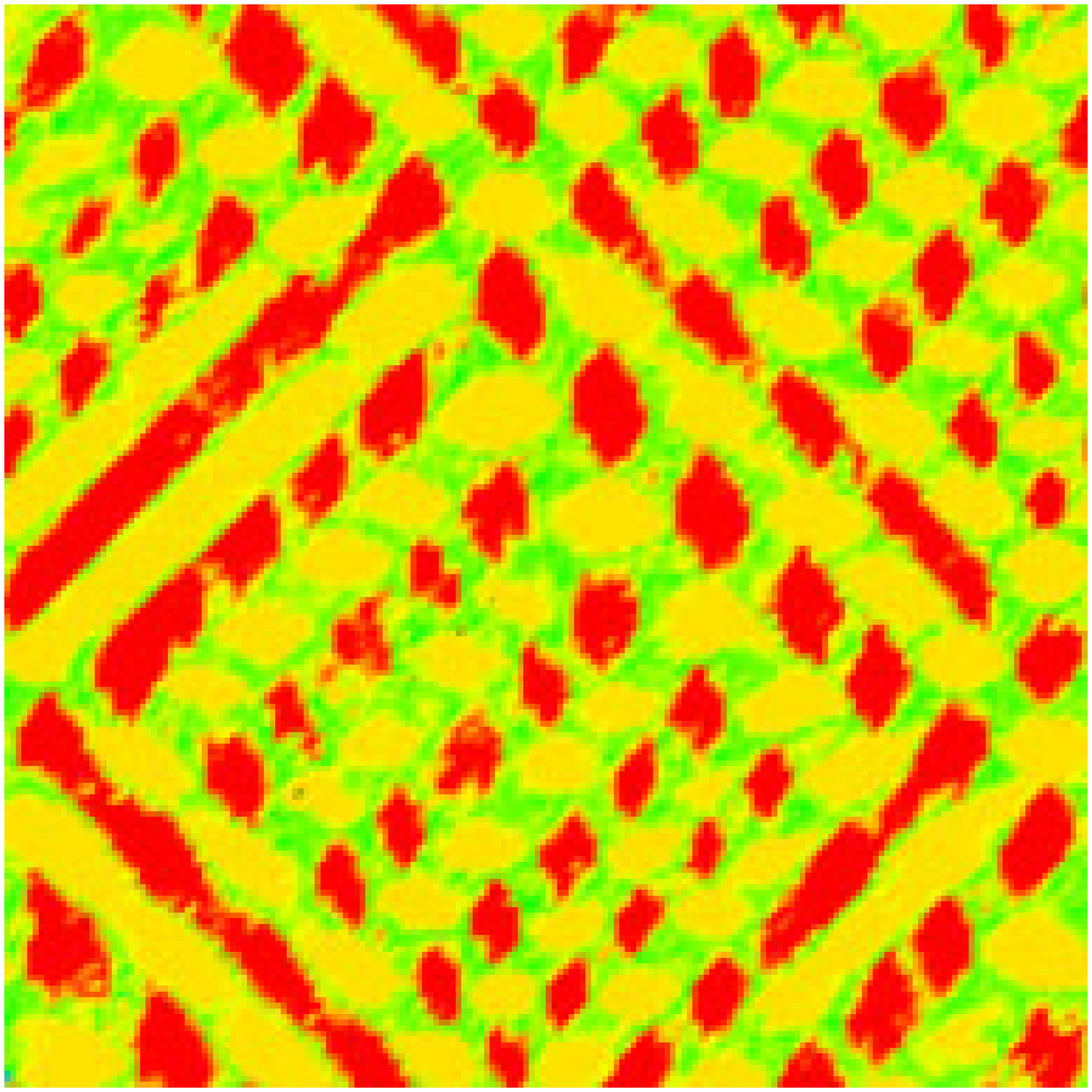}
	\put(-0.65, 1.55){\bf(k)}
}\subfigure{
    \label{early-085_0-1500}
    \includegraphics[height=2.cm]{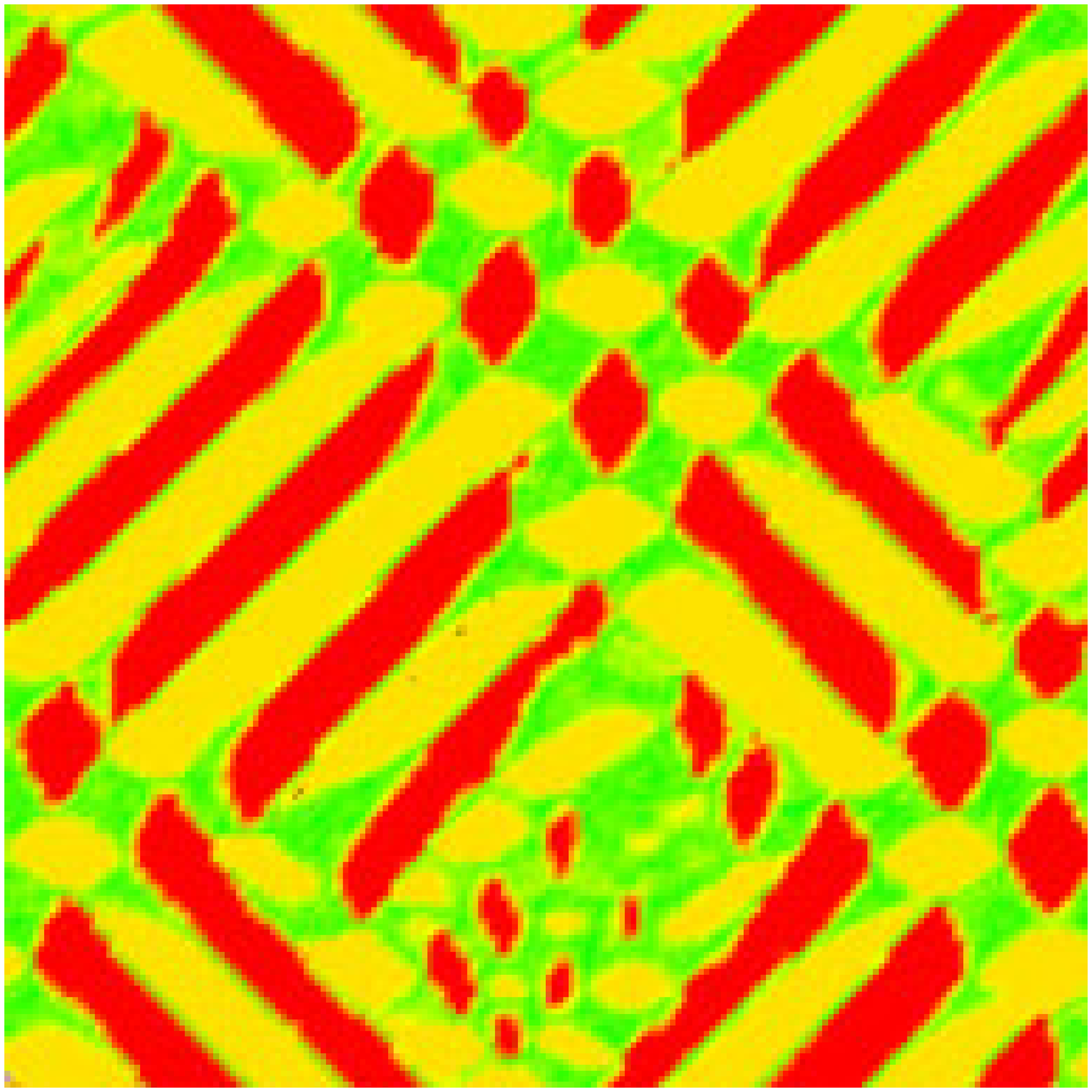}
	\put(-0.65, 1.55){\bf(l)}
}\vspace{-.2cm}\\
\subfigure{
	\put(-0.25, 0.5){\rotatebox{90}{$x_{12}=1$}}
    \label{early-1_0-850}
    \includegraphics[height=2.cm]{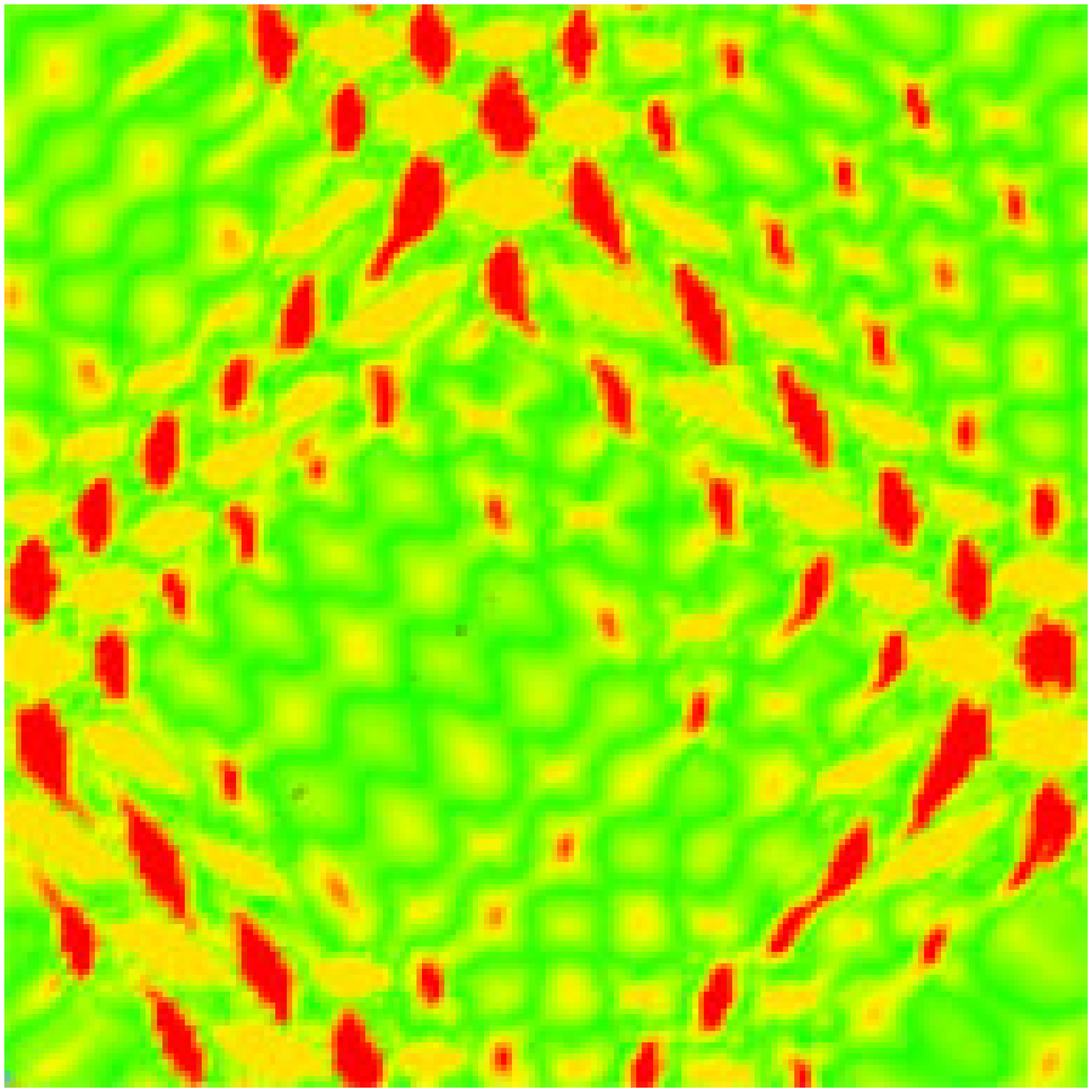}
	\put(-0.65, 1.55){\bf(m)}
}\subfigure{
    \label{early-1_0-900}
    \includegraphics[height=2.cm]{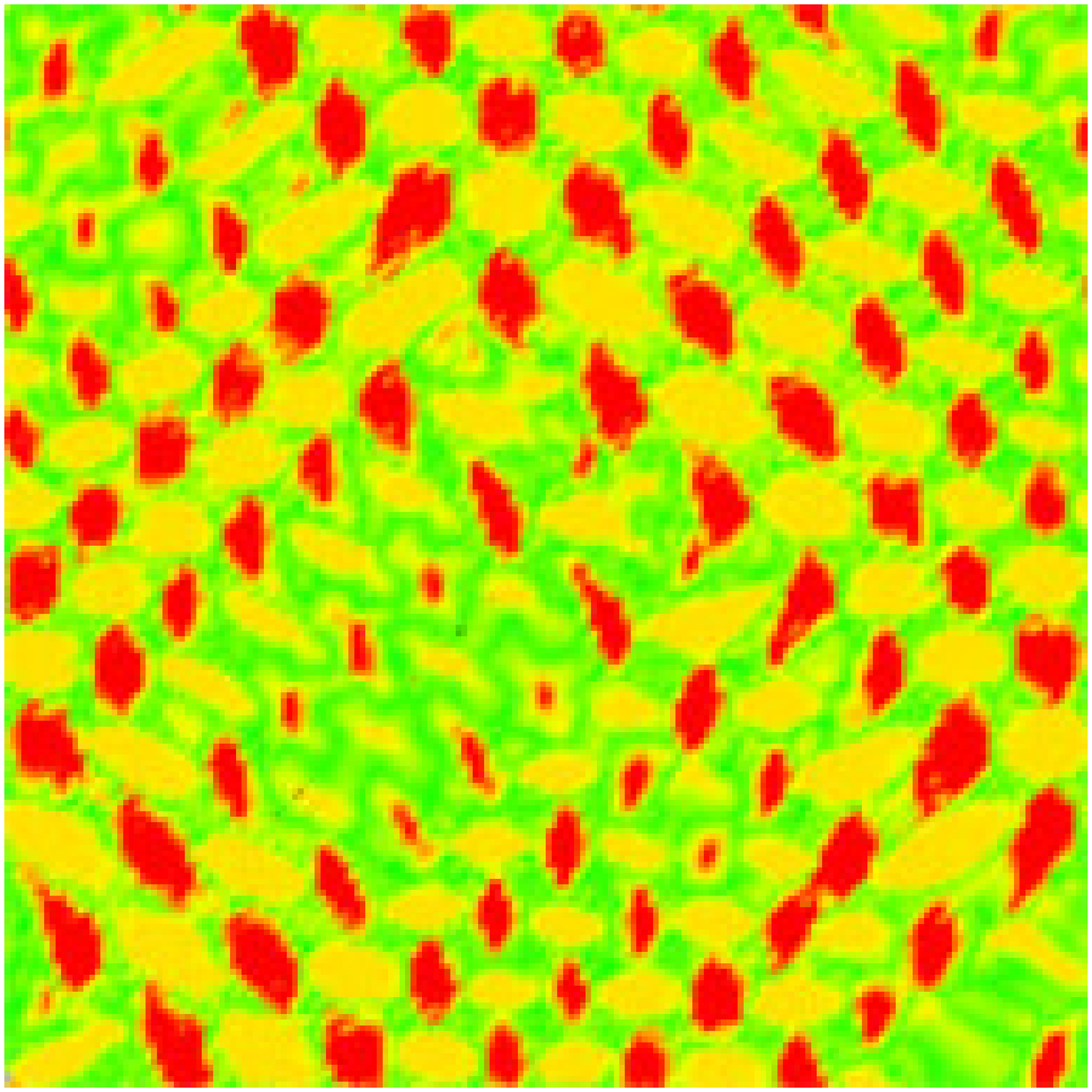}
	\put(-0.65, 1.55){\bf(n)}
}\subfigure{
    \label{early-1_0-1000}
    \includegraphics[height=2.cm]{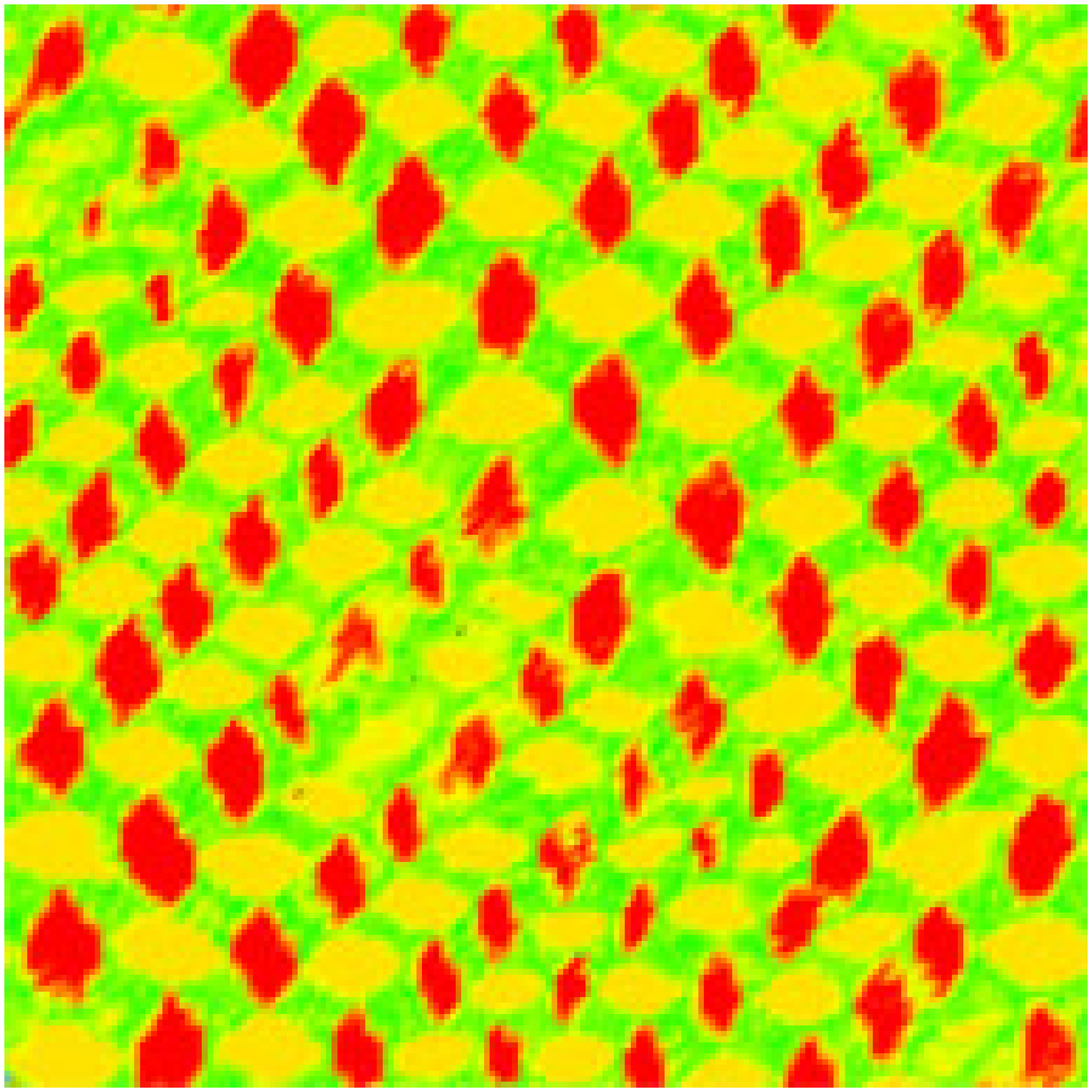}
	\put(-0.65, 1.55){\bf(o)}
}\subfigure{
    \label{early-1_0-1500}
    \includegraphics[height=2.cm]{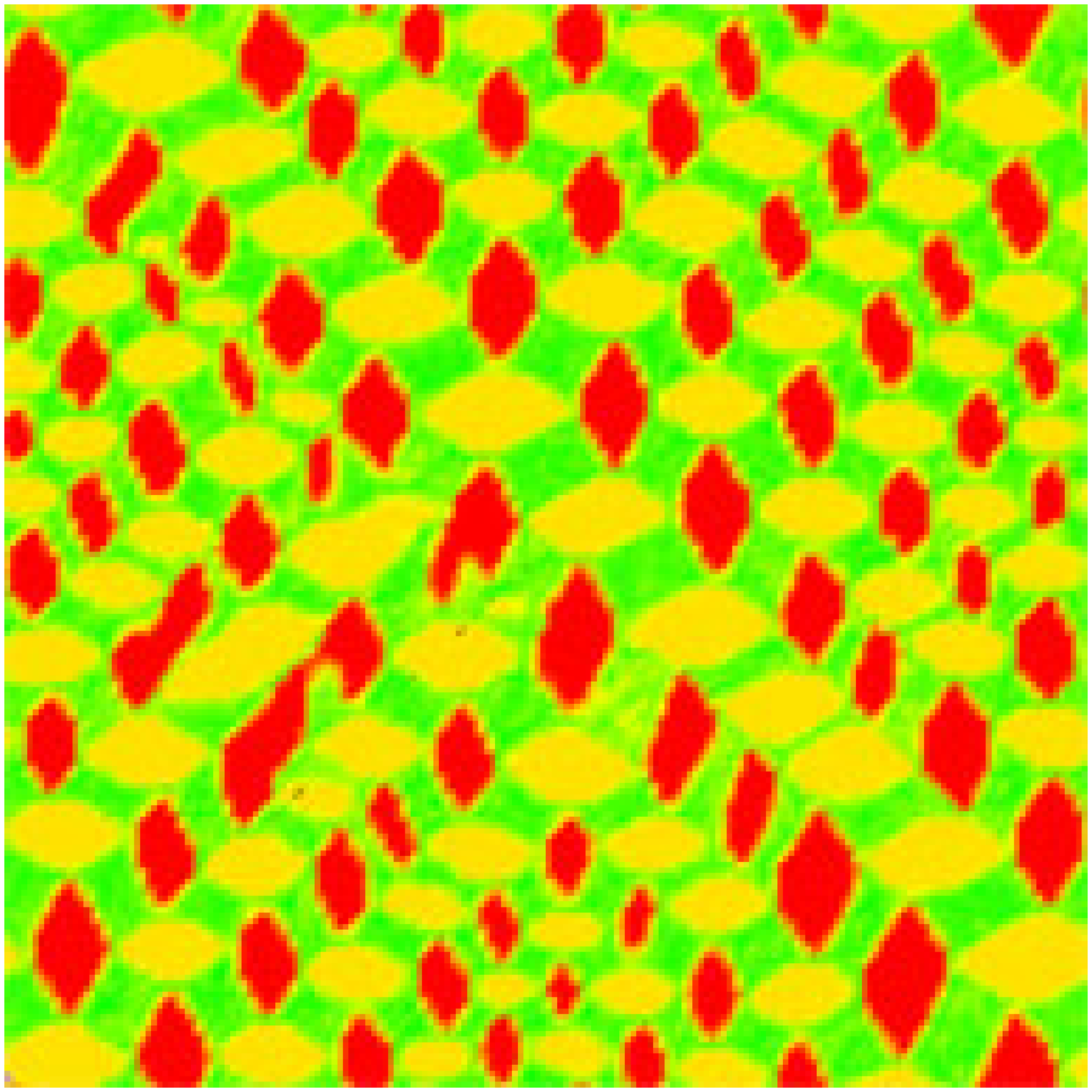}
	\put(-0.65, 1.55){\bf(p)}
}\vspace{-.2cm}\\
\subfigure{
	\put(-0.2, 0.3){\rotatebox{90}{$x_{12}=1.5$}}
    \label{early-15_0-850}
    \includegraphics[height=2.cm]{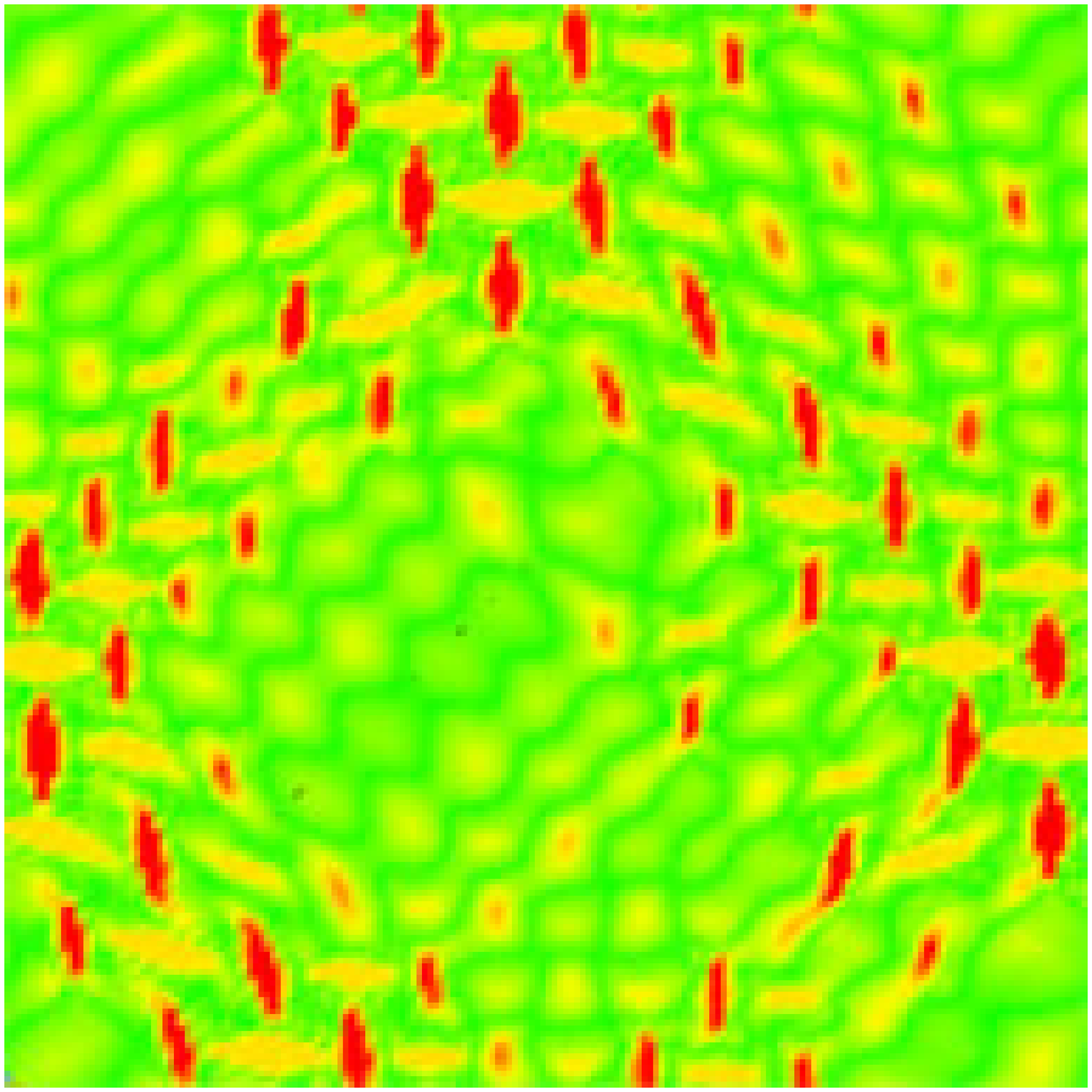}
	\put(-0.65, 1.55){\bf(q)}
}\subfigure{
    \label{early-15_0-900}
    \includegraphics[height=2.cm]{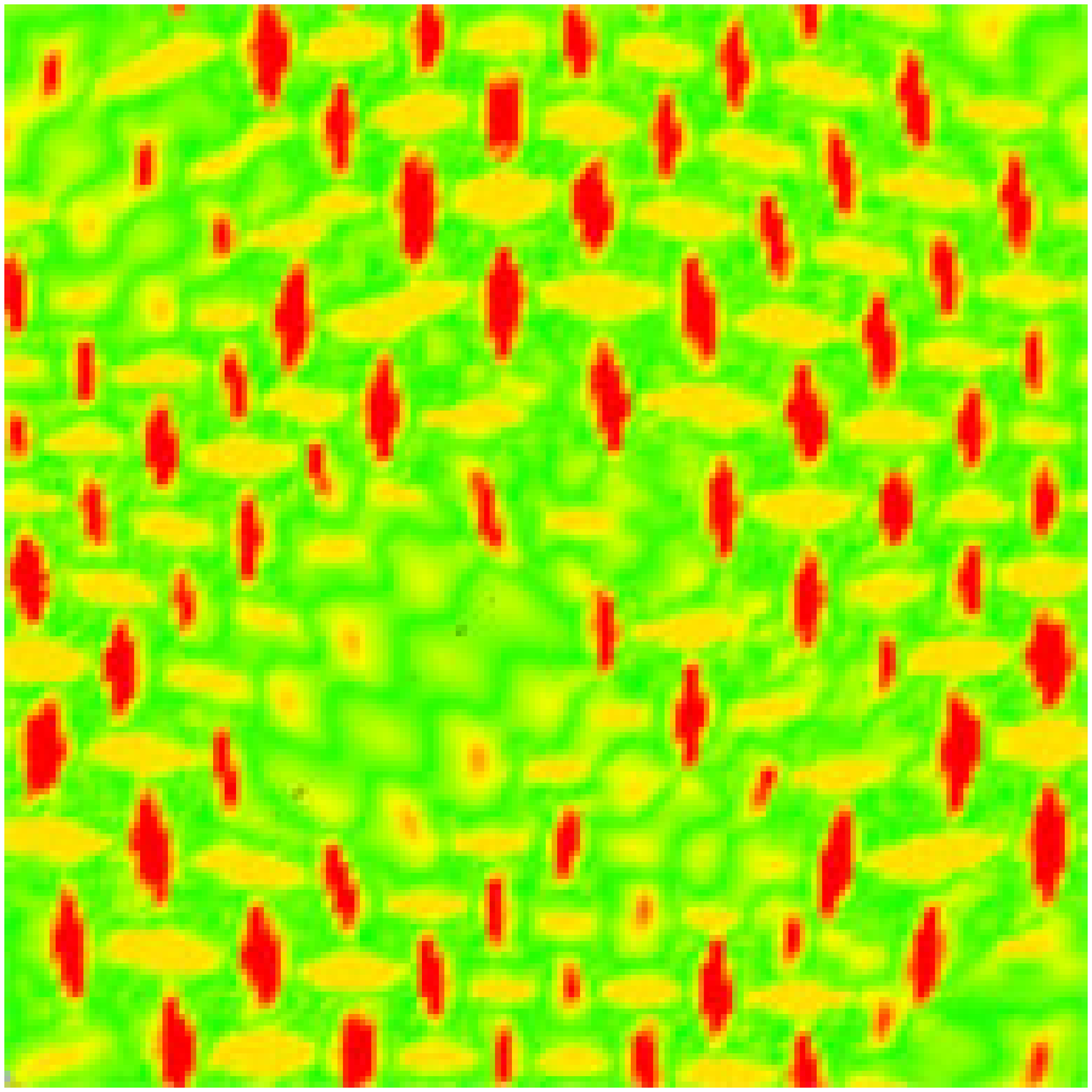}
	\put(-0.65, 1.55){\bf(r)}
}\subfigure{
    \label{early-15_0-1000}
    \includegraphics[height=2.cm]{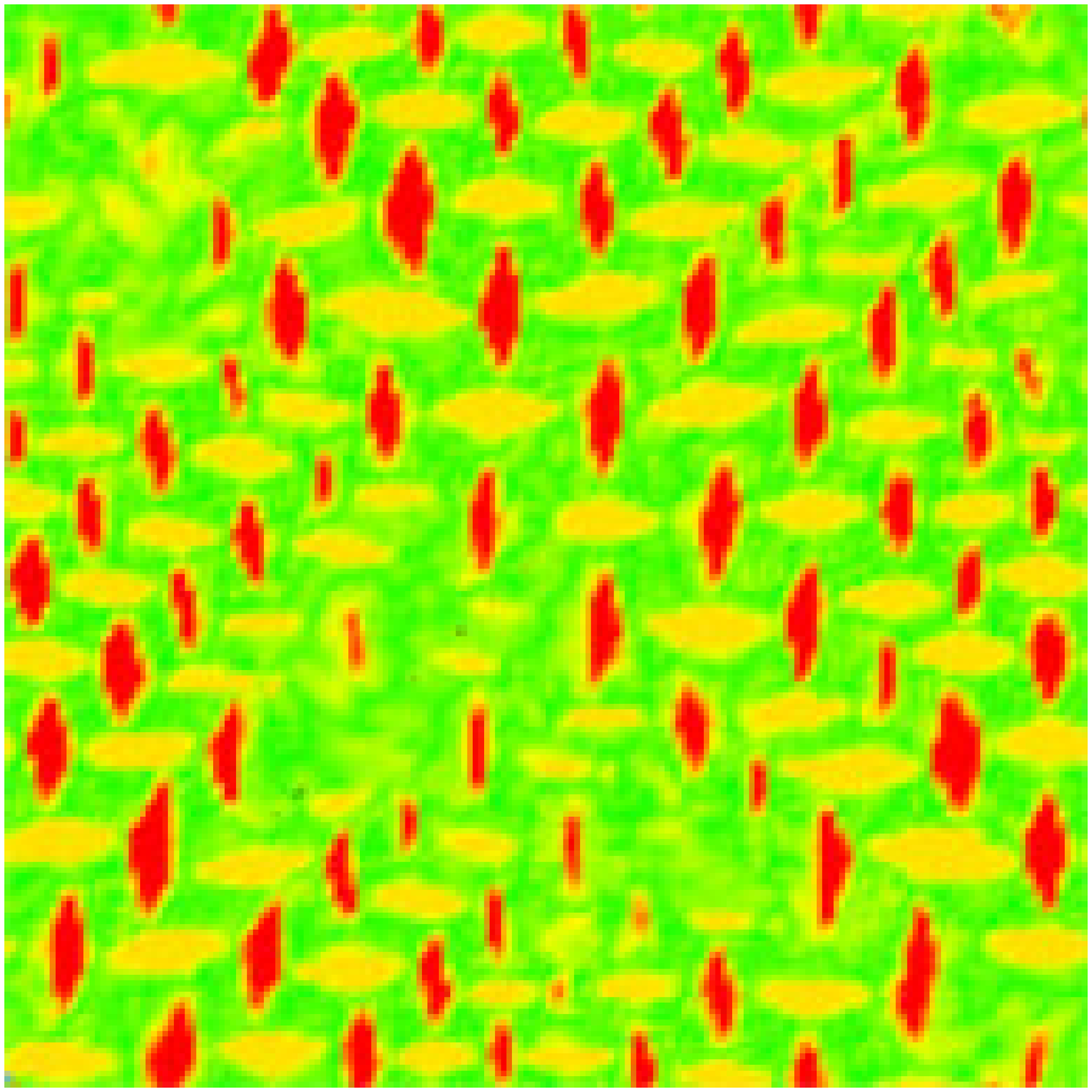}
	\put(-0.65, 1.55){\bf(s)}
}\subfigure{
    \label{early-15_0-1500}
    \includegraphics[height=2.cm]{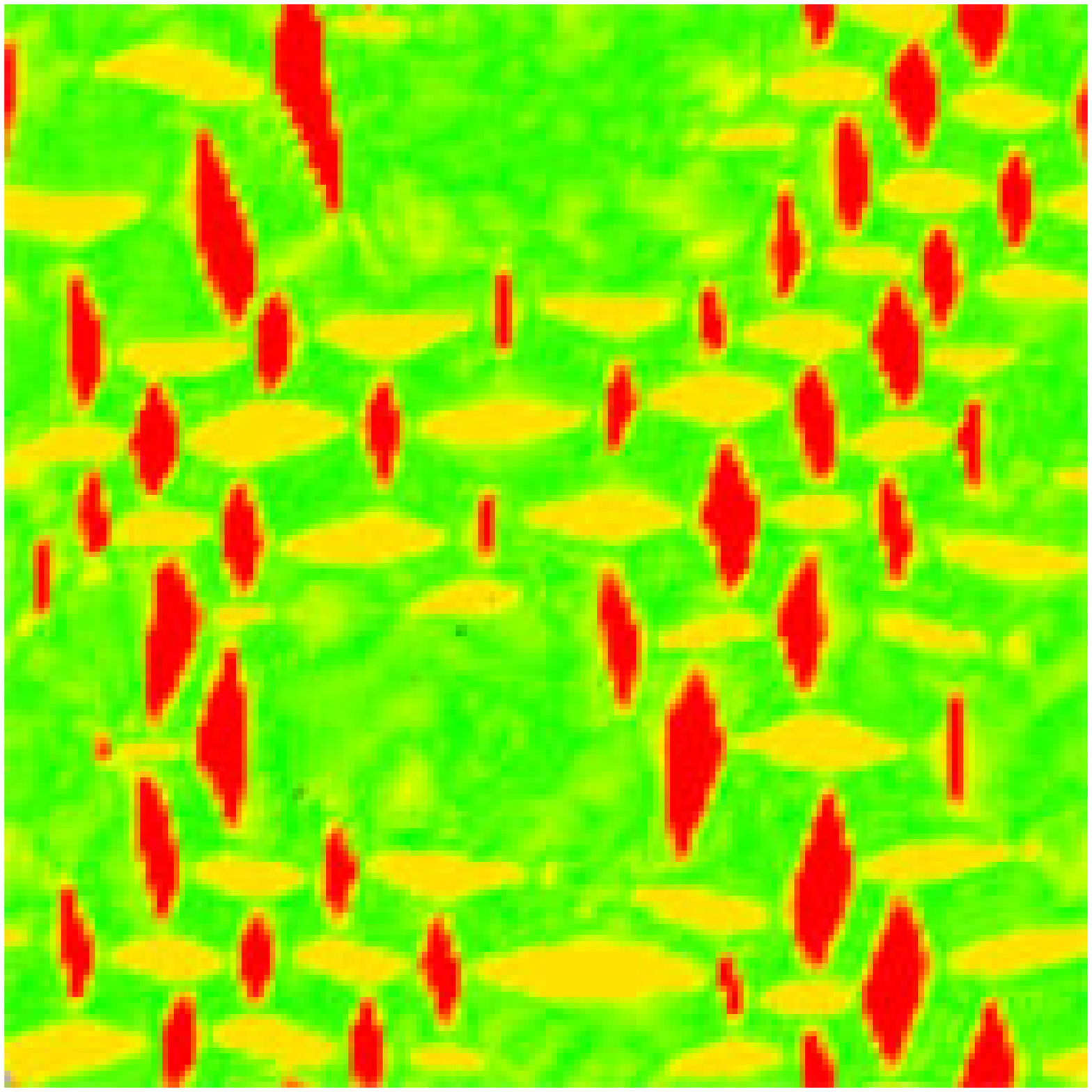}
	\put(-0.65, 1.55){\bf(t)}
}}
\end{picture}
	\caption{\label{early}(color online) Microstructures at $T=0.49$ and $x_{1c}=0$ for various times and values of $x_{12}$. 
Red and yellow: martensite; green: austenite.}
\end{figure}

If $x_{1c}\,c + x_{12}\,(e_2)^2$ is not equal to zero, neither is the equilibrium value of $e_1$.
Each of the two phases may then have a different value of $e_1$ and $e_2$. Calling $\delta e_1$ and $\delta e_2$ the jumps of $e_1$ and $e_2$ across the interface, Eqs.~(\ref{derivatives}) and~(\ref{compatibility}) give
\begin{equation*}
	\delta e_1\cos^2\theta + \delta e_1\sin^2\theta - \delta e_2\cos^2\theta  + \delta e_2\sin^2\theta = 0.
\end{equation*}
\noindent This condition is satisfied if
\begin{equation}
	\cos 2\theta = \frac{\delta e_1}{\delta e_2}.
	\label{compatibility4}
\end{equation}
\noindent For an interface between two martensite variants, $e_1$ is constant ($\delta e_1 = 0$) and one recovers Eq.~(\ref{compatibility2}). For martensite--austenite and martensite--pearlite interfaces $\delta e_1$ needs not be zero. In the case of Fig.~\ref{128-1_1-T0490_t000050}, for instance, the values of $e_1$ and $e_2$ give $\theta \approx \pm 30^\text{o}$ or $\theta \approx \pm 60^\text{o}$, which is consistent with the interface orientations observed in the figure.

In the case of pearlite--pearlite interfaces (i.e.\ ferrite--cementite interfaces), $e_2=0$ everywhere and only $e_1$ varies spatially if $x_{1c} \ne 0$. Equation~(\ref{compatibility4}) then gives $\cos 2\theta = 0$, i.e.\ the interfaces are along $\langle 1\,1 \rangle$, as seen in Figs.~\ref{128-0_1-T0850_t002000} and~\ref{128-1_1-T0800_t010000}.  Although this result is identical to Eq.~(\ref{compatibility2}), the latter does not apply to pearlite as it assumes that $e_1 = 0$ and $e_2 \ne 0$, which is the opposite of the pearlite case. If $x_{1c} = 0$ then both $e_1$ and $e_2$ are zero everywhere and there is no constraint on the orientation of the interfaces, as shown by Figs.~\ref{fine-pearlite} and~\ref{coarse-pearlite}.

\subsection{\label{sec-early}Early stages of martensite formation}
\begin{figure}
\centering
\setlength{\unitlength}{1cm}
\begin{picture}(8.5,4.55)(-.1,0)
\shortstack[c]{
\subfigure{
	\put(-0.25, 0.25){\rotatebox{90}{$x_{12}=0.75$}}
	\put(0.75, 2.1){{$t=3$}}
    \label{128-075_0-T0490_t000003}
    \includegraphics[height=2.cm]{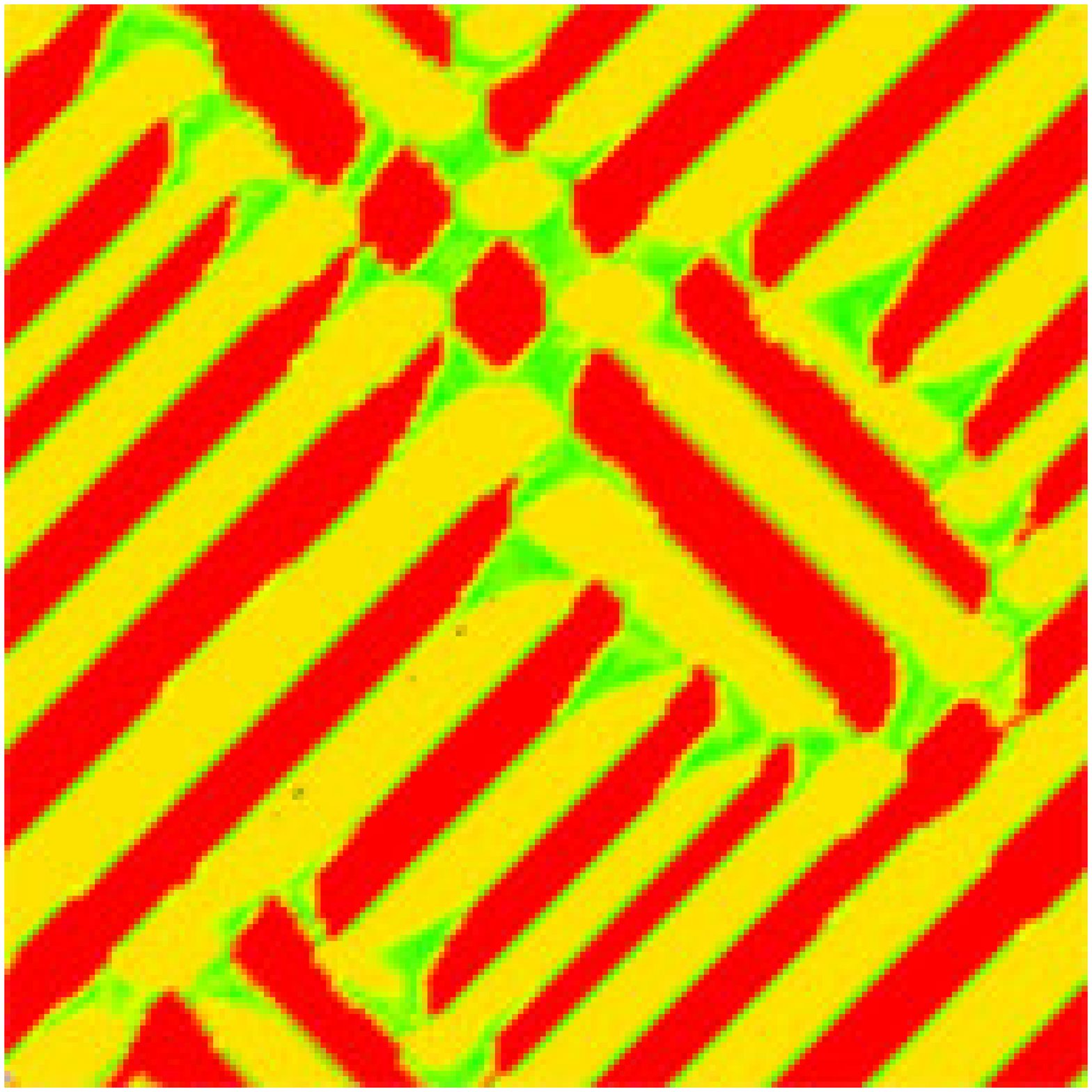}
	\put(-0.6, 1.55){\bf(a)}
}\subfigure{
	\put(0.75, 2.1){{$t=4$}}
    \label{128-075_0-T0490_t000004}
    \includegraphics[height=2.cm]{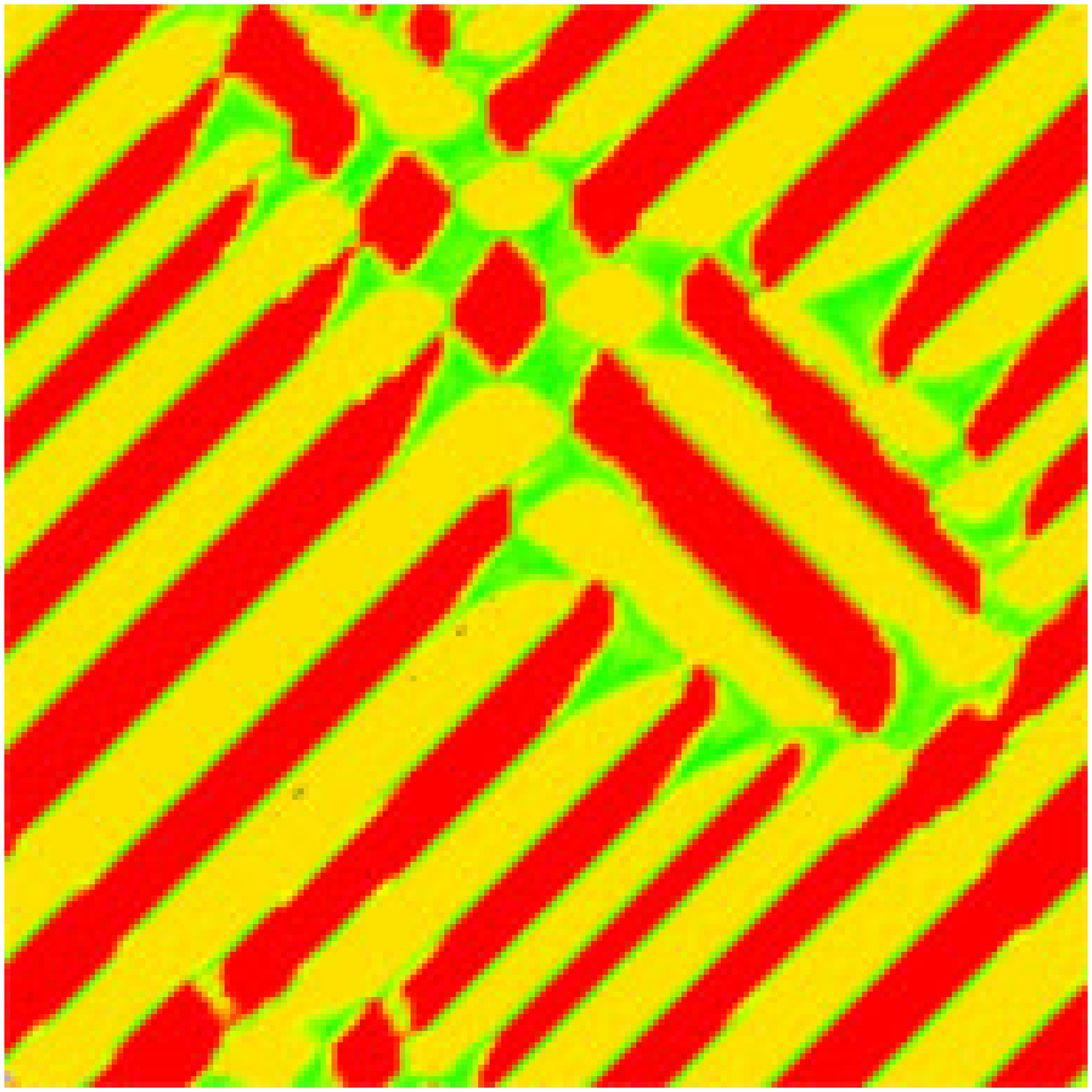}
	\put(-0.6, 1.55){\bf(b)}
}\subfigure{
	\put(0.75, 2.1){{$t=5$}}
    \label{128-075_0-T0490_t000005}
    \includegraphics[height=2.cm]{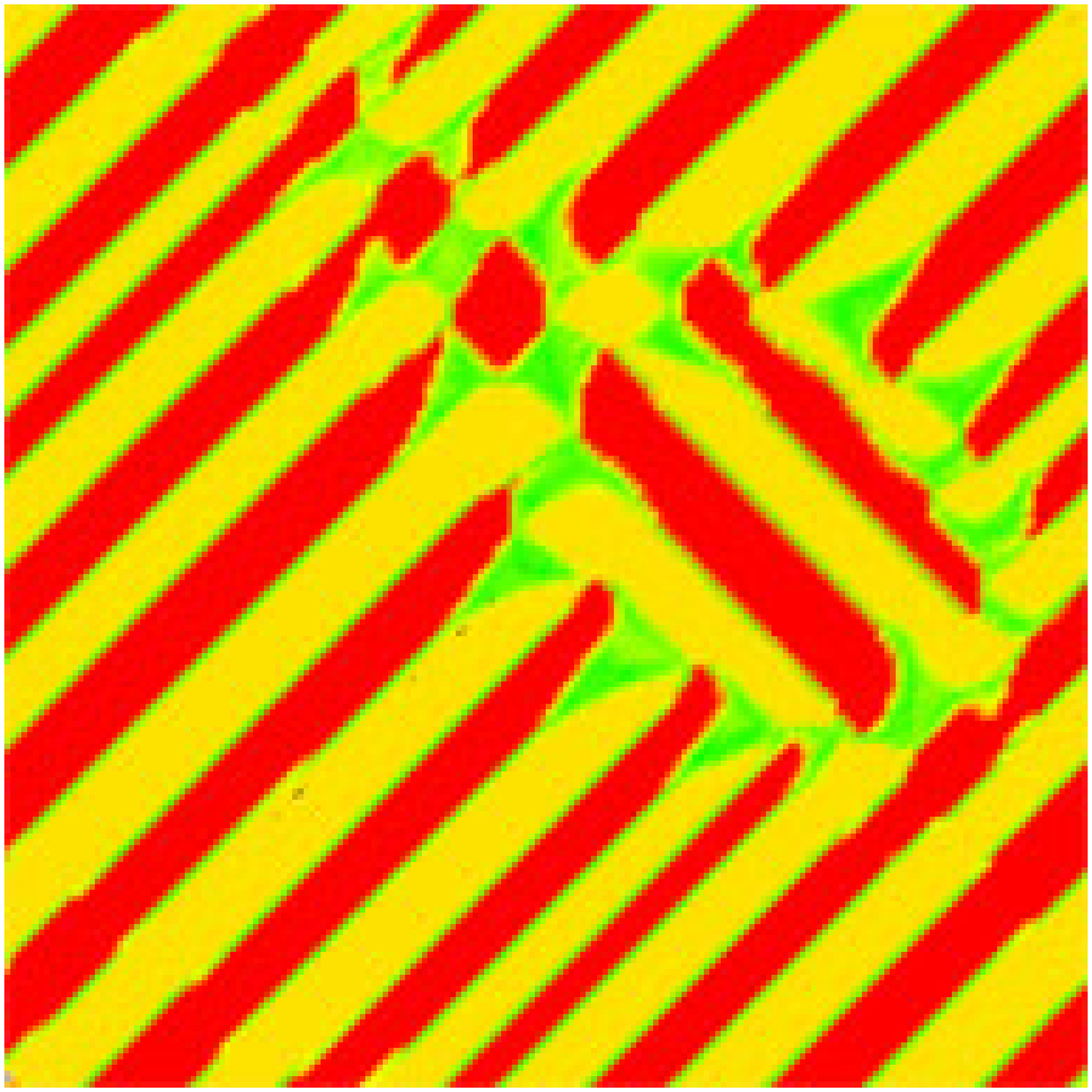}
	\put(-0.6, 1.55){\bf(c)}
}\subfigure{
	\put(0.65, 2.1){{$t=12$}}
    \label{128-075_0-T0490_t000012}
    \includegraphics[height=2.cm]{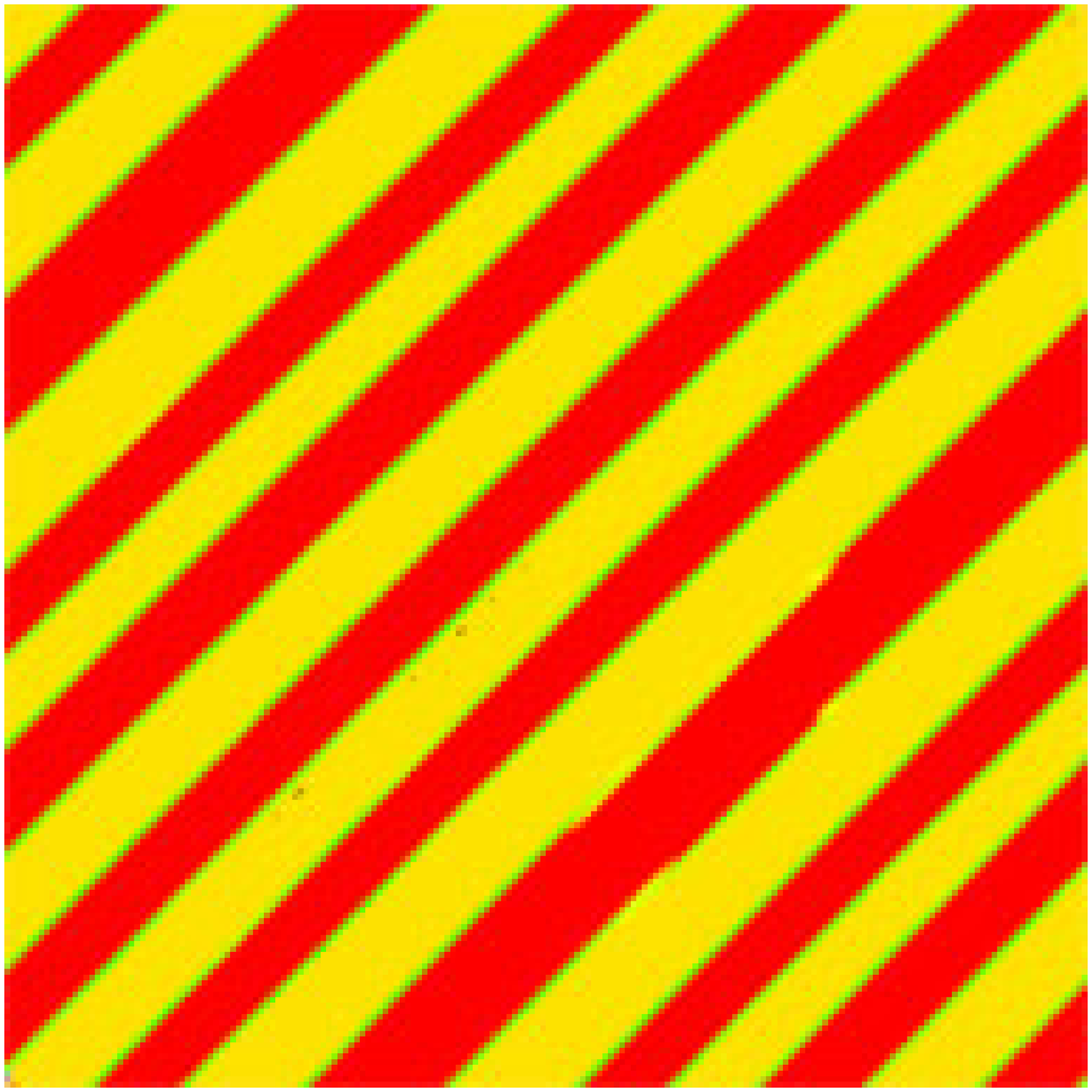}
	\put(-0.6, 1.55){\bf(d)}
}\vspace{-.2cm}\\
\subfigure{
	\put(-0.25, 0.25){\rotatebox{90}{$x_{12}=0.85$}}
    \label{128-085_0-T0490_t000003}
    \includegraphics[height=2.cm]{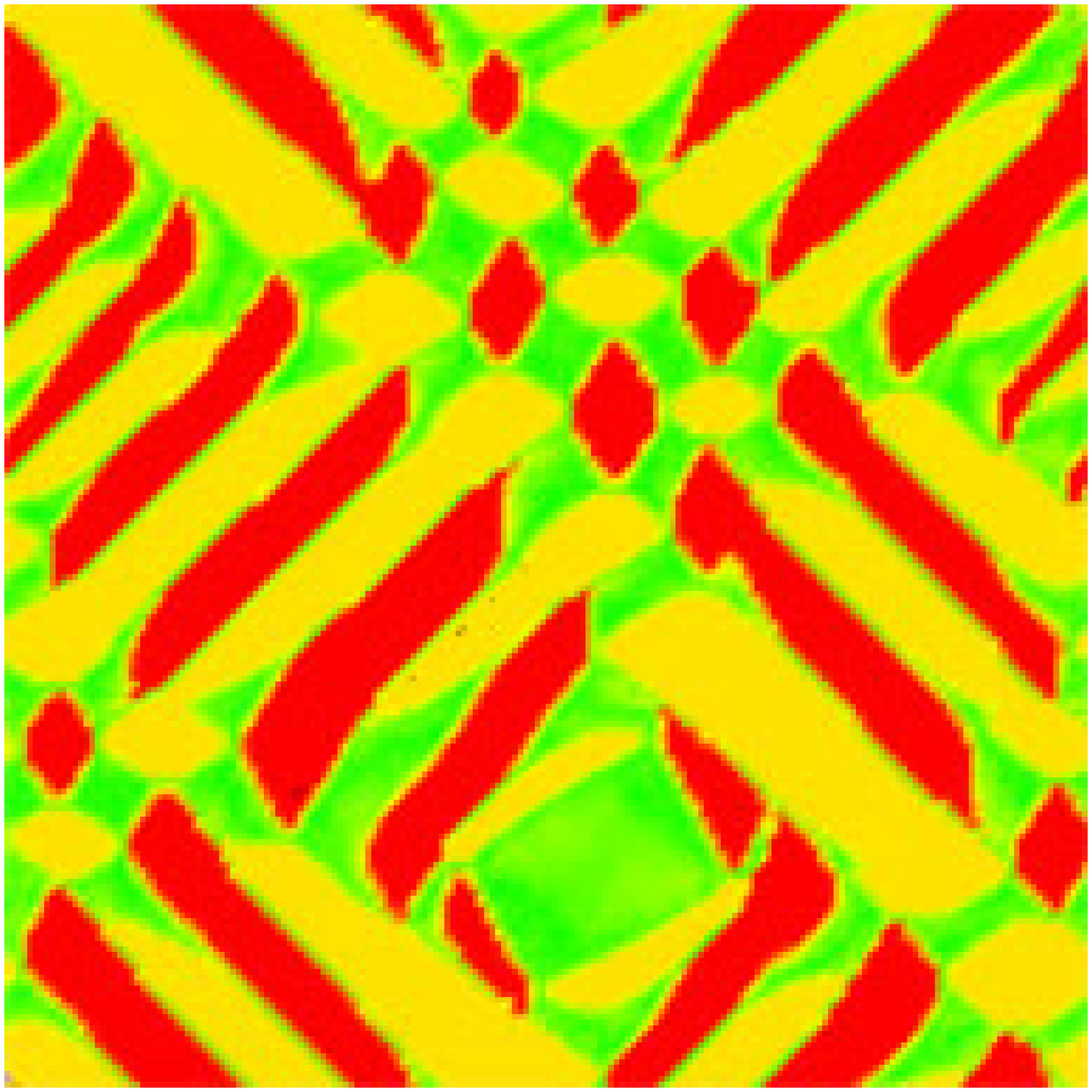}
	\put(-0.6, 1.55){\bf(e)}
}\subfigure{
    \label{128-085_0-T0490_t000004}
    \includegraphics[height=2.cm]{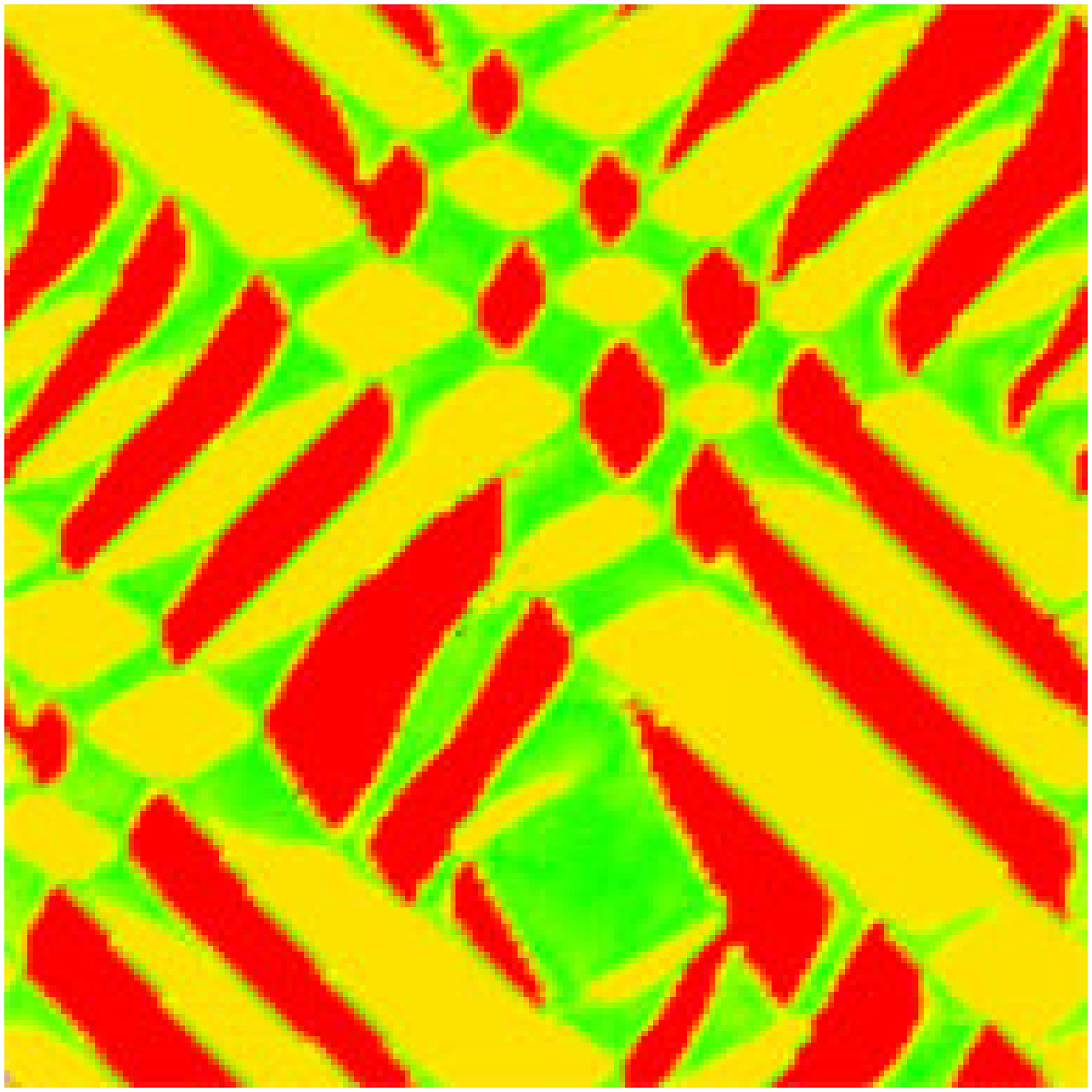}
	\put(-0.6, 1.55){\bf(f)}
}\subfigure{
    \label{128-085_0-T0490_t000005}
    \includegraphics[height=2.cm]{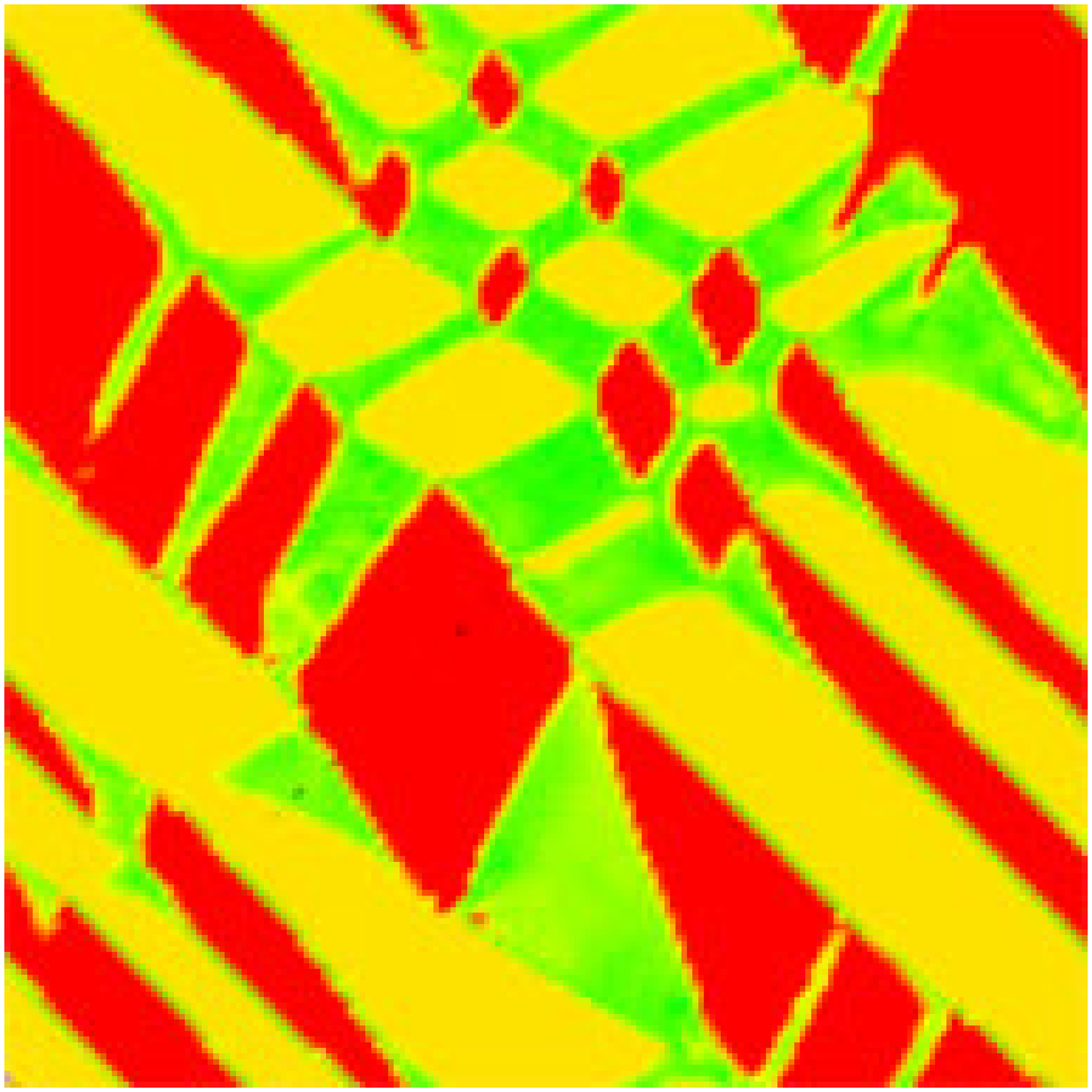}
	\put(-0.6, 1.55){\bf(g)}
}\subfigure{
    \label{128-085_0-T0490_t000012}
	\includegraphics[height=2.cm]{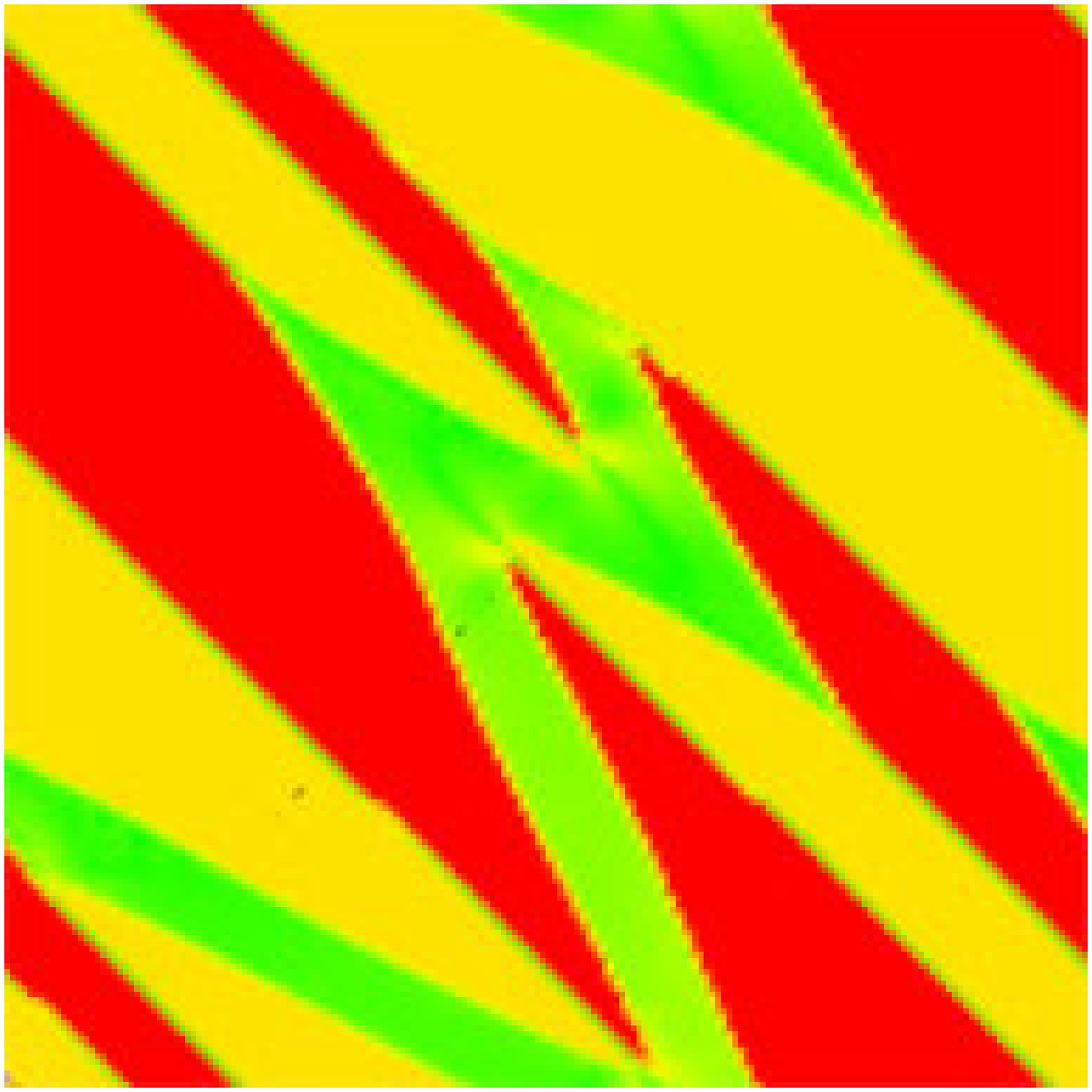}
	\put(-0.6, 1.55){\bf(h)}
}}
\end{picture}
	\caption{\label{early-2}(color online) Microstructures at $T=0.49$ and $x_{1c}=0$ for $x_{12}=0.75$ (top) and $x_{12}=0.85$ (bottom). 
Red and yellow: martensite; green: austenite.}
\end{figure}

Figures~\ref{128-0_0-T0490_t000050}--\ref{128-15_0-T0490_t000050} show four different microstructures: pure martensite, mostly martensitic with small austenitic regions, patterned (checkerboard) martensite--austenite mixture, and isolated martensite grains in an austenite matrix.
Figure~\ref{early} shows the early stages of martensite formation leading to these microstructures.\footnote{As was already mentioned, $x_{1c}$ plays a minor role at short times, therefore we simply use $x_{1c} = 0$ to obtain Figs.~\ref{early} and~\ref{early-2}. All cases shown in Figs.~\ref{early} and~\ref{early-2} have the same (random) initial conditions.}

The initial stages of martensite formation are very similar for all values of $x_{12}$ and the differentiation between these various types of microstructures occurs rather late. 
At $t=0.85$ and $t=0.9$ the differences between $x_{12} = 0.75$
, $x_{12} = 0.85$
, and $x_{12} = 1$ 
are very minor. Differentiation occurs at $t=1$: continuous strips form at $x_{12} = 0.75$ and $x_{12} = 0.85$ [Figs.~\ref{early-075_0-1000} and~\ref{early-085_0-1000}] but not at $x_{12} = 1$, Fig.~\ref{early-1_0-1000}.

\begin{figure*}
\centering
\setlength{\unitlength}{1cm}
\begin{picture}(17,6.2)(.1,0)
\shortstack[c]{
\subfigure{
	\put(1,2.95){$x_{1c}=0$}
	\put(-0.25, 0.85){\rotatebox{90}{$x_{12}=0$}}
    \label{128-0_0-T0490_t020000}
    \includegraphics[height=2.8cm]{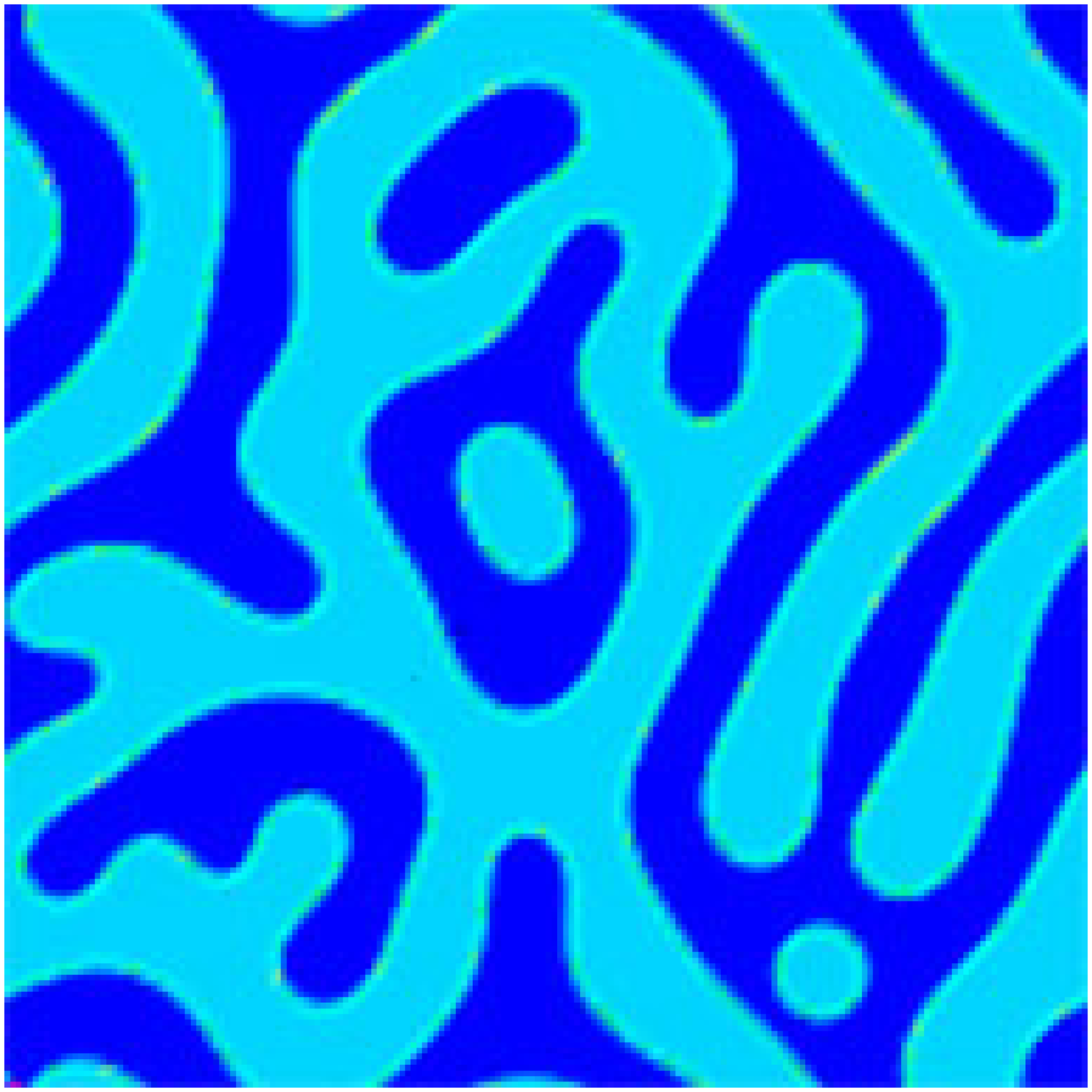}
	\put(-0.65, 2.35){\color{white}\bf(a)}
}\subfigure{
	\put(.8,2.9){$x_{1c}=0.25$}
    \label{128-0_025-T0490_t020000}
    \includegraphics[height=2.8cm]{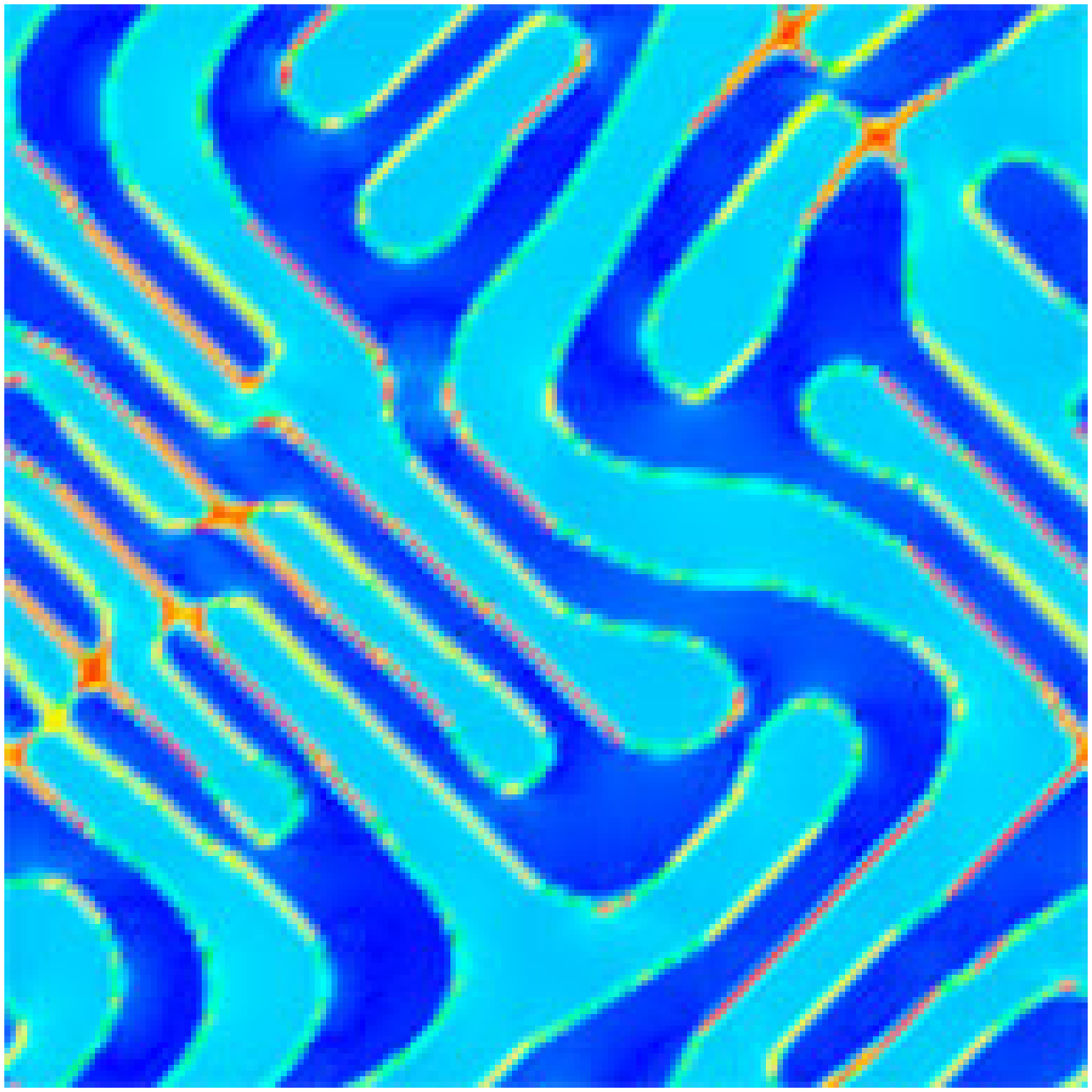}
	\put(-0.65, 2.35){\color{white}\bf(b)}
}\subfigure{
	\put(.85,2.9){$x_{1c}=0.5$}
    \label{128-0_05-T0490_t020000}
    \includegraphics[height=2.8cm]{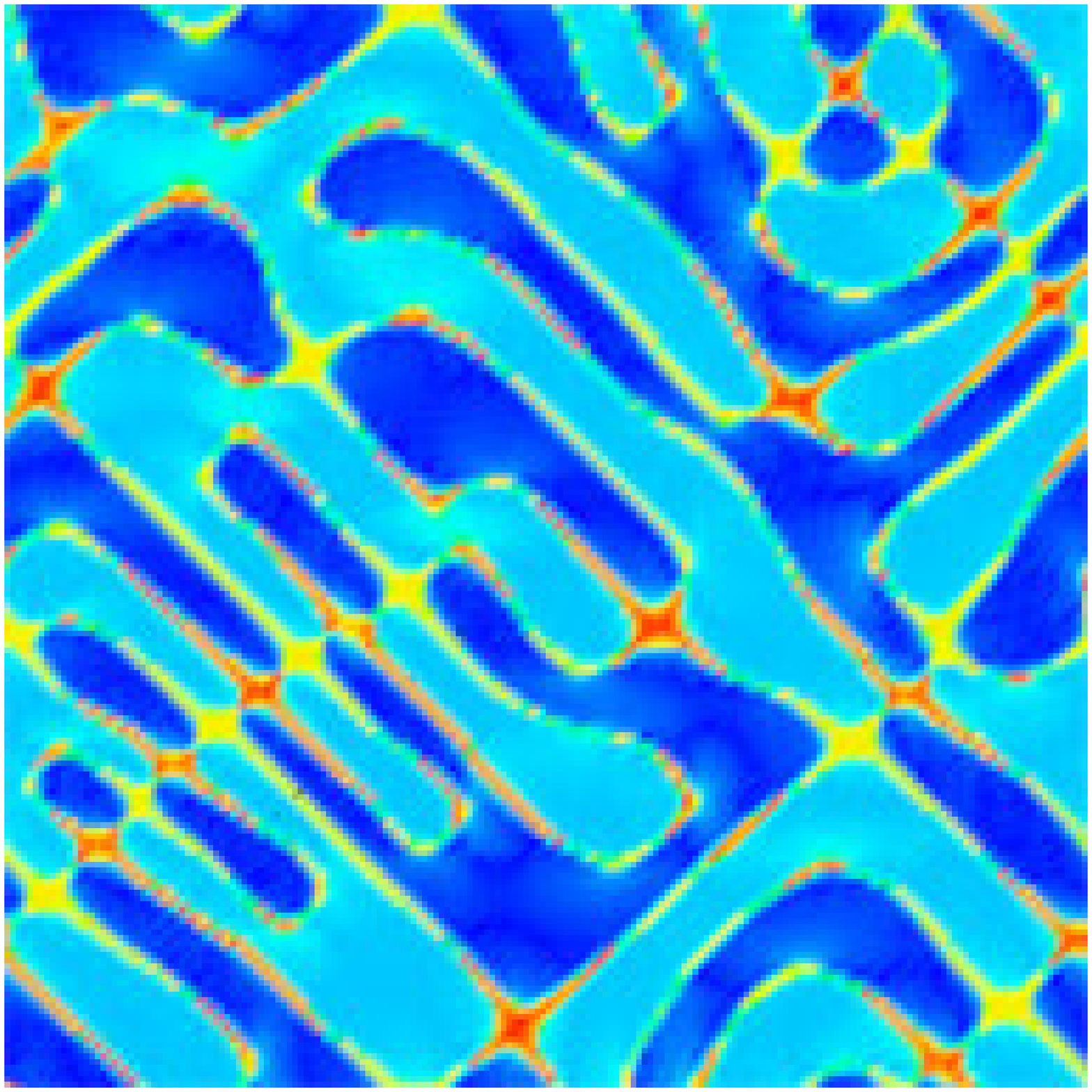}
	\put(-0.6, 2.35){\color{white}\bf(c)}
}\subfigure{
	\put(.8,2.9){$x_{1c}=0.75$}
    \label{128-0_075-T0490_t020000}
    \includegraphics[height=2.8cm]{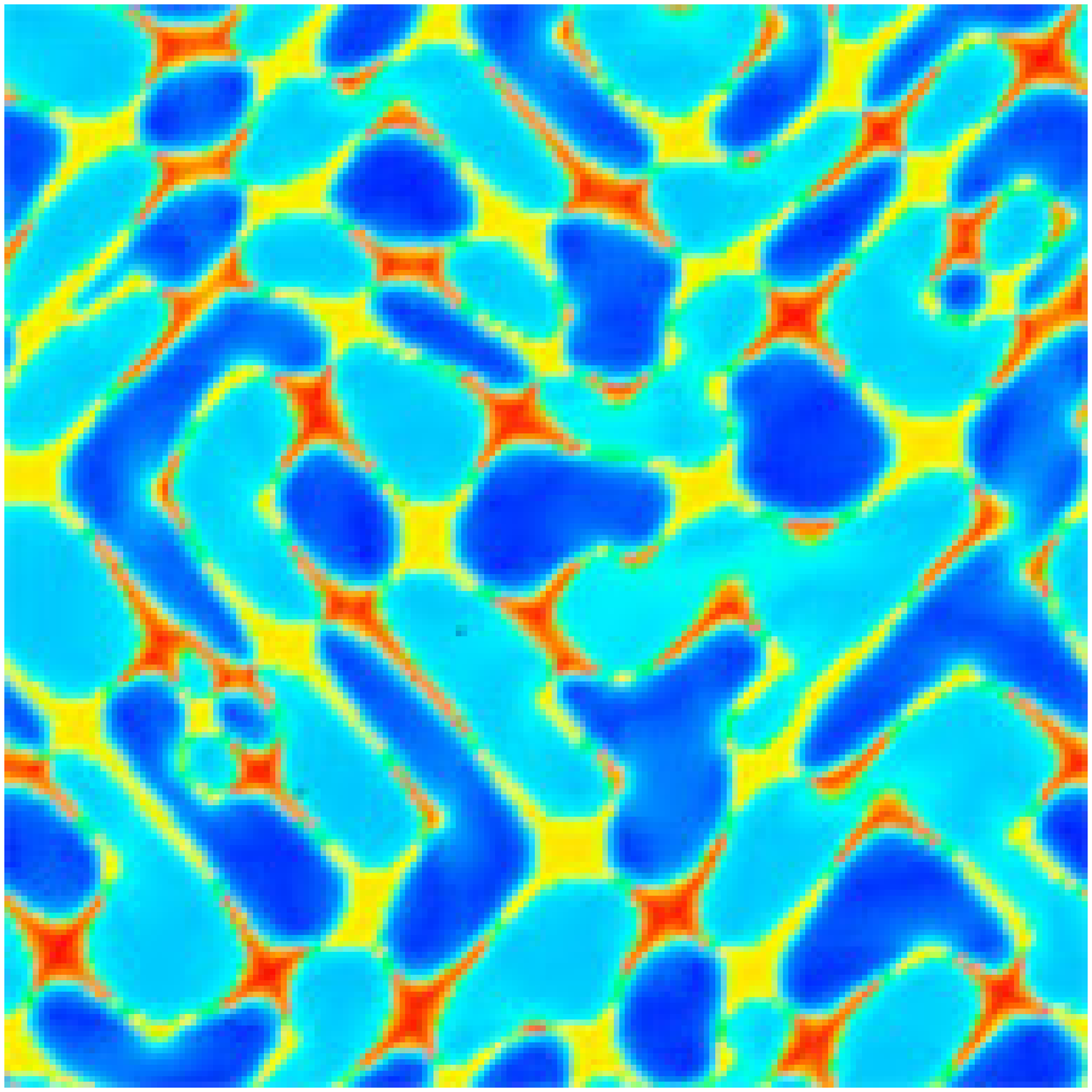}
	\put(-0.55, 2.35){\color{white}\bf(d)}
}\subfigure{
	\put(1,2.95){$x_{1c}=1$}
    \label{128-0_1-T0490_t020000}
    \includegraphics[height=2.8cm]{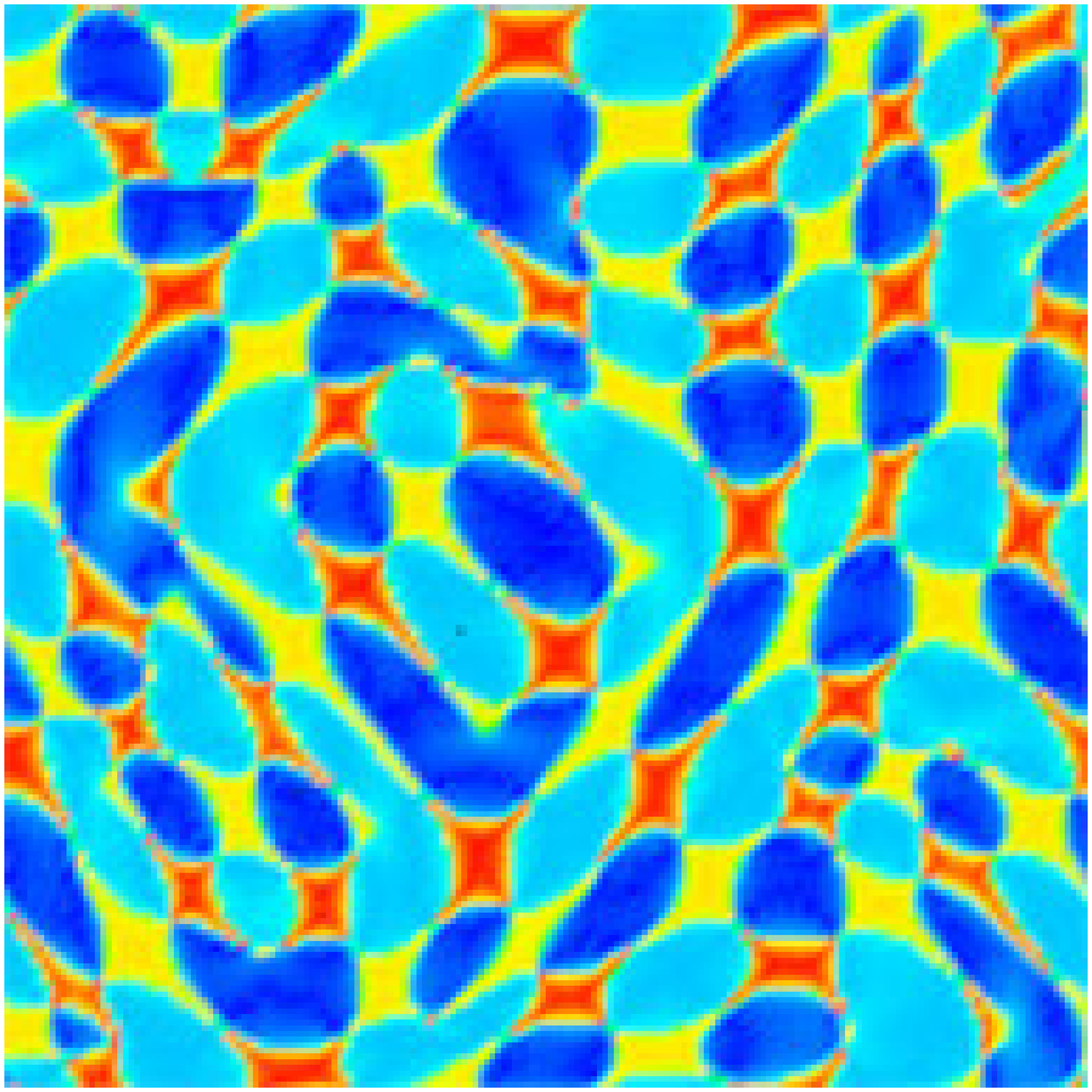}
	\put(-0.7, 2.35){\color{white}\bf(e)}
}\subfigure{
	\put(.85,2.9){$x_{1c}=1.5$}
    \label{128-0_15-T0490_t020000}
    \includegraphics[height=2.8cm]{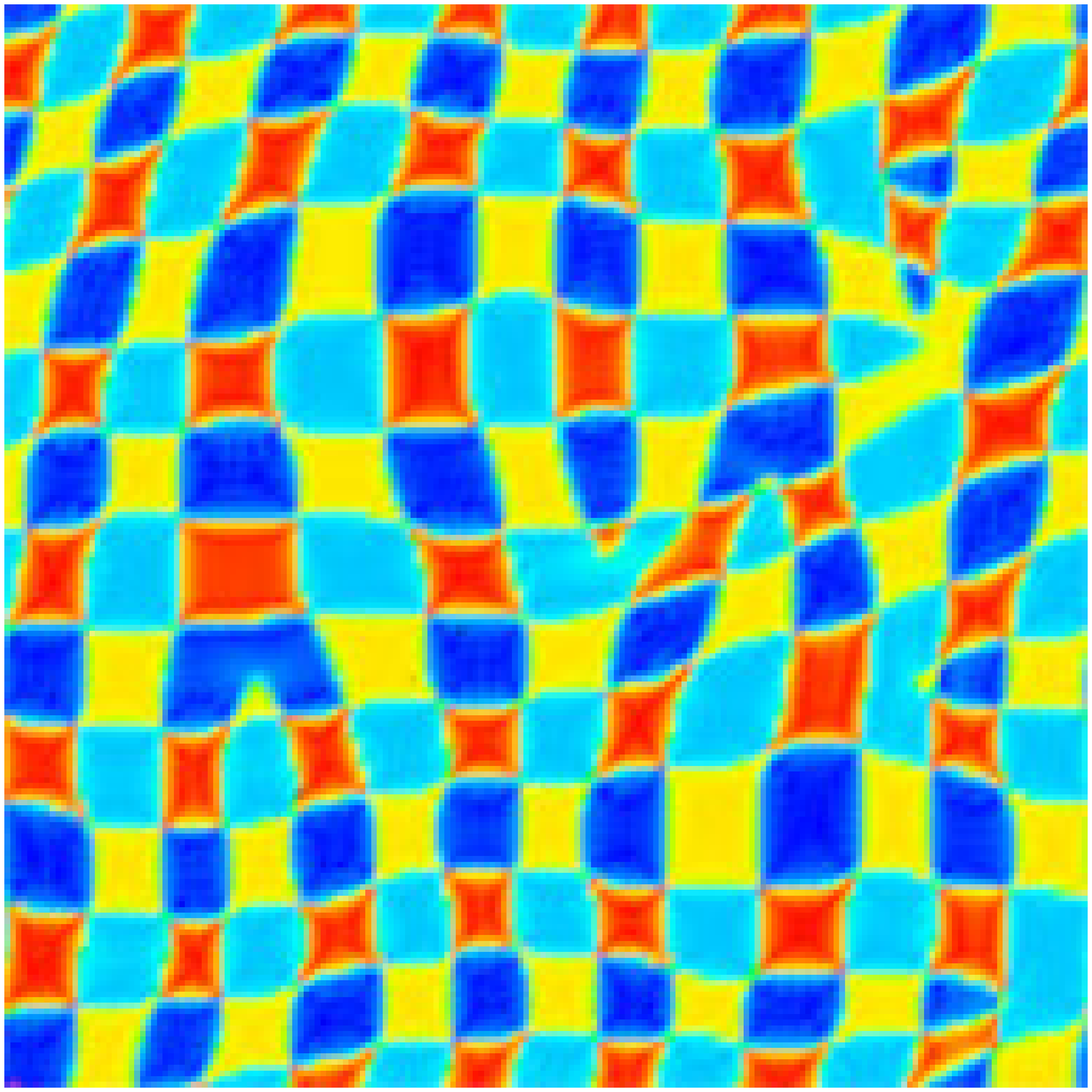}
	\put(-0.65, 2.35){\color{white}\bf(f)}
}\vspace{-.25cm}\\
\subfigure{
	\put(-0.25, 0.85){\rotatebox{90}{$x_{12}=1$}}
    \label{128-1_0-T0490_t020000}
    \includegraphics[height=2.8cm]{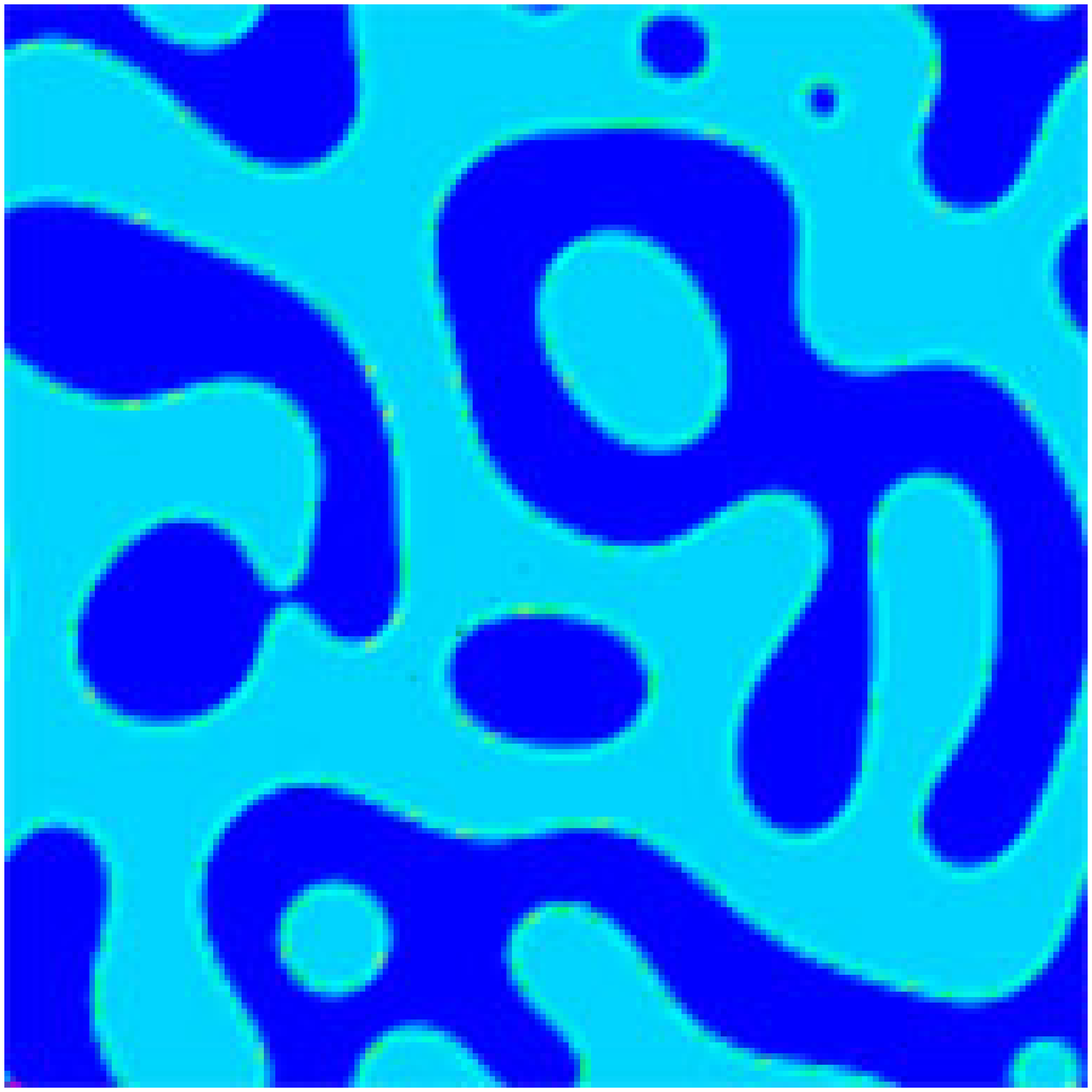}
	\put(-0.65, 2.35){\color{white}\bf(g)}
}\subfigure{
    \label{128-1_025-T0490_t020000}
    \includegraphics[height=2.8cm]{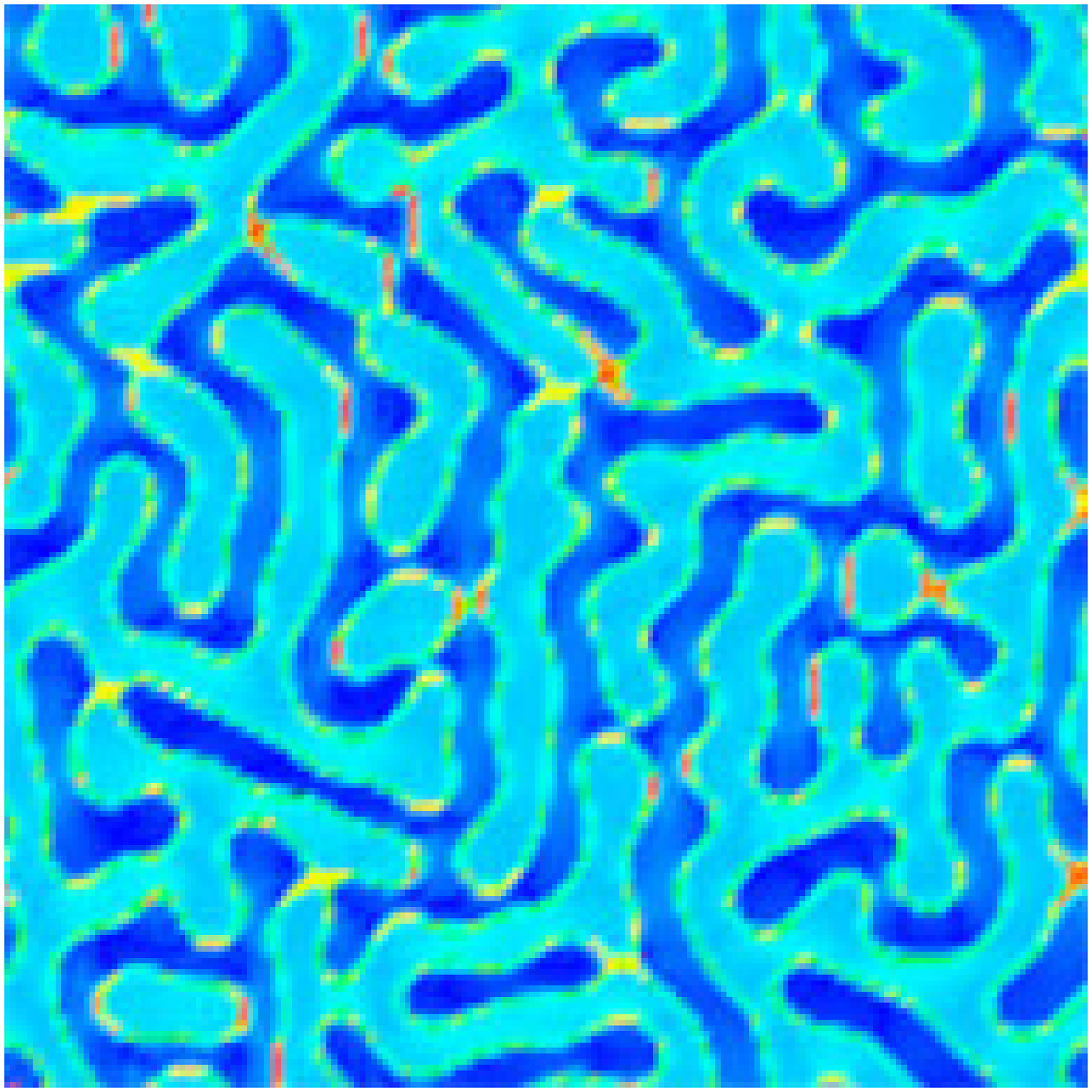}
	\put(-0.65, 2.35){\color{white}\bf(h)}
}\subfigure{
    \label{128-1_05-T0490_t020000}
    \includegraphics[height=2.8cm]{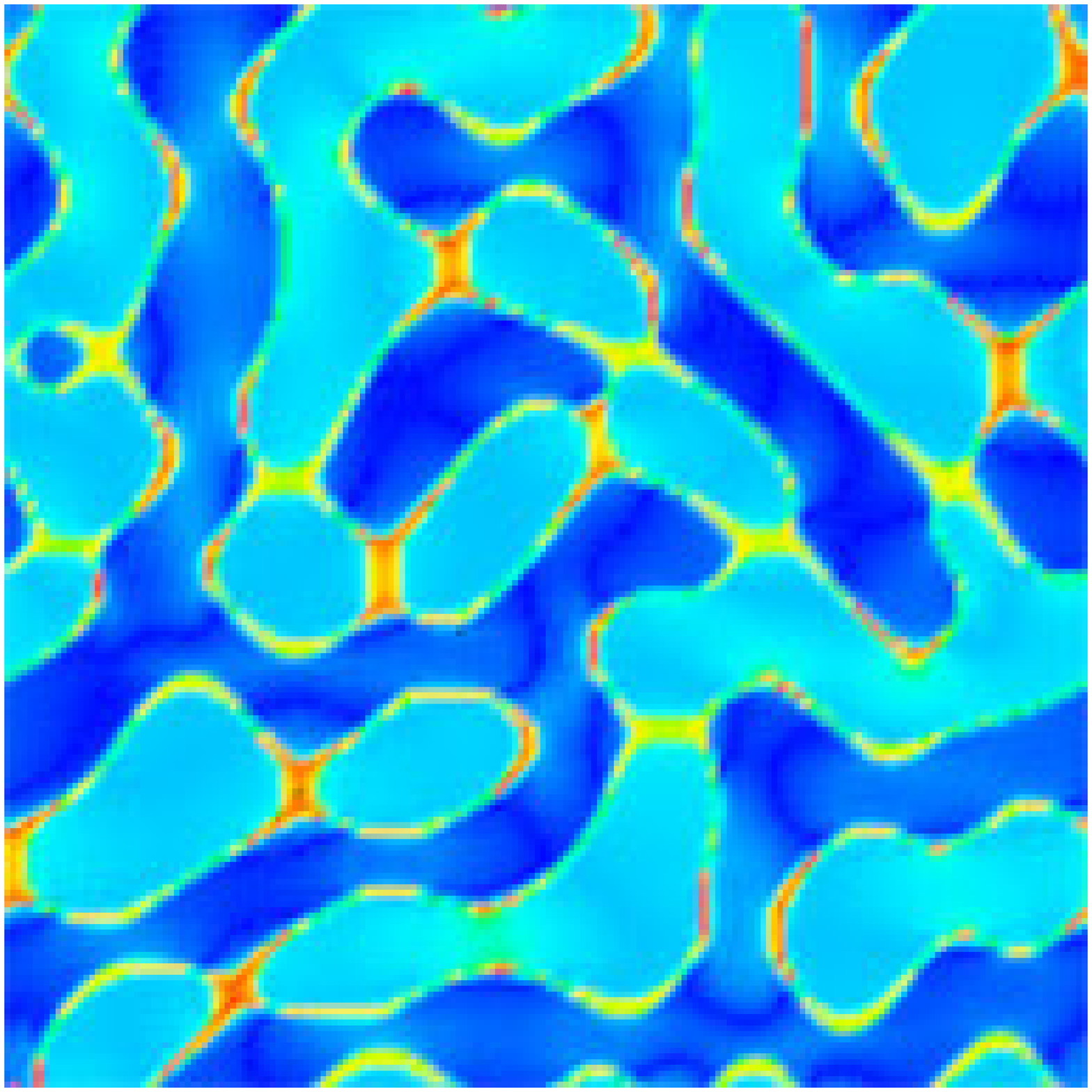}
	\put(-0.65, 2.35){\color{white}\bf(i)}
}\subfigure{
    \label{128-1_075-T0490_t020000}
    \includegraphics[height=2.8cm]{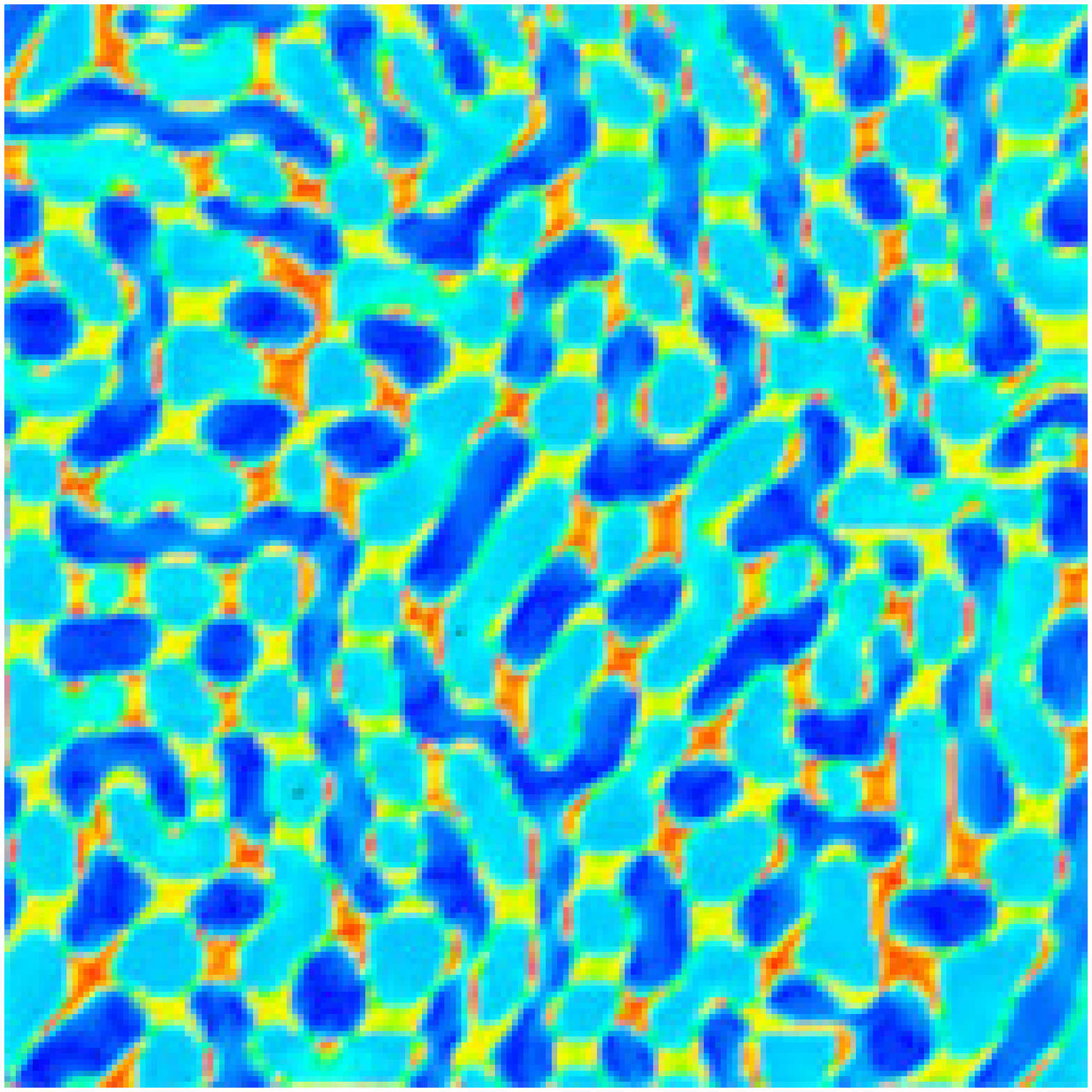}
	\put(-0.65, 2.35){\bf(j)}
}\subfigure{
    \label{128-1_1-T0490_t020000}
    \includegraphics[height=2.8cm]{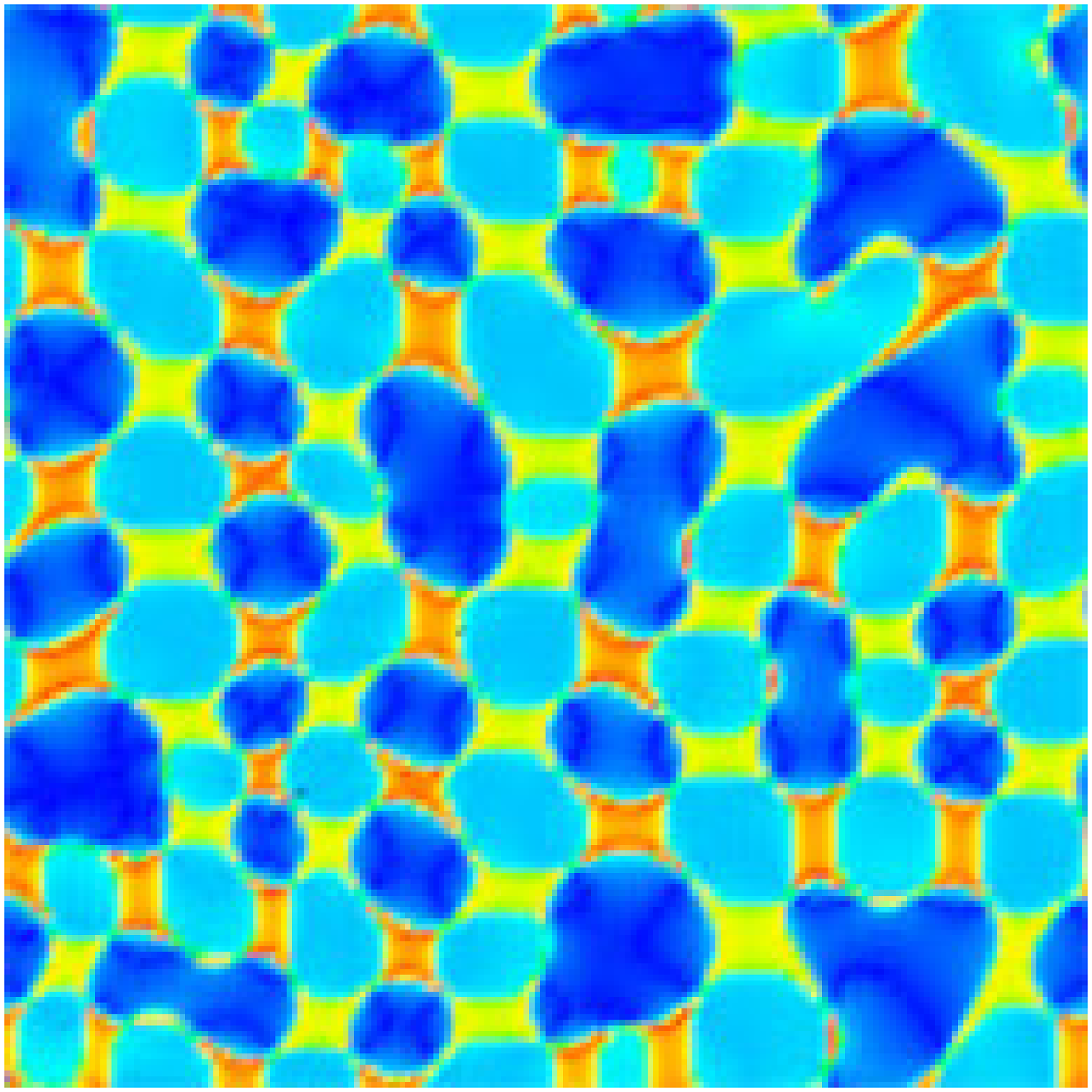}
	\put(-0.7, 2.35){\color{white}\bf(k)}
}\subfigure{
    \label{128-1_15-T0490_t020000}
    \includegraphics[height=2.8cm]{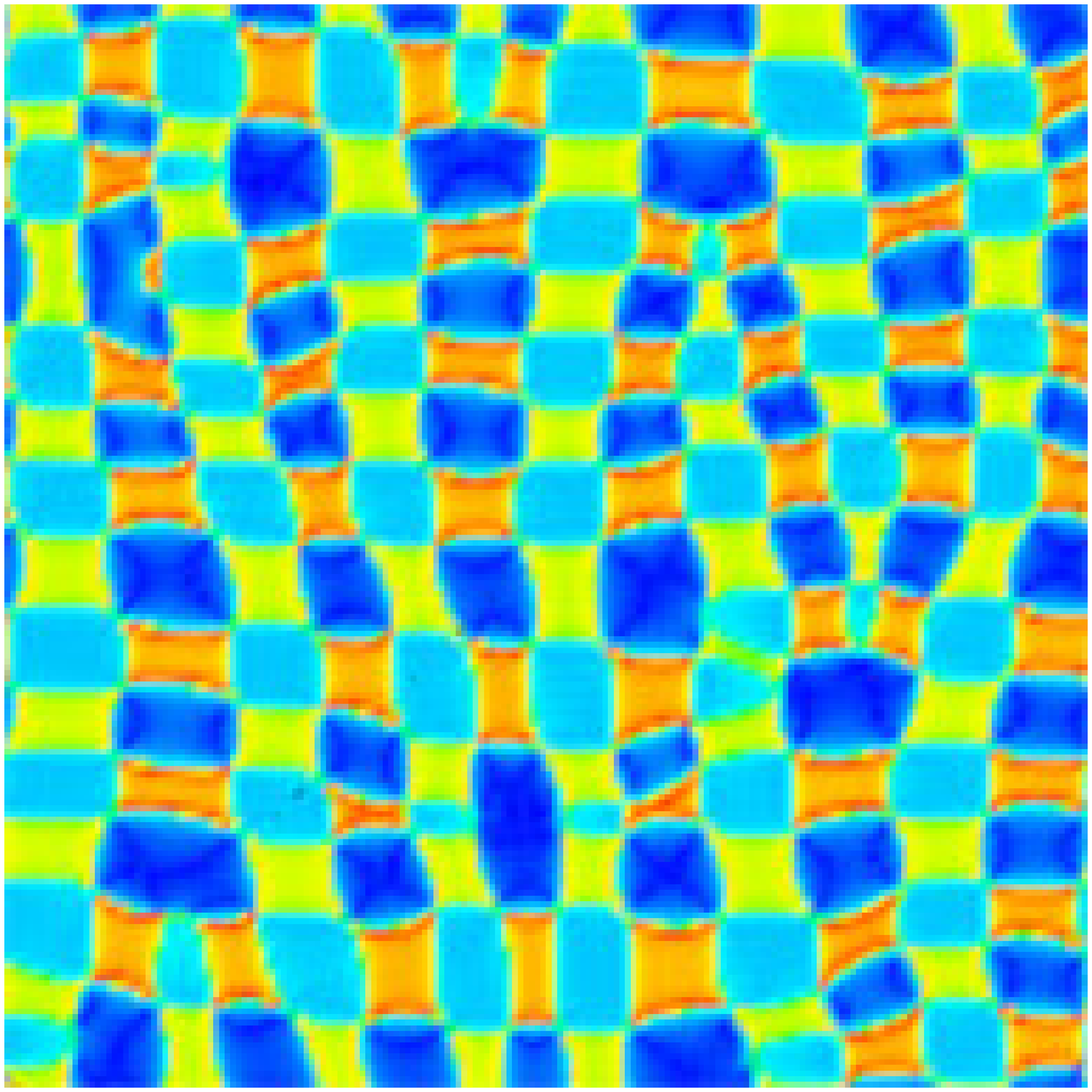}
	\put(-0.65, 2.35){\color{white}\bf(l)}
}}
\end{picture}
	\caption{\label{128-T0490_t020000}(color online) Microstructures at $T=0.49$ and $t=20~000$ for several values of $x_{12}$ and $x_{1c}$. 
Red and yellow: martensite; light and dark blue: pearlite. Unlike in Fig.~\ref{128-T0490_t000050}, $x_{1c}$ varies from left to right.}
\end{figure*}

Differentiation between $x_{12} = 0.75$ and $x_{12} = 0.85$ occurs even later (see Fig.~\ref{early-2}). Not only will the final martensite fraction be different but the orientation too will differ. At $t=1.5$, $x_{12} = 0.75$ and $x_{12} = 0.85$ are very similar, Figs.~\ref{early-075_0-1500} and~\ref{early-085_0-1500}, but for $x_{12} = 0.75$ martensite will finally align along $[1\,1]$, Fig.~\ref{128-075_0-T0490_t000012}, whereas in the case of $x_{12} = 0.85$ some austenite is retained and martensite is aligned along $[1\,\bar{1}]$, Fig.~\ref{128-085_0-T0490_t000012}.

If $x_{12}$ is smaller than about 0.8, the system will transform to pure martensite. Up to $x_{12} \approx 0.65$, the martensitic transformation completes very rapidly after it starts (by $t=1$). For larger values of $x_{12}$ (e.g.\ $x_{12}=0.75$) there is a two-step process: first $80\text{--}90\%$ martensite forms
, as in Fig.~\ref{early-075_0-1500}, and the remaining austenite transforms to martensite only much later, Fig.~\ref{128-075_0-T0490_t000012}.

At $t=0.85$, for $0.75 \le x_{12} \le 1.5$ small martensite grains are aligned along $\langle 1\,1 \rangle$.
When $x_{12} = 1.5$ this alignment is lost at later times (unlike what can be observed at lower values of $x_{12}$) but the shape of the grains is roughly conserved [compare Fig.~\ref{128-15_0-T0490_t000050} to Fig.~\ref{early-15_0-1500}].

In spite of these differences, several features seem independent of the value of $x_{12}$: (i) the transformation starts around $t=0.85$, (ii) the final martensite fraction is reached very quickly (except for $x_{12}=0.75$), and (iii) there is some form of alignment along $\langle 1\,1 \rangle$ at the initial stages. Although all martensite formation occurs between $t=0.85$ and $t=1$, there may be non-negligible microstructural evolution up to $t \approx 12$, i.e.\ there can be two steps: first martensite forms very fast (within about 0.15 time units) and then its structure evolves at constant martensite fraction (which can take up to 10 time units).

\section{\label{sec-x1c}Effect of pearlite lattice mismatch \texorpdfstring{($x_{1c}\ne0$)}{(x{1c} > 0)}}
The term $x_{1c}\,c$ in Eq.~(\ref{eq-g_el}) corresponds to a lattice mismatch between ferrite and cementite: one expands and the other contracts; these strains cancel out and there is no net volume change associated with the austenite--pearlite transformation. Nevertheless, the coherency strains generated by this volume change crucially influence the microstructures and TTT diagrams.

\subsection{TTT diagram}
Figure~\ref{128-0_1-TTT} shows the TTT diagram corresponding to $x_{12}=0$ and $x_{1c}=1$. The main difference from the TTT diagram obtained with $x_{12}=x_{1c}=0$, Fig.~\ref{128-0_0-TTT}, is that martensite can be found above $T=0.5$. There is then a possibility to observe a coexistence of martensite and pearlite. Indeed, due to the coherency strains and elastic compatibility, a mixture of pearlite and martensite is more stable than pearlite on its own (this remains true up to $T \approx 0.7$). 

This splitting between pure pearlite and pearlite--martensite mixture of what is the pearlite region in Fig.~\ref{128-0_0-TTT} is akin to the pearlite--bainite transition in steel. However, the corresponding microstructure, shown in Fig.~\ref{128-0_1-T0490_t020000}, is clearly different from that of bainite: in bainite structural distortion and diffusion occur in the same domain, whereas in the microstructure shown here diffusive and displacive mechanisms are involved in physically distinct regions.

\subsection{\label{x1c-microstructures}Microstructures}
Figure~\ref{128-T0490_t020000} shows the effect of $x_{1c}$ on the microstructure. As mentioned in Sec.~\ref{sec-x12}, $x_{12}$ helps to retain austenite at short times: Figs.~\ref{128-0_0-T0490_t000050} and~\ref{128-0_1-T0490_t000050} (which correspond to $x_{12}=0$) show pure martensite whereas in Figs.~\ref{128-085_0-T0490_t000050}--\ref{128-15_0-T0490_t000050} and~\ref{128-085_1-T0490_t000050}--\ref{128-15_1-T0490_t000050} (corresponding to non-zero $x_{12}$) there is a mixture of martensite and austenite. The long-time microstructures on the other hand are controlled by the value of $x_{1c}$: in Figs.~\ref{128-0_0-T0490_t020000} and~\ref{128-1_0-T0490_t020000}, $x_{1c}=0$ and there is only pearlite whereas Figs.~\ref{128-0_025-T0490_t020000}--\ref{128-0_15-T0490_t020000} and~\ref{128-1_025-T0490_t020000}--\ref{128-1_15-T0490_t020000} ---which correspond to $x_{1c}>0$--- show both pearlite and martensite. These pearlite--martensite microstructures do not disappear in longer simulations. $x_{1c}$ stabilizes martensite at long times for a similar reason why $x_{12}$ stabilizes austenite at short times: when $x_{1c}=1$ a purely pearlitic system is unstable as the lattice mismatch strains generated by the phase separation tend to cause a stress-induced martensitic transformation.

\begin{figure}
\centering
\setlength{\unitlength}{1cm}
\begin{picture}(5.9,2.8)(.2,0)
\subfigure{
	\label{128-0_1-T0680_t040000}
    \includegraphics[height=2.8cm]{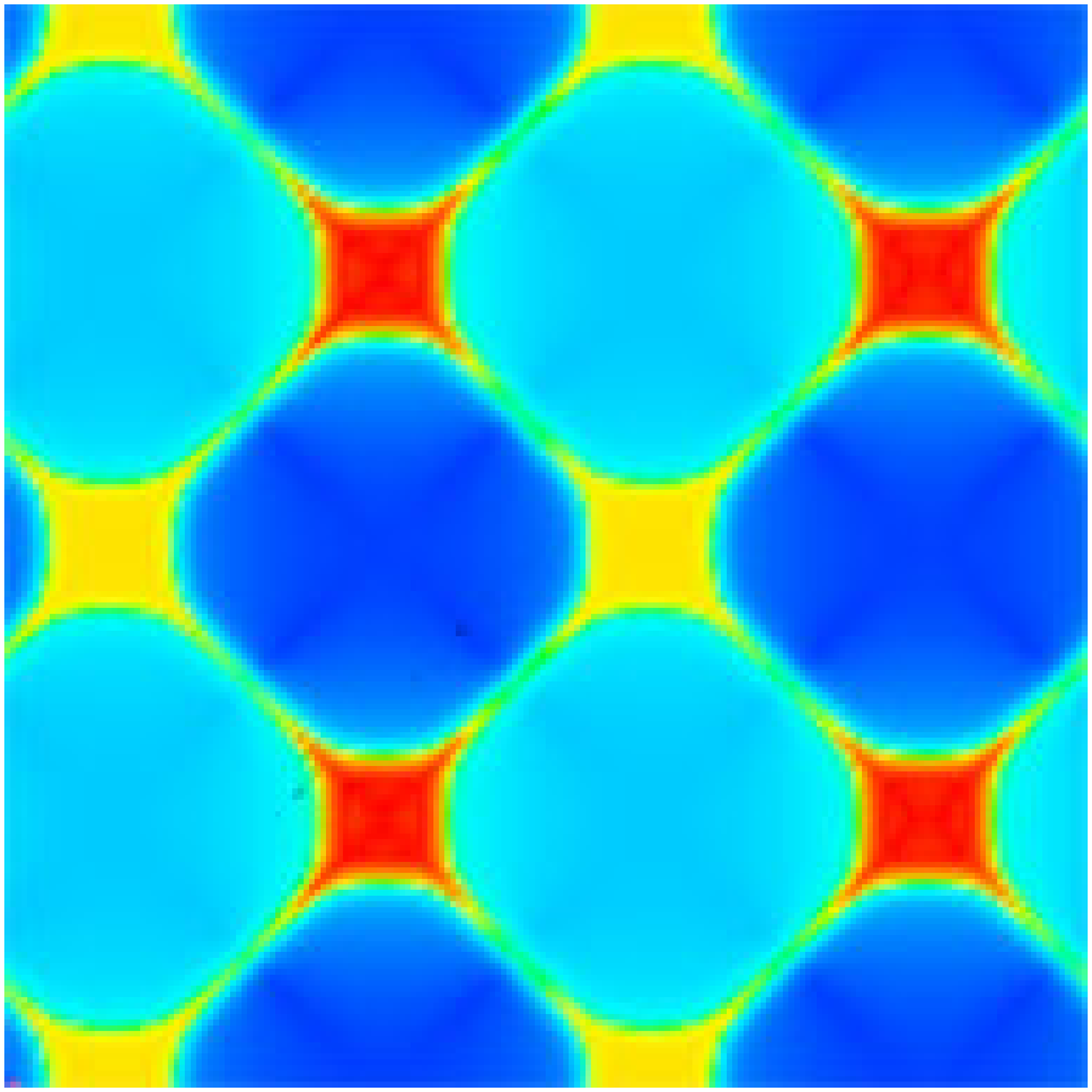}
	\put(-0.65,2.35){\color{white}\bf(a)}
}
\subfigure{
	\label{128-1_1-T0680_t200000}
    \includegraphics[height=2.8cm]{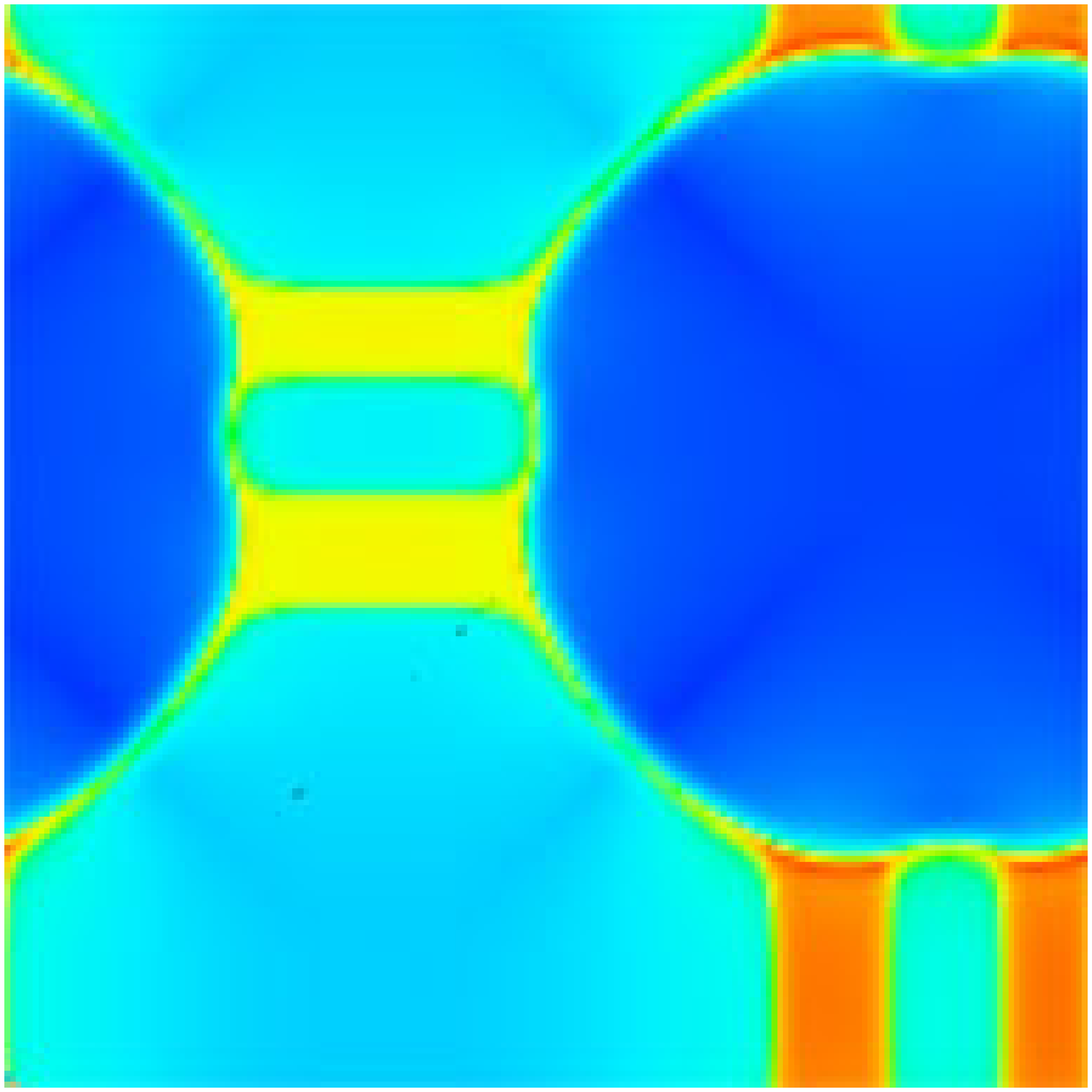}
	\put(-0.65,2.35){\color{white}\bf(b)}
}
\end{picture}
	\caption{\label{128-x_1-T0680}(color online) Microstructure at $T=0.68$ and $x_{1c}=1$. (a): $x_{12}=0$ and $t=40~000$, (b): $x_{12}=1$ and $t=200~000$.}
\end{figure}

Mixed microstructures can exist up to $T \approx 0.7$. Figure~\ref{128-x_1-T0680} shows that at $T=0.68$, the pearlite--martensite mixtures are more regular than at lower temperature, Fig.~\ref{128-0_1-T0490_t020000} and~\ref{128-1_1-T0490_t020000}. The greater diffusion allows the system to evolve towards lower minima, i.e.\ more regular states. However, one should also note that this kind of self-assembly is due in part to the periodic boundaries which allow only discrete periods (128, 64, etc.).

Figures~\ref{128-0_1-T0850_t002000} and~\ref{128-1_1-T0800_t010000} show that if $x_{1c}=1$, pearlite is textured [compare to Fig.~\ref{coarse-pearlite}]. If $x_{1c} \ne 0$, one component of pearlite has a positive value of $e_1$ and the other a negative value. Elastic compatibility then dictates that the interfaces be along $\langle 1\,1 \rangle$ as it is the case for martensite (see Sec.~\ref{sec-orientation}).

\subsection{Both \texorpdfstring{$x_{12}\ne0$ and $x_{1c}\ne0$}{x{12} > 0 and x{1c} > 0}}
Figure~\ref{128-1_1-TTT} shows the TTT diagram for $x_{12}=x_{1c}=1$. It includes the features from the TTT diagrams obtained with $x_{12}=1$ and $x_{1c}=0$ [Fig.~\ref{128-1_0-TTT}], and with $x_{12}=0$ and $x_{1c}=1$ [Fig.~\ref{128-0_1-TTT}]: there exist martensite start and finish temperatures, martensite can be found above $T=0.5$, and there is no discontinuity across ${T=0.5}$.

\begin{figure}
\centering
\setlength{\unitlength}{1cm}
\begin{picture}(8.5,2.75)(.1,0)
\subfigure{
    \label{128-1_1-T0490_t001000}
    \includegraphics[height=2.75cm]{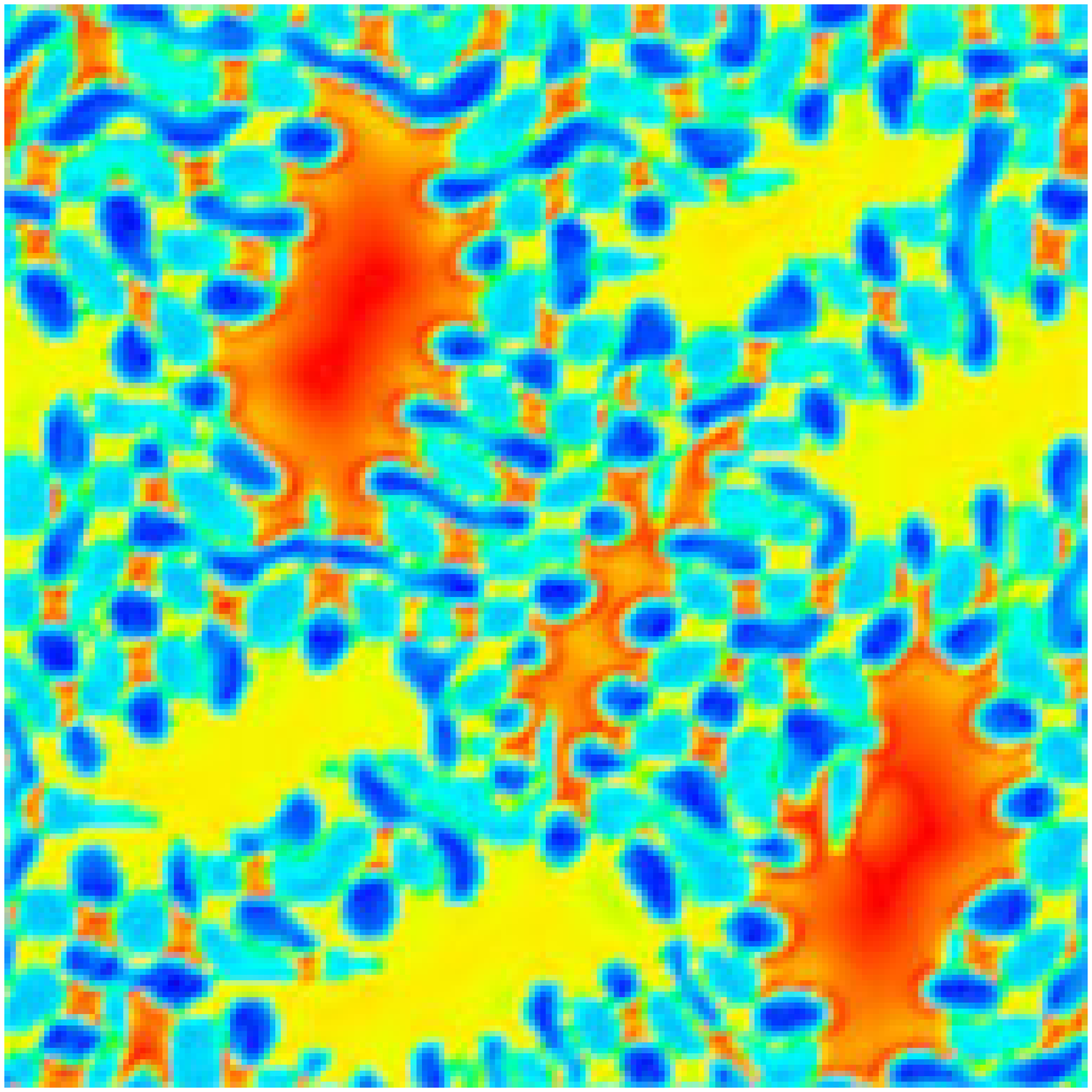}
	\put(-0.65,2.3){\bf(a)}
}\subfigure{
    \label{128-1_1-T0490_t001500}
    \includegraphics[height=2.75cm]{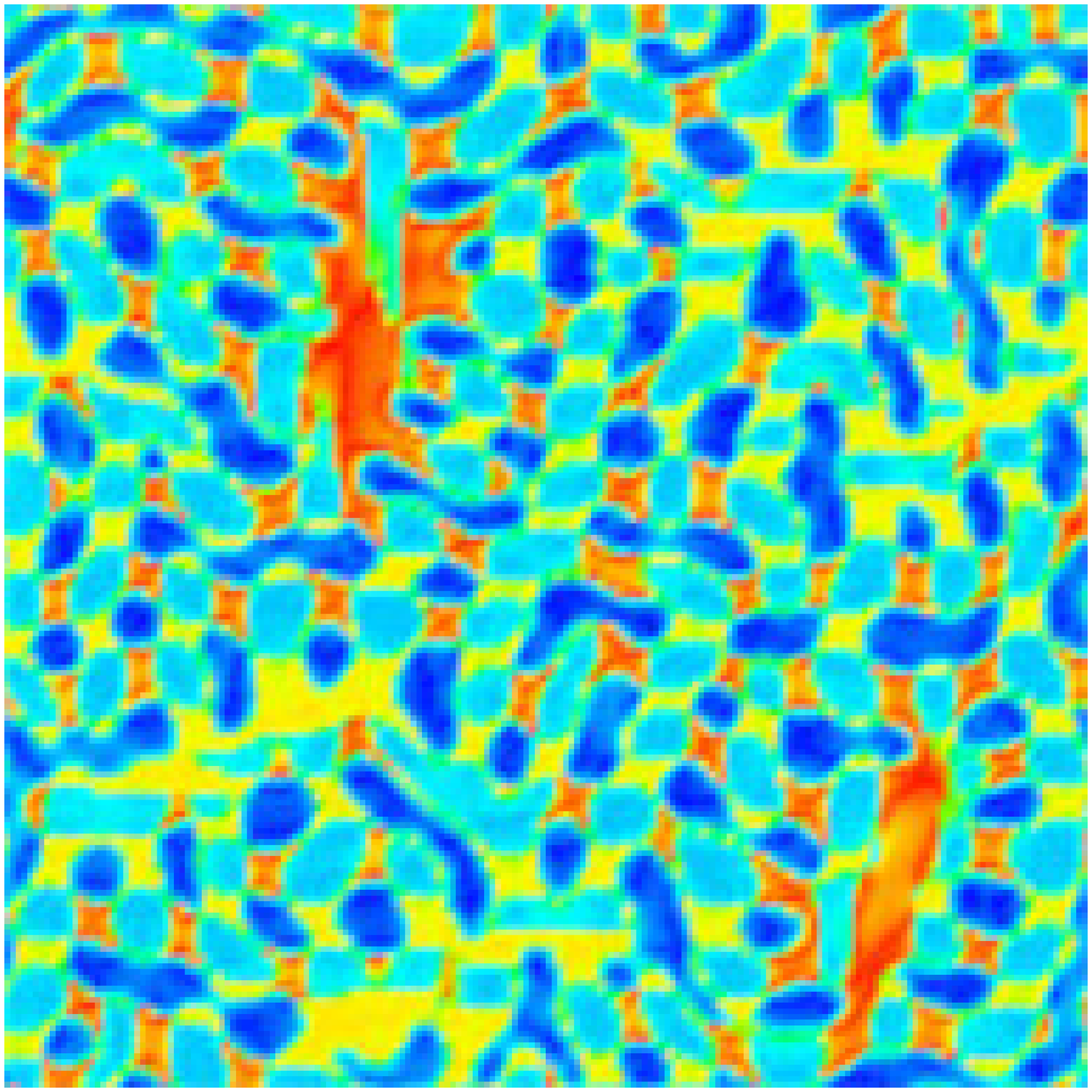}
	\put(-0.65,2.3){\bf(b)}
}\subfigure{
    \label{128-1_1-T0490_t002000}
    \includegraphics[height=2.75cm]{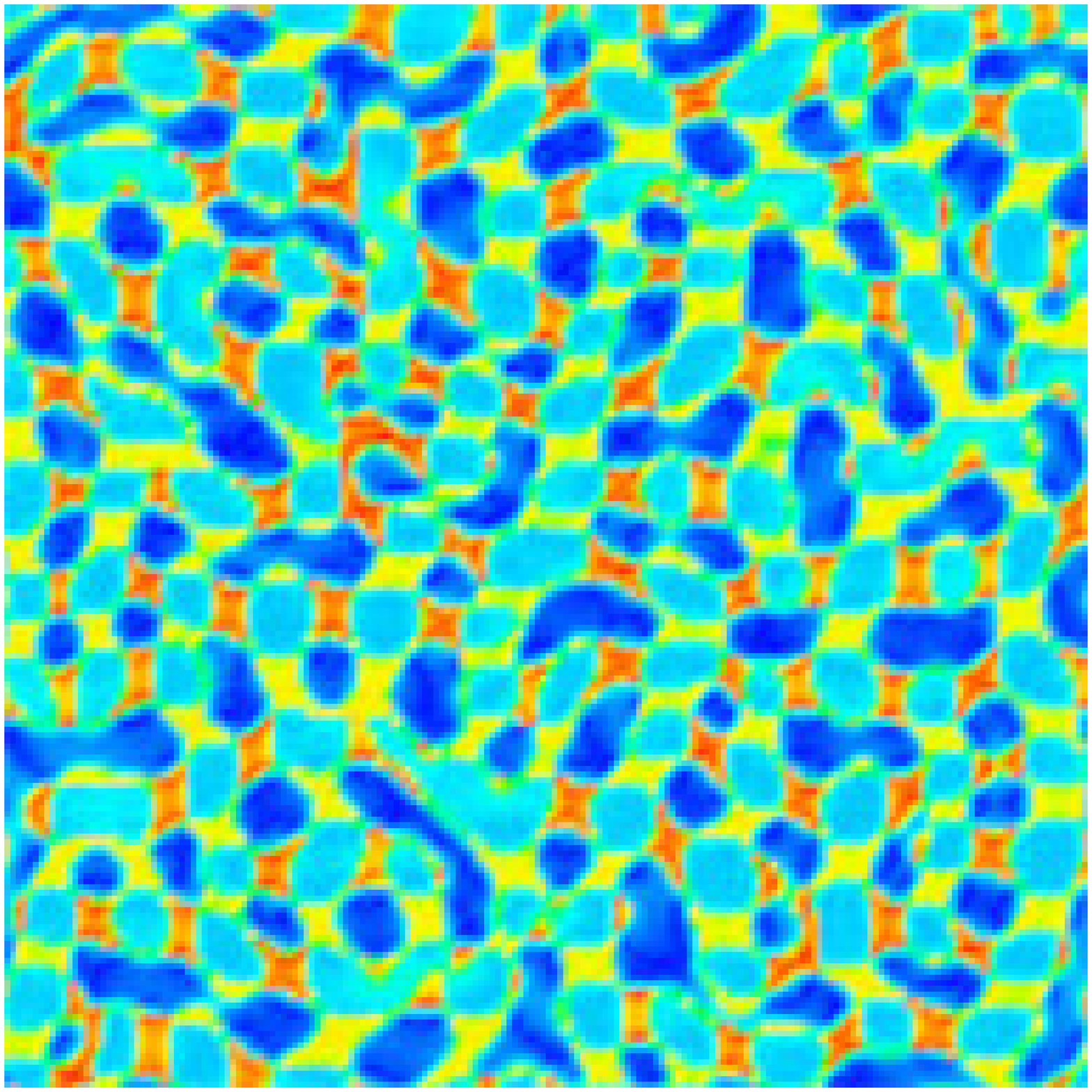}
	\put(-0.7,2.3){\color{white}\bf(c)}
}
\end{picture}
	\caption{\label{128-1_1}(color online) Microstructures for $x_{12}=x_{1c}=1$ at $T=0.49$: (a) $t=1~000$, (b) $t=1~500$, and (c) $t=2~000$. They correspond to points~k, l, and m in Fig.~\ref{128-1_1-TTT}. Red and yellow: martensite; light and dark blue: pearlite.}
\end{figure}

Figures~\ref{128-1_1-T0490_t000050}, \ref{128-1_1-T0490_t001000}--\ref{128-1_1-T0490_t002000}, and~\ref{128-1_1-T0490_t020000} show the time evolution of the microstructure at $T=0.49$ for $x_{12}=x_{1c}=1$. By $t = 1~000$, pearlite nucleated in the retained austenite and new martensite formed where there was none before (pure pearlite cannot form  due to the lattice mismatch stresses it generates), Fig.~\ref{128-1_1-T0490_t001000}. After austenite disappears pearlite grows at the expense of the large grains of `primary' martensite
, Fig.~\ref{128-1_1-T0490_t001500}, which finally disappear, Fig.~\ref{128-1_1-T0490_t002000}. The microstructure then coarsens, Fig.~\ref{128-1_1-T0490_t020000}.

From large martensite grains alternating with areas completely devoid of martensite at $t=50$ [Fig.~\ref{128-1_1-T0490_t000050}], the system evolves to a state where martensite is more homogeneously distributed [Fig.~\ref{128-1_1-T0490_t002000}] through a double mechanism of martensite formation and martensite destruction, and finally to a coarsened microstructure, Fig.~\ref{128-1_1-T0490_t020000}. The feature size thus goes from large to small to medium with increasing time. Consequently the `final' structure is independent of the initial one, both in terms of grain size and interface orientation [compare Fig.~\ref{128-1_1-T0490_t020000} to Fig.~\ref{128-1_1-T0490_t000050}].

\begin{figure}
\centering
\setlength{\unitlength}{1cm}
\begin{picture}(8.5,2.75)(.1,0)
\subfigure{
    \label{128-1_1-T0600_t000050} 
    \includegraphics[height=2.75cm]{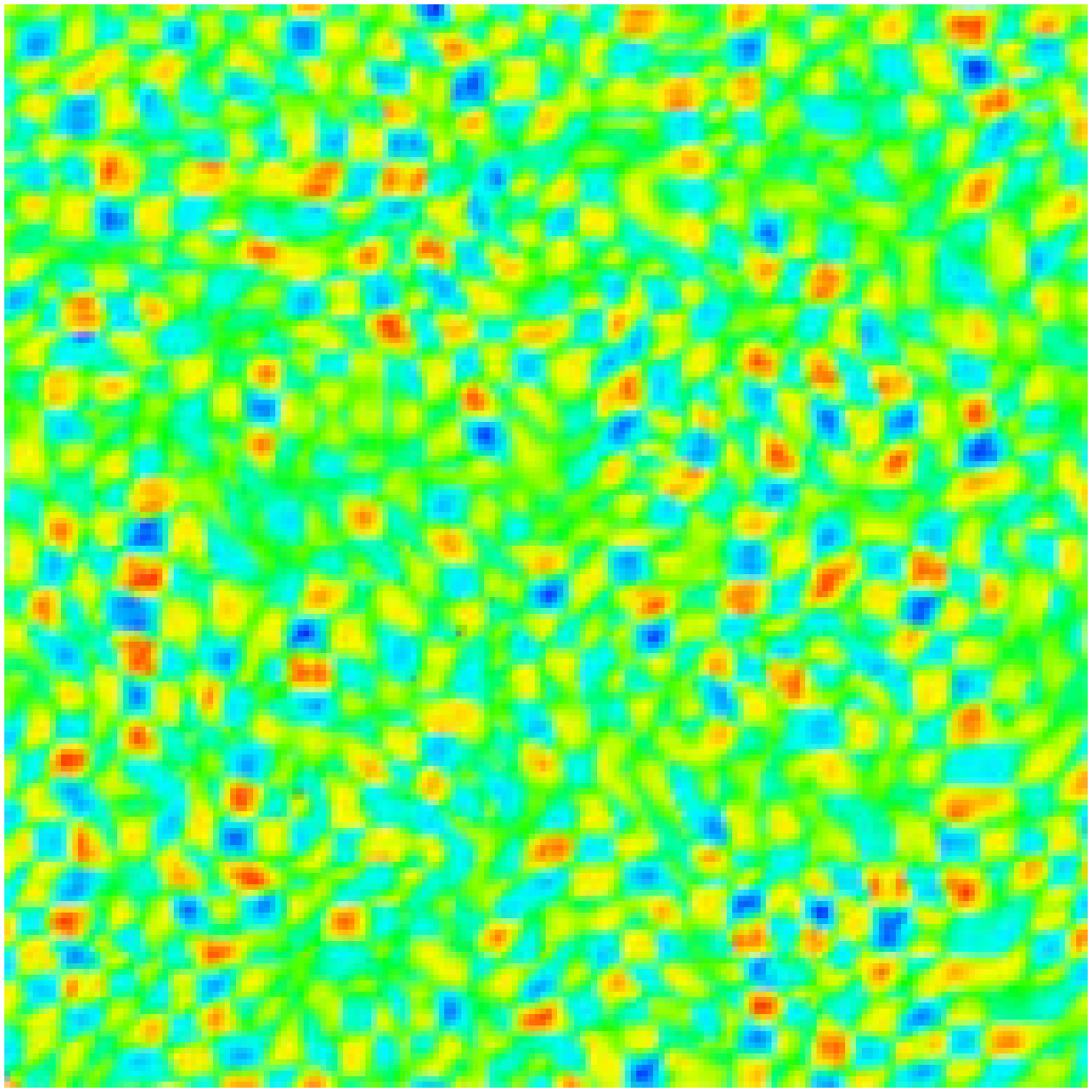}
	\put(-0.65, 2.3){\bf(a)}
}\subfigure{
    \label{128-1_1-T0600_t000100}
    \includegraphics[height=2.75cm]{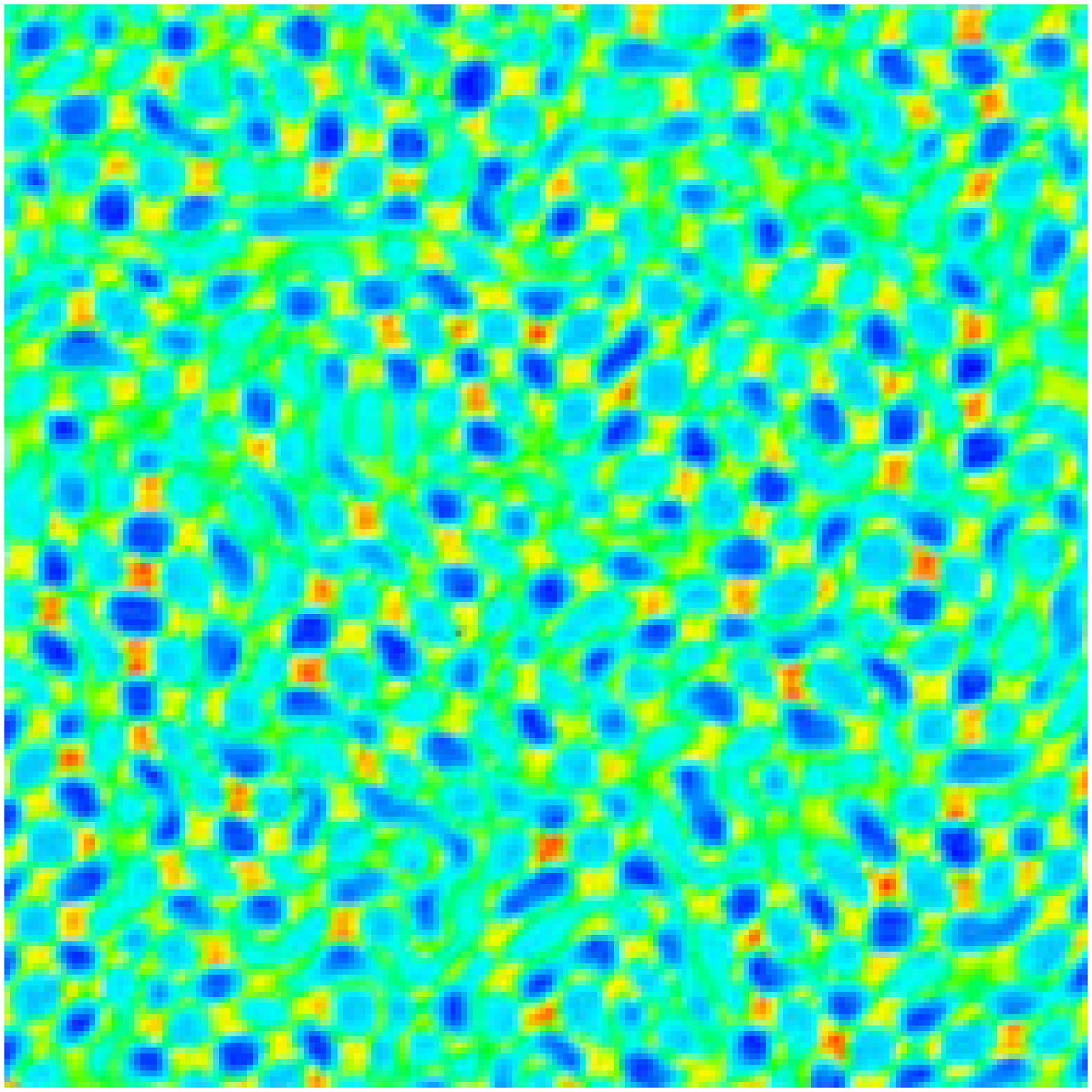}
	\put(-0.65, 2.3){\bf(b)}
}\subfigure{
    \label{128-1_1-T0600_t000200}
    \includegraphics[height=2.75cm]{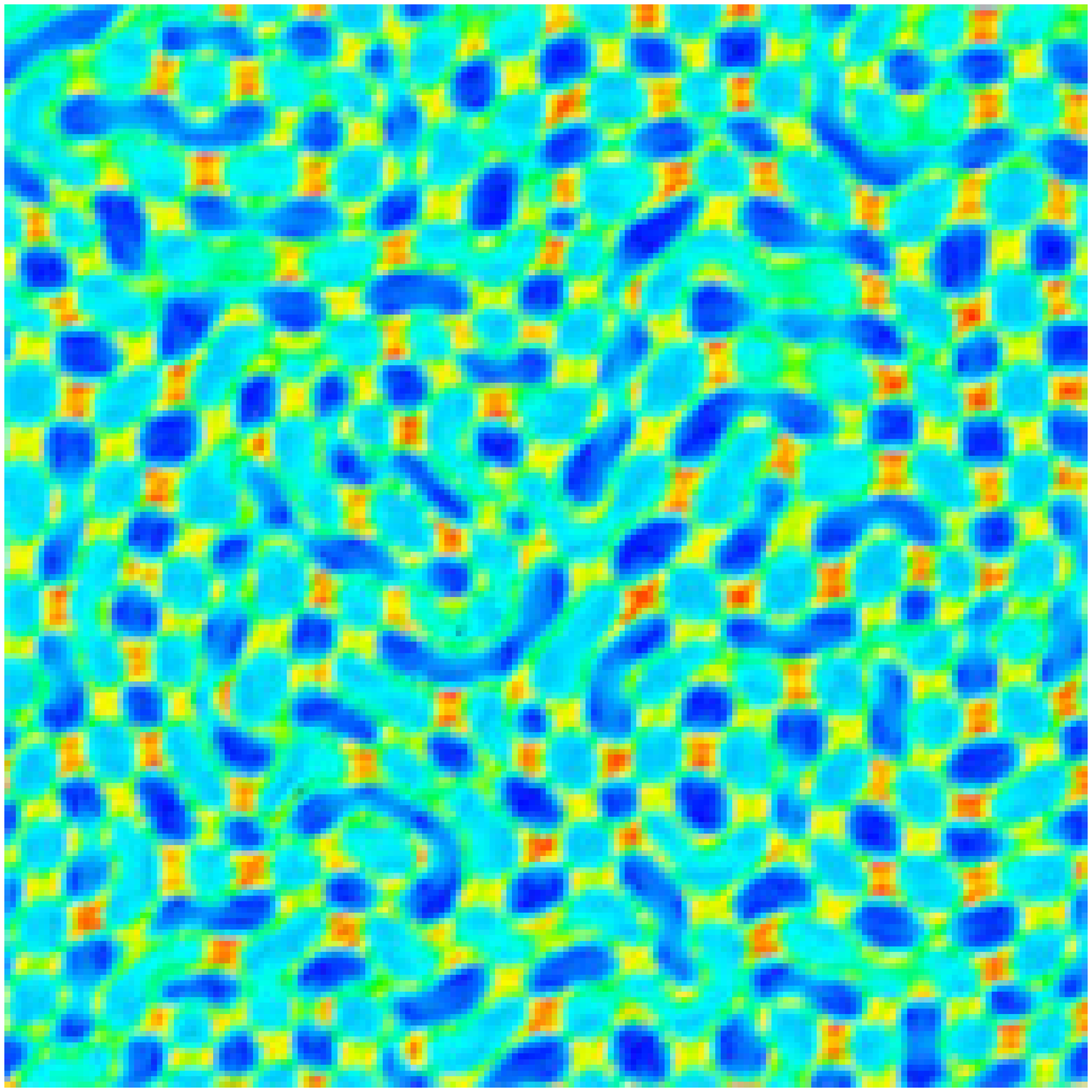}
	\put(-0.65, 2.3){\color{white}\bf(c)}
}
\end{picture}
	\caption{\label{128-1_1-T0600}(color online) Microstructures at $T=0.6$ for $x_{12} = x_{1c} = 1$. (a) $t=50$, (b): $t=100$, and (c): $t=200$. They correspond to points~h, i, and~j in Fig.~\ref{128-1_1-TTT}. Red and yellow: martensite; light and dark blue: pearlite; green: austenite.}
\end{figure}

\subsection{Cooperative \texorpdfstring{pearlite--martensite}{pearlite-martensite} nucleation}
Below $T=0.5$ martensite formation is followed by pearlite nucleation in a clearly sequential process. If $x_{1c}=1$, at temperatures between $T=0.5$ and $T \approx 0.7$ the nucleations of pearlite and martensite are simultaneous. In the case of $x_{12}=x_{1c}=1$, at $T=0.6$ and $t=50$ the system is mostly made of austenite, with a few pearlite and martensite nuclei, Fig.~\ref{128-1_1-T0600_t000050}. At $t=100$, the system is made of small pearlite and martensite grains, Fig.~\ref{128-1_1-T0600_t000100}. By $t=200$ austenite has disappeared, Fig.~\ref{128-1_1-T0600_t000200}, and the microstructure is similar to what was obtained at lower temperature, Fig.~\ref{128-1_1-T0490_t002000}, but the path is noticeably different. At lower temperature martensite forms first then pearlite nucleates at a later time and at higher temperature pearlite forms first and martensite nucleates after the system is mostly pearlitic.


\begin{figure*}
\centering
\setlength{\unitlength}{1cm}
\begin{picture}(17.6, 13.)(-.5,0)
\shortstack[c]{
\subfigure{
	\put(-0.5, 2.7){\rotatebox{90}{$x_{1c}=0$}}
	\put(4, 6.3){{$x_{12}=0$}}
    \label{128-0_0-CCT}
    \includegraphics[width=8cm]{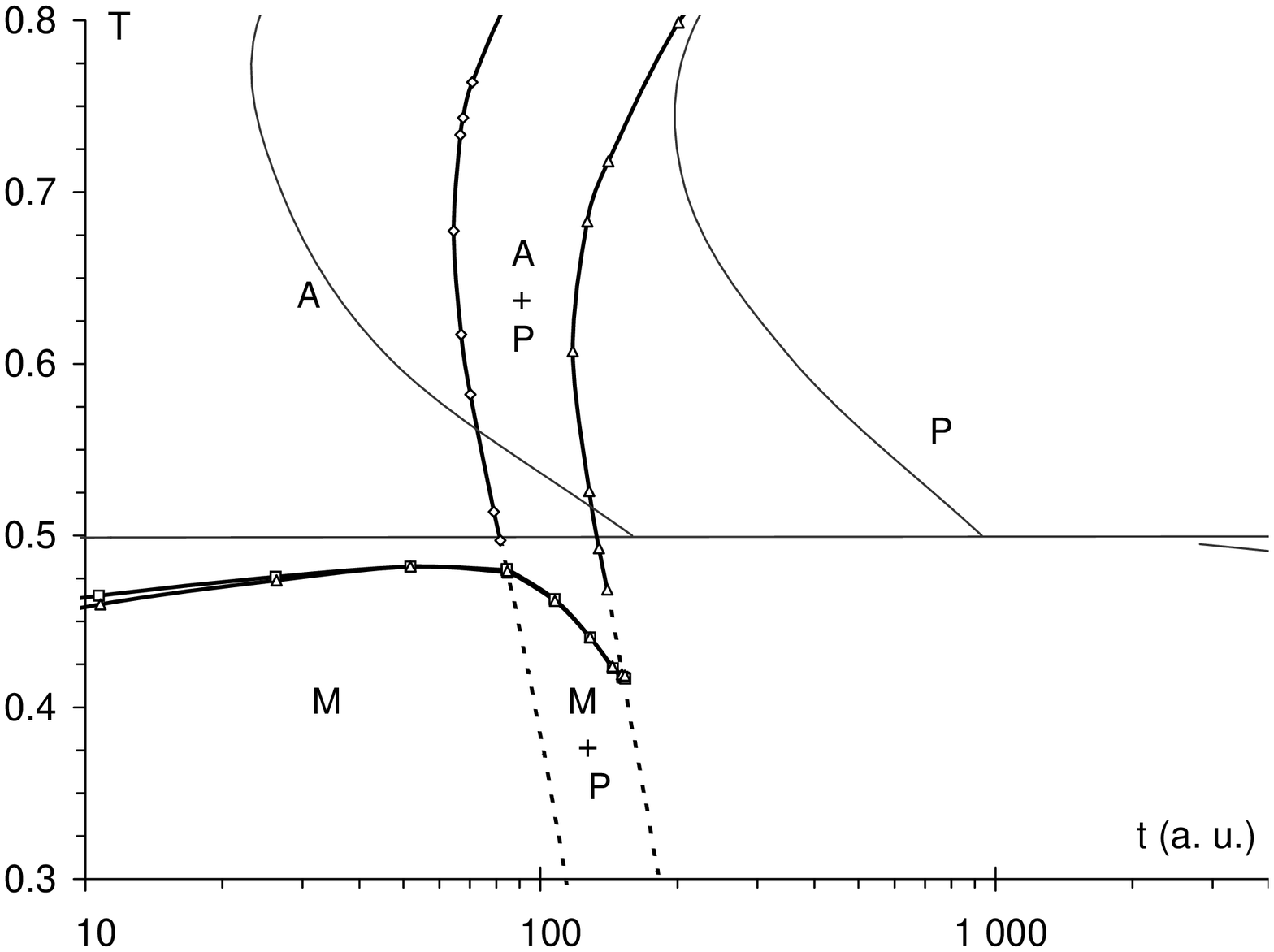}
	\put(-0.8, 5.5){\bf(a)}
}\quad
\subfigure{
	\put(4, 6.3){{$x_{12}=1$}}
    \label{128-1_0-CCT}
    \includegraphics[width=8cm]{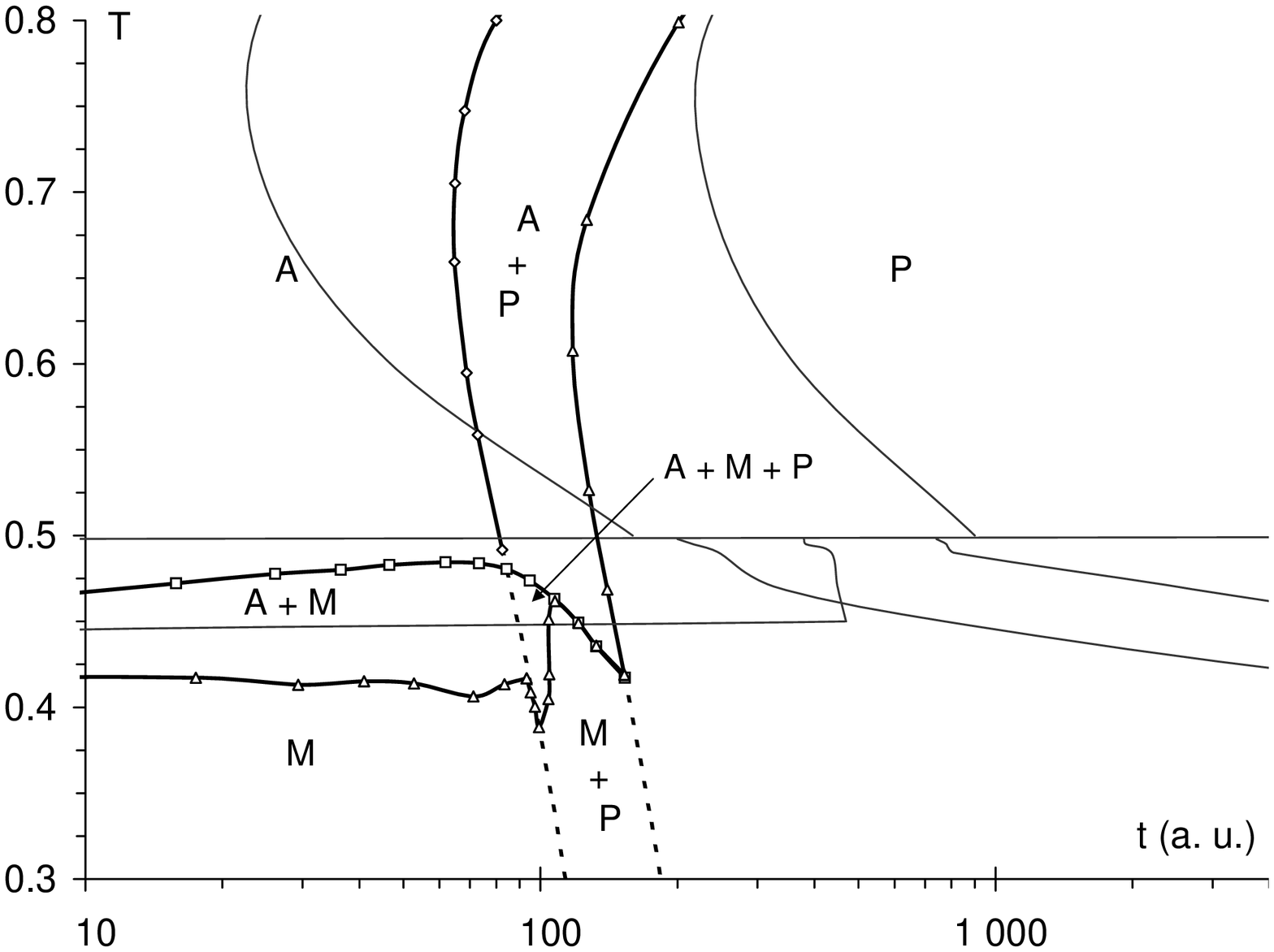}
	\put(-0.8, 5.5){\bf(b)}
}\\
\subfigure{
	\put(-0.5, 2.7){\rotatebox{90}{$x_{1c}=1$}}
    \label{128-0_1-CCT}
    \includegraphics[width=8cm]{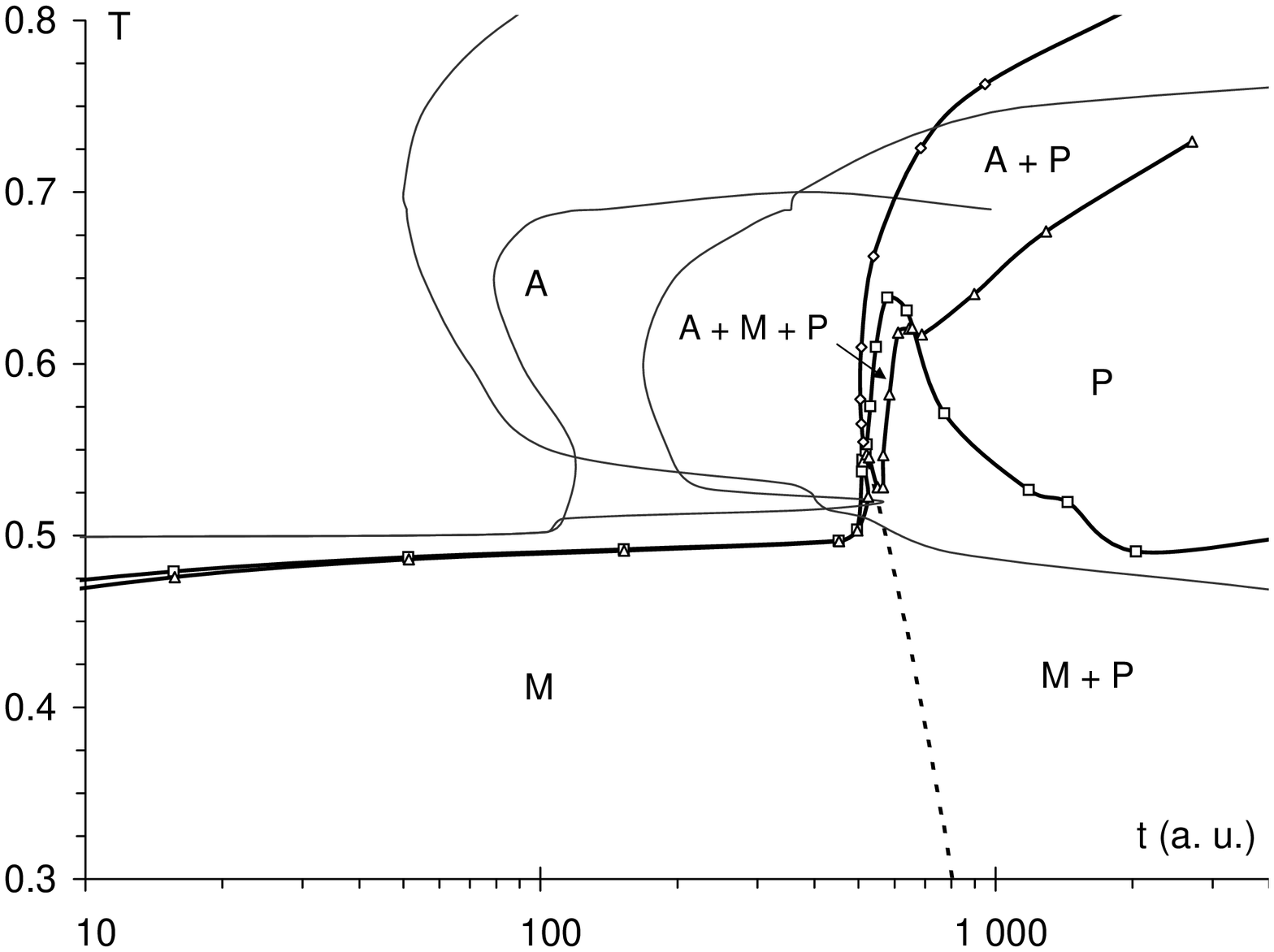}
	\put(-0.8, 5.5){\bf(c)}
}\quad
\subfigure{
    \label{128-1_1-CCT}
    \includegraphics[width=8cm]{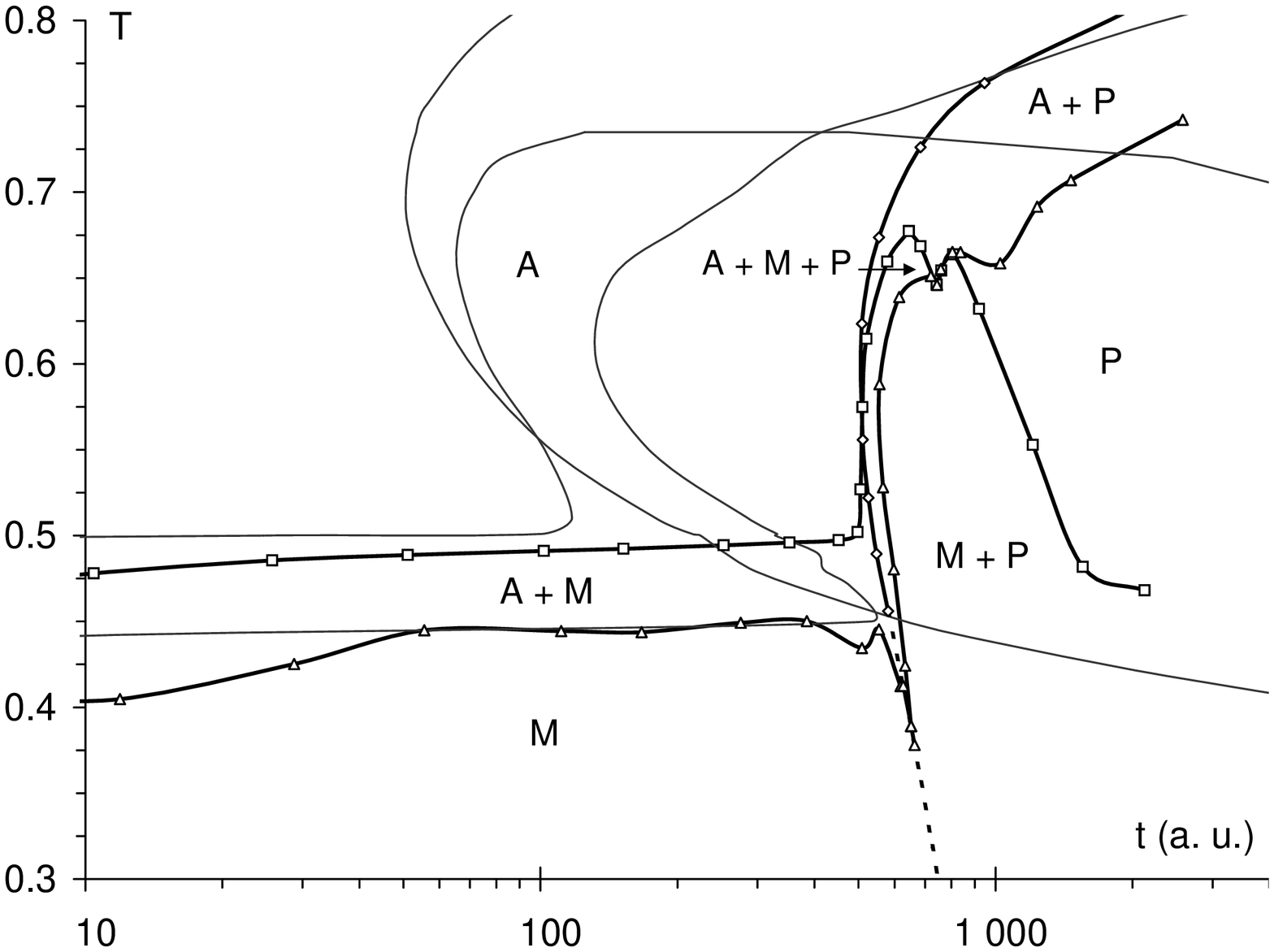}
	\put(-0.8, 5.5){\bf(d)}
}}
\end{picture}
	\caption{\label{128-CCT}Continuous cooling transformation diagrams for several values of $x_{12}$ and $x_{1c}$. A: at least 10\% austenite, M: at least 10\% martensite, and P: at least 10\% pearlite. The thin lines correspond to the TTT diagrams of Fig.~\ref{128-TTT}.}
\end{figure*}

\section{\label{CCT}Continuous cooling transformations}
So far, we have focused on isothermal transformations (TTT diagrams). However, these are not the most practical heat treatments. Instead of quenching the system and then holding the temperature constant, it is more natural to continuously decrease the temperature. If this procedure is repeated for various constant cooling rates, one obtains continuous cooling transformation (CCT) diagrams. The initial temperature is taken to be $T=1$ in all simulations. As for TTT diagrams, we record and plot the times and temperatures corresponding to 10\% martensite, 10\% pearlite, and 10\% austenite. Figure~\ref{128-CCT} shows the resulting CCT diagrams for several values of $x_{12}$ and $x_{1c}$.

Slow cooling rates, which correspond to a quasistatic situation, give rise to pearlite. On faster cooling, pearlite does not have time to form and martensite forms instead. One can notice that martensite forms at a temperature lower than the martensite start temperature of the TTT diagrams. This is because no martensite will form until $T$ becomes smaller than 0.5. Then it takes a finite amount of time for martensite to form, during which temperature continues to decrease. This effect is more noticeable for very fast cooling and the discrepancy becomes smaller and smaller as the cooling rate decreases.

The two dotted lines in Fig.~\ref{128-0_0-CCT} correspond to the critical cooling rates required to obtain pure martensite and pure pearlite, respectively. At intermediate cooling rates, cooling is slow enough for pearlite to form but too fast to allow the transformation of austenite to pearlite to complete. At low temperature this retained austenite transforms to martensite, leading to the martensite--pearlite region seen in Fig.~\ref{128-0_0-CCT}. One can notice that for long times, the austenite-to-martensite and austenite-to-pearlite lines tend asymptotically to their isothermal counterparts.


For $x_{12}=1$ and $x_{1c}=0$ [Fig.~\ref{128-1_0-CCT}], there is a martensite finish line, as was already observed in the case of isothermal transformations. 
One can also notice an austenite--pearlite--martensite mixture at intermediate cooling rates. No such mixture exists in the isothermal case.

Figures~\ref{128-0_1-CCT} and~\ref{128-1_1-CCT} show CCT diagrams for $x_{1c} = 1$. Like the TTT diagrams shown in Figs.~\ref{128-0_1-TTT} and~\ref{128-1_1-TTT}, they exhibit several mixed microstructures. However, they are shifted to lower temperatures and longer times. For slow cooling, first pearlite forms, leading to an austenite--pearlite mixture. Then austenite disappears, leaving pure pearlite. Finally martensite forms, which gives rise to a pearlite--martensite mixture. On faster cooling, martensite and pearlite form cooperatively, as was already observed in the isothermal case. Soon after pearlite and martensite form, austenite disappears and the final microstructure is again pearlite plus martensite.

\section{\label{sec-conclusion}Conclusion}
We presented a phase-field model which can be used to study alloys which can undergo displacive as well as diffusive transformations. It captures the important features of time--temperature--transformation diagrams, continuous cooling transformation diagrams, and microstructures. It also sheds some light on the role of the interplay between the two types of transformations in stabilizing mixed microstructures.
The existence of a martensite finish temperature (i.e.\ of retained austenite) is due to a hydrostatic strain associated with the martensitic transformation. When a strain is associated with pearlite formation, martensite and pearlite form cooperatively at intermediate temperatures, i.e.\ in the region of the TTT diagram where bainite is typically found in steel. The model also shows that in these mixed microstructures the habit planes are different from the pure martensite case and that small differences in volume changes can have noticeable effects on the early stages of martensite formation and on the resulting microstructures.
CCT diagrams show a shift of the transformations towards longer times, consistent with experiments. They can also exhibit mixed microstructures, which cannot exist in isothermal transformations.

\acknowledgments
The authors thank Srikanth Vedantam for useful discussions and the A*Star Visiting Investigator Program for financial support.

\bibliography{mart}
\bibliographystyle{apsrev}

\end{document}